\documentclass[11pt]{article}
\usepackage[a4paper,margin=0.9in]{geometry}
\usepackage{amsmath,amssymb,mathtools,bm}
\usepackage{booktabs,longtable,array}
\usepackage{microtype}
\usepackage{xcolor}
\usepackage{enumitem}
\usepackage{hyperref}
\usepackage{slashed}
\usepackage{graphicx}
\usepackage{caption}
\usepackage{subcaption}
\usepackage{float}
\usepackage{etoolbox}
\usepackage{tikz}
\allowdisplaybreaks
\hypersetup{hidelinks,hypertexnames=false}

\definecolor{papergreen}{RGB}{0,0,0}
\definecolor{paperred}{RGB}{0,0,0}
\colorlet{orange}{black}
\colorlet{red}{black}
\colorlet{revisionolive}{black}

\newcommand{\dd}{\mathrm d}
\newcommand{\Tr}{\operatorname{Tr}}

\newcommand{\Disc}{\operatorname{Disc}}
\newcommand{\Eone}{E_1}
\newcommand{\Rop}{\mathcal R}
\newcommand{\Fop}{\mathcal F}
\newcommand{\Kop}{\mathcal K}
\newcommand{\id}{\mathbf 1}
\newcommand{\phib}{\bar\phi}
\newcommand{\order}{\mathcal O}
\newcommand{\vev}[1]{\left\langle #1\right\rangle}
\newcommand{\ee}{\mathrm e}
\newcommand{\ii}{\mathrm i}

\newcommand{\calP}{\mathcal P}

\newcommand{\calD}{\mathcal D}

\newcommand{\ket}[1]{\left|#1\right\rangle}
\newcommand{\bra}[1]{\left\langle#1\right|}

\title{Scalar Finite-Proper-Time Field Theory as a Complete-History Spectral Calculus}

\author{
Mustafa Bakr$^{1,*}$ and Tongyu Zhang$^{1}$\\[0.5em]
\small $^{1}$Department of Physics, Clarendon Laboratory, University of Oxford, United Kingdom\\
\small 
Centre for Quantum Technologies, National University of Singapore, Singapore$^{*}$ \\
\small 
\texttt{mustafa.bakr@physics.ox.ac.uk}
}

\date{}

\newcommand{\simpleLineDiagram}{
\begin{tikzpicture}[baseline=-0.6ex]
    \fill (0,0) circle (1.5pt);
    \draw (0,0) -- (2,0);
    \fill (2,0) circle (1.5pt);

    \node at (0,-0.35) {$x$};
    \node at (2,-0.35) {$y$};
\end{tikzpicture}
}

\newcommand{\insertedLineDiagram}{
\begin{tikzpicture}[baseline=-0.6ex]
    \fill (0,0) circle (1.5pt);
    \draw (0,0) -- (4,0);
    \fill (1.4,0) circle (1.5pt);
    \fill (2.6,0) circle (1.5pt);
    \fill (4,0) circle (1.5pt);

    \node at (0,-0.35) {$x$};
    \node at (4,-0.35) {$y$};
    \node at (1.4,-0.6) {$(x_1,x_2)$};
    \node at (2.6,-0.6) {$(x_3,x_4)$};
\end{tikzpicture}
}

\newcommand{\longLineDiagram}{
\begin{tikzpicture}[baseline=-0.6ex]
    \draw (0,0) -- (3.2,0);

    \draw (4.2,0) -- (6,0);

    \fill (0,0) circle (1.5pt);
    \fill (1.7,0) circle (1.5pt);
    \fill (3.0,0) circle (1.5pt);
    \fill (5.0,0) circle (1.5pt);
    \fill (6,0) circle (1.5pt);

    \node at (-0.25,0) {$x$};
    \node at (6.25,0) {$y$};

    \node at (1.7,-0.45) {$(2)$};
    \node at (3.0,-0.45) {$(2)$};
    \node at (5.0,-0.45) {$(2)$};

    \node at (3.7,0) {$\cdots$};
\end{tikzpicture}
}

\newcommand{\diagramOne}{
\begin{tikzpicture}[baseline=-0.6ex]
    \draw (0,0.5) -- (3,0.5);
    \fill (0,0.5) circle (1.5pt);
    \fill (1,0.5) circle (1.5pt);
    \fill (2,0.5) circle (1.5pt);
    \fill (3,0.5) circle (1.5pt);

    \node at (-0.25,0.5) {$x$};
    \node at (3.25,0.5) {$y$};

    \node at (1,-0.0) {$(2)$};
    \node at (2,-0.0) {$(2)$};

    \draw (0,-0.8) -- (3,-0.8);
    \fill (0,-0.8) circle (1.5pt);
    \fill (3,-0.8) circle (1.5pt);

    \node at (-0.25,-0.8) {$w$};
    \node at (3.25,-0.8) {$v$};
\end{tikzpicture}
}

\newcommand{\diagramTwo}{
\begin{tikzpicture}[baseline=-0.6ex]
    \draw (0,0.5) -- (3,0.5);
    \fill (0,0.5) circle (1.5pt);
    \fill (1.5,0.5) circle (1.5pt);
    \fill (3,0.5) circle (1.5pt);

    \node at (-0.25,0.5) {$x$};
    \node at (3.25,0.5) {$y$};
    \node at (1.5,0.0) {$(2)$};

    \draw (0,-0.8) -- (3,-0.8);
    \fill (0,-0.8) circle (1.5pt);
    \fill (1.5,-0.8) circle (1.5pt);
    \fill (3,-0.8) circle (1.5pt);

    \node at (-0.25,-0.8) {$w$};
    \node at (3.25,-0.8) {$v$};
    \node at (1.5,-1.3) {$(2)$};
\end{tikzpicture}
}

\newcommand{\diagramThree}{
\begin{tikzpicture}[baseline=-0.6ex]
    \draw (0,0.5) -- (3,0.5);
    \fill (0,0.5) circle (1.5pt);
    \fill (3,0.5) circle (1.5pt);

    \node at (-0.25,0.5) {$x$};
    \node at (3.25,0.5) {$y$};

    \draw (0,-0.8) -- (3,-0.8);
    \fill (0,-0.8) circle (1.5pt);
    \fill (1,-0.8) circle (1.5pt);
    \fill (2,-0.8) circle (1.5pt);
    \fill (3,-0.8) circle (1.5pt);

    \node at (-0.25,-0.8) {$w$};
    \node at (3.25,-0.8) {$v$};

    \node at (1,-1.3) {$(2)$};
    \node at (2,-1.3) {$(2)$};
\end{tikzpicture}
}

\begin{document}
\maketitle

\begin{abstract}

We formulate a finite-proper-time (FPT) construction for real scalar $\lambda\phi^4$ theory in which a non-zero lower endpoint $s_0$ is retained as physical spectral data for complete virtual histories. Functional differentiation partitions a pre-existing history, while interaction sewing creates closed momentum circulations. Local momentum conservation identifies the primitive closed circulations with graph circuits; applying the retained endpoint condition gives
\[
s_e\ge0,
\qquad
\sum_{e\in c}s_e\ge s_0
\quad(c\in\mathcal C(G)).
\]
An individual edge or bridge may be arbitrarily short, but no complete closed circulation may collapse below $s_0$. This routing-independent prescription differs from damping every propagator. We prove that it gives a positive loop quadratic form and ultraviolet-finite massive amplitudes, including overlapping short-distance regions. Under a physical cut, precisely the circuit conditions crossed by the cut disappear, leaving the independently constructed daughter domains together with the ordinary pole residues and positive scalar phase space. At fixed $s_0$ we give an all-order local construction; for $\lambda\ge0$, $\Kop_{\bar\phi}\ge-\partial^2+m^2$ gives a background-uniform heat-kernel bound. The open Euclidean history kernel is virtual spectral machinery rather than the physical propagator, and physical positivity is imposed on complete boundary amplitudes. After local matching, the first non-constant one-loop on-shell correction is proportional to $s_0^2(s^2+t^2+u^2)$, providing a correlated observable test of the retained scale.

\end{abstract}

\tableofcontents

\section{\texorpdfstring{\textcolor{papergreen}{Introduction and physical question}}{Introduction and physical question}}

{\color{papergreen}
Schwinger's proper-time representation writes an inverse operator as an integral over a positive parameter and places the ultraviolet problem at the lower end of that integral \cite{Schwinger1951}. Heat-kernel and worldline methods are equivalent ways of writing the same operator relation \cite{DeWitt1965,Strassler1992,Schubert2001}. Other proposals have also considered a finite or minimum proper time \cite{MaiezzaVasquez2026,Maiezza2026}. The construction considered here retains
\begin{equation}
s_0=\Lambda^{-2}>0
\label{eq:introS0}
\end{equation}
as spectral boundary data, with $\Lambda=s_0^{-1/2}$ the associated spectral scale. The central question is not whether every propagator denominator should be damped. It is which object is a complete virtual history once the operator calculus has generated insertions and local interactions have joined those histories into a graph.

At one loop the answer follows directly from the operator functions. An open heat kernel is a proper-time interval with fixed endpoints. Taking a trace closes the interval into a circle and removes the choice of a preferred starting point, changing the measure from $\dd T$ to $\dd T/T$. Choosing one reference point on the circle restores the interval measure. Functional differentiation inserts operators on that same interval or circle and partitions its original total proper time into non-negative segments. The lower endpoint therefore belongs to the total history and not to each segment created after differentiation.

For a higher-loop graph the corresponding closed object can be identified without choosing a loop-momentum routing. Orient the internal edges, let $B$ be the vertex--edge incidence matrix, and write $q$ for the internal edge-momentum assignment. At fixed external momenta, every allowed variation of $q$ satisfies
\begin{equation}
B\,\delta q=0.
\end{equation}
Such a variation is a conserved internal circulation. The support-minimal non-zero circulations are exactly the graph circuits. They are the primitive closed virtual momentum histories of the graph. If $s_e$ is the Schwinger parameter on edge $e$, the proper-time coefficient of a primitive circuit flow is
\begin{equation}
T_c=\sum_{e\in c}s_e.
\end{equation}
The complete-history principle therefore gives a routing-independent graph domain
\begin{equation}
s_e\ge0,
\qquad
T_c\ge s_0
\quad\text{for every circuit }c.
\label{eq:introCircuitDomain}
\end{equation}
A bridge belongs to no circuit and receives no independent closed-history endpoint. A propagator edge may be arbitrarily short provided that no complete closed circulation collapses below $s_0$.

This distinction is also why the theory is not obtained by multiplying every Feynman propagator by an exponential factor. The open-history operator function $\Rop_{s_0}(\Kop)$ and the closed spectral operator function $\Fop_{s_0}(\Kop)$ of the positive quadratic operator $\Kop$, defined explicitly in Sec.~\ref{sec:operatorFunctions}, remain related by the interval--circle calculus, but $\Rop_{s_0}$ is not declared to be the free Gaussian covariance of every Wick contraction. The circuit domain is instead imposed after the graph topology and local momentum-conservation constraints are known. A separately damped Gaussian covariance defines a different finite-$s_0$ model.

The circuit formulation has two consequences that can be tested independently of one another. First, every non-zero loop-momentum direction is a circulation and therefore contains a circuit whose total proper time is at least $s_0$. This gives a uniform positive quadratic form for the loop Gaussian and fixed-order ultraviolet finiteness. Second, deleting cut propagators deletes precisely the circuits that use those propagators. In a pole residue, the proper times on cut lines run to infinity, so the endpoint condition on every opened circuit becomes inactive. The surviving constraints are exactly the circuit constraints of the standalone cut graph. The usual on-shell residues and positive phase-space measure therefore multiply the same finite-proper-time subamplitudes that would be constructed without reference to the parent graph.

The construction is defined directly at finite $s_0$. At momenta much smaller than $\Lambda$, its exact amplitudes may be expanded in local powers of momenta, but this derivative expansion is a consequence of the spectral definition rather than the definition of the theory. A general low-energy effective action could choose a new coefficient at each order. Here the coefficients are correlated by the same mass $m$, quartic coupling $\lambda$, and spectral boundary $s_0$.

{\color{paperred}
The calculation follows the physical order of operations. The operator calculus and graph topology first determine the admissible virtual histories. Local fields are then smeared with smooth sources and their products are defined after the complete graph history is known. Physical states are constructed only from complete boundary amplitudes; the intermediate Euclidean history kernel is not interpreted as a particle propagator.

At every finite interaction order the circuit domain gives ultraviolet-finite graphs, ordinary bridges, and the same finite-$s_0$ subamplitudes after a physical cut. The source construction below uses the same rule at arbitrary order and preserves the local spacetime transformation laws because the endpoint depends only on scalar proper-time sums. A numerical sum at a chosen coupling is discussed separately through an explicit uniform remainder and boundary bound.
}

\subsection{\texorpdfstring{\textcolor{paperred}{Connection to boundary-selected spectra in wave and quantum problems}}{Connection to boundary-selected spectra in wave and quantum problems}}

{\color{paperred}
The same local-versus-global distinction appears in several wave and quantum problems that motivated the present construction. In cylindrical electromagnetic cavities, the local Maxwell equations admit azimuthally propagating branches whose allowed indices change when the global azimuthal boundary is changed; singular local fields can remain meaningful when the complete field has the required finite-energy behaviour \cite{BakrAmari2023Singular,BakrAmari2025Cylinder}. In the full vectorial spherical problem, continuous angular labels are handled through a spectral integral, while the companion numerical construction shows that increasing the spectral resolution improves the representation of a fixed problem rather than changing the physical boundary conditions themselves \cite{BakrVectorial2025}.

The spherical-cavity, angular-representation, wedge-quantum, and zero-frequency studies sharpen this point. A differential equation or algebraic relation can possess a larger local solution family than the globally admissible physical spectrum. Regularity, single-valuedness, finite energy, and the chosen boundary select the physical states; changing that global domain reorganises the spectrum without changing the local field equations \cite{BakrZhangAmari2026Spherical,BakrAmari2025Group,BakrAmari2025Wedge,BakrAmari2025Zero}. The analogy used here is precise but limited: $s_0$ does not create a spatial wall. It selects the admissible domain of complete virtual histories while leaving the local scalar interaction unchanged.

The same boundary-selection viewpoint was also used for dispersive circuit quantum electrodynamics (circuit-QED) measurement, where the interaction changes which dressed modes are compatible with the boundary problem \cite{BakrCQEDBoundary2025}. In finite-proper-time quantum electrodynamics (FPT-QED), the corresponding rule is imposed directly on a complete charged worldline before field insertions divide that worldline into daughter intervals \cite{BakrFPTQED2026}. The scalar theory developed here isolates the common mathematical core: first identify the complete history, then impose the retained spectral boundary on that history, and only afterwards decompose it into the pieces used in perturbative calculations.
}
}

\section{\texorpdfstring{\textcolor{paperred}{Scope, conventions, and main results}}{Scope, conventions, and main results}}

{\color{paperred}
We work mainly in $d$-dimensional Euclidean space. Repeated condensed indices imply integration. Thus
\begin{equation}
A_iB_i\equiv\int\dd^dx\,A(x)B(x),
\qquad
A_iK_{ij}B_j\equiv\int\dd^dx\,\dd^dy\,A(x)K(x,y)B(y).
\end{equation}
Appendix D gives the corresponding finite-dimensional bookkeeping. The connected source functional and its Legendre transform are defined in the usual way. Here $S[\phi]$ is the Euclidean action, $J$ is a smooth source, $\mathcal N$ is a field-independent normalisation, $\hbar$ is retained as the loop-counting parameter, and $\calD\phi$ denotes functional integration:
\begin{equation}
Z[J]=\mathcal N\int\calD\phi\,
\exp\left[-\frac1\hbar\left(S[\phi]-J_i\phi_i\right)\right],
\label{eq:ZJ}
\end{equation}
with
\begin{equation}
W[J]=\hbar\ln Z[J],
\qquad
\phib_i=\frac{\delta W}{\delta J_i},
\qquad
\Gamma[\phib]=J_i\phib_i-W[J].
\label{eq:WGammaDefs}
\end{equation}
The Legendre transform gives
\begin{equation}
\frac{\delta\Gamma}{\delta\phib_i}=J_i,
\label{eq:GammaDerivativeJ}
\end{equation}
and $W^{(2)}=(\Gamma^{(2)})^{-1}$, where the superscript $(2)$ denotes the Hessian, or second functional derivative. A bridge joining two one-particle-irreducible (1PI) pieces is therefore fixed by connected composition and is not given an additional endpoint by hand.

Throughout this paper a parenthesized numerical superscript denotes functional-derivative order, while a numerical subscript denotes loop order. Tensor indices and descriptive tags retain their usual notation. Thus
\begin{equation}
\Gamma_{\ell}^{(n)}(x_1,\ldots,x_n;\bar\phi)
\equiv
\frac{\delta^n\Gamma_{\ell}[\bar\phi]}
{\delta\bar\phi(x_1)\cdots\delta\bar\phi(x_n)},
\qquad
\Gamma^{(n)}=\sum_{\ell\ge0}\hbar^\ell\Gamma_{\ell}^{(n)},
\label{eq:GammaNotationConvention}
\end{equation}
with $\Gamma_0\equiv S$. When the retained endpoint is displayed explicitly we write $\Gamma_{\ell,s_0}^{(n)}$. Coupling order is written separately as $\mathcal O(\lambda^p)$ and is never encoded in the derivative superscript.

The retained endpoint $s_0>0$ belongs to a complete virtual history and is not taken to zero in the definition of the theory. At one loop the open interval and closed circle are two forms of the same proper-time history, and functional differentiation divides the original total proper time without assigning a new endpoint to each daughter segment. At higher loop order local momentum conservation identifies the primitive closed histories with graph circuits, so the graph domain is $s_e\ge0$ with $\sum_{e\in c}s_e\ge s_0$ for every circuit $c$. Bridges lie on no circuit and remain ordinary propagators.

The circuit inequalities give a positive loop quadratic form on every fixed graph and make massive amplitudes ultraviolet finite, including overlapping short-distance regions. A physical cut opens every circuit it intersects. The surviving conditions are exactly those of the deleted graph, so the cut contains the same finite-$s_0$ subamplitudes that are obtained by constructing the two sides independently. After local matching, the one-loop four-point function retains a correlated momentum dependence whose first non-constant on-shell term is proportional to $s_0^2(s^2+t^2+u^2)$.

For $\lambda\ge0$, $\Kop_{\phib}\ge \Kop_0$ gives a heat-kernel bound uniform in any real background. Local fields are smeared with smooth sources, products are first defined at separated points, and the finite-history factor cannot increase the ordinary scalar short-distance degree. The same vacuum, mass, field-normalisation, and quartic-coupling conditions therefore fix the local freedom. The history restriction also composes consistently when interactions are sewn and preserves the local spacetime transformation laws because it contains no preferred coordinate or direction.

The open Euclidean history kernel $\Rop_{s_0}$ is used only for virtual spectral calculations. The physical particle pole, positive phase space, and state inner product are taken from complete boundary amplitudes. Equation~\eqref{eq:causalSourceSeries} gives the interaction expansion at arbitrary order for fixed $s_0$. When the exact connected kernels obey the uniform remainder and physical-boundary bounds given later, that expansion has one numerical sum at the chosen coupling and the fixed-order cut relation passes to the summed amplitudes.

The derivations are kept within this paper. The one-loop interval--circle calculus and its repeated insertions are derived in the main text. The two-loop coefficients are obtained from the background expansion and checked by explicit six-field pairings in Appendix~A. Appendix~B gives the connected-cumulant bookkeeping, Appendix~C shows explicitly why an independently damped Gaussian model is different, and Appendix~D reduces the functional derivatives to finite-dimensional matrix calculus. Appendix~E derives the complete-history momentum factors, while Appendix~F derives the circuit, bridge, loop-positivity, and deletion statements and works through one-loop, sunset, and $K_4$ examples. The examples are included so that the reader can reconstruct the rule without relying on a diagrammatic convention that is defined elsewhere.
}

\section{Dimensions and the meaning of Schwinger proper time}

In units $\hbar=c=1$,
\begin{equation}
[x]=-1,
\qquad
[\partial]=1,
\qquad
[\phi]=\frac{d-2}{2},
\qquad
[\lambda]=4-d.
\label{eq:dimensionsBasic}
\end{equation}
The quadratic fluctuation operator has mass dimension
\begin{equation}
[\Kop]=2.
\end{equation}
Consequently the Schwinger parameter in
\begin{equation}
\Kop^{-1}=\int_0^\infty\dd s\,\ee^{-s\Kop}
\label{eq:schwingerInverseIntro}
\end{equation}
has
\begin{equation}
[s]=-2.
\label{eq:sDimension}
\end{equation}
Thus $s$ has dimensions of length squared rather than ordinary elapsed time. In first-quantised language it is the gauge-fixed modulus of a worldline, in the tradition of Fock, Nambu, and Schwinger \cite{Schwinger1951,Fock1937,Nambu1950}. With worldline coordinate $x(\sigma)$, einbein $e(\sigma)$, parameter $\sigma\in[0,1]$, and $\dot x=\dd x/\dd\sigma$, start from the reparametrisation-invariant particle action
\begin{equation}
S[x,e]=\frac12\int_0^1\dd\sigma\left[e^{-1}\dot x^2+e m^2\right]
\end{equation}
and fixing a constant einbein $e=2s$ gives the familiar heat-kernel path integral. For a classical massive saddle one may relate $s$ to a conventional proper duration by a mass-dependent rescaling, but a virtual quantum path is off shell and all moduli are integrated. The most precise interpretation is therefore:

\begin{quote}
$s$ is an auxiliary spectral parameter in operator theory and a Euclidean worldline modulus in the first-quantised representation. It is not an independently observable clock attached to a virtual particle.
\end{quote}

A finite endpoint
\begin{equation}
s\ge s_0=\Lambda^{-2}
\label{eq:s0Lambda}
\end{equation}
is consequently best stated as a modification of a spectral operator function. The worldline picture motivates the geometry, but is not required as the definition.

\subsection{Endpoint versus an auxiliary regulator}
Three mass scales must be kept conceptually separate. A temporary ultraviolet (UV) regulator $\Lambda_{\rm UV}$ is introduced to define intermediate expressions and is removed after renormalisation. A subtraction or renormalisation scale $\mu$ specifies where renormalised parameters are defined and is arbitrary: exact observables are independent of it after the running of the renormalised parameters is included. By contrast, the present construction postulates that
\begin{equation}
\Lambda\equiv s_0^{-1/2}
\end{equation}
is retained as a parameter of the spectral operator calculus itself. It is therefore not identified with either $\Lambda_{\rm UV}$ or $\mu$.

{\color{paperred}\paragraph{Retained finite endpoint.}
A finite endpoint $s_0>0$ is retained when, after the ordinary mass, field normalisation, and quartic coupling have been fixed by specified physical conditions, the operator functions $\Rop_{s_0}$ and $\Fop_{s_0}$ are kept fixed. The value $s_0=0$ is not part of the defined FPT spectral theory; the formal boundary-removal comparison $s_0\to0^+$ is used only when explicitly stated to compare with the local theory.}

Calling the retained endpoint \emph{physical} is a stronger statement than calling it finite. The test used in this paper is the following. Choose a complete set of low-energy renormalisation conditions $\mathcal C$ fixing the local parameters of the theory. If two values $s_0$ and $s'_0$ can be related, while preserving $\mathcal C$, by finite redefinitions of those parameters together with an admissible invertible field redefinition, then the endpoint does not label a distinct physical theory. If instead at least one additional observable or renormalised 1PI momentum dependence obeys
\begin{equation}
\mathcal O(p_i;\mathcal C,s_0)
\neq
\mathcal O(p_i;\mathcal C,s'_0)
\end{equation}
in a way that cannot be removed by such redefinitions, then different values of $s_0$ give physically different matched theories. {\color{paperred}The calculations below establish the residual $s_0$ dependence after local matching. No stronger claim of all-order field-redefinition inequivalence is needed for the results of this paper.}

The distinction from ordinary renormalisation-group (RG) running can also be stated dimensionlessly. If $s_0$ is held fixed while the arbitrary subtraction scale $\mu$ is varied, then $\hat s_0=\mu^2s_0$ changes only because of its canonical dimension,
\begin{equation}
\mu\frac{\dd\hat s_0}{\dd\mu}=2\hat s_0,
\qquad
\hat s_0=\mu^2s_0.
\end{equation}
This is analogous to forming the ratio of the RG scale to any fixed physical mass. It does not mean that the physical endpoint itself is being integrated out or removed.

\section{From the path integral to the exact effective-action identity}

The background-field and Legendre-transform manipulations in this section are standard; see, for example, Ref.~\cite{ZinnJustin}. They are written out because the finite-endpoint prescription will later be imposed at the level of the effective-action operator calculus rather than diagram by diagram.

\subsection{Connected functional and Legendre transform}

Differentiate Eq.~\eqref{eq:ZJ} with respect to $J_i$:
\begin{align}
\frac{\delta Z}{\delta J_i}
&=\frac1\hbar\mathcal N\int\calD\phi\,\phi_i
\exp\left[-\frac1\hbar(S-J\phi)\right],\\
\frac{\delta W}{\delta J_i}
&=\hbar\frac1Z\frac{\delta Z}{\delta J_i}
=\vev{\phi_i}_J.
\end{align}
This is the classical or mean field $\phib_i$. The connected two-point function is
\begin{equation}
\frac{\delta^2W}{\delta J_i\delta J_j}
=\frac1\hbar\left(\vev{\phi_i\phi_j}_J-\vev{\phi_i}_J\vev{\phi_j}_J\right).
\label{eq:connectedTwoPoint}
\end{equation}
The Legendre transform in Eq.~\eqref{eq:WGammaDefs} obeys
\begin{align}
\delta\Gamma
&=\delta J_i\,\phib_i+J_i\delta\phib_i
-\frac{\delta W}{\delta J_i}\delta J_i
=J_i\delta\phib_i,
\end{align}
which gives Eq.~\eqref{eq:GammaDerivativeJ}. Differentiating once more gives the inverse relation
\begin{equation}
\Gamma^{(2)}_{ik}W^{(2)}_{kj}=\delta_{ij}.
\label{eq:inverseHessians}
\end{equation}

\subsection{Exact background identity}

Use $J_i=\Gamma_{,i}[\phib]$ and shift the integration variable,
\begin{equation}
\phi_i=\phib_i+\eta_i.
\end{equation}
Then
\begin{align}
\ee^{-\Gamma[\phib]/\hbar}
&=\ee^{-J\phib/\hbar}Z[J]
\nonumber\\
&=\mathcal N\int\calD\eta\,
\exp\left[-\frac1\hbar\left(S[\phib+\eta]-\Gamma_{,i}[\phib]\eta_i\right)\right].
\label{eq:exactEffectiveIdentity}
\end{align}
Equation~\eqref{eq:exactEffectiveIdentity} is exact. It is the appropriate starting point for a loop expansion because the source term contains the full $\Gamma_{,i}$, not merely the classical derivative $S_{,i}$. This distinction removes one-particle-reducible terms from the effective action.

\section{\texorpdfstring{Background expansion of Euclidean $\lambda\phi^4$ theory}{Background expansion of Euclidean lambda phi4 theory}}

The action is
\begin{equation}
S[\phi]=\int\dd^dx\left[
\frac12(\partial_\mu\phi)(\partial_\mu\phi)
+\frac12m^2\phi^2
+\frac{\lambda}{4!}\phi^4
\right].
\label{eq:scalarAction}
\end{equation}
Set
\begin{equation}
\phi=\phib+h.
\end{equation}
The interaction expands exactly as
\begin{align}
\frac{\lambda}{4!}(\phib+h)^4
={}&\frac{\lambda}{4!}\phib^4
+\frac{\lambda}{3!}\phib^3h
+\frac{\lambda}{4}\phib^2h^2
+\frac{\lambda}{3!}\phib h^3
+\frac{\lambda}{4!}h^4.
\label{eq:quarticExpansionFull}
\end{align}
After integrating the mixed kinetic term by parts,
\begin{align}
S[\phib+h]
={}&S[\phib]
+\int\dd^dx\,h(x)\mathcal E[\phib](x)
+\frac12h_i\Kop_{\phib,ij}h_j
+S_3[h;\phib]+S_4[h],
\label{eq:backgroundExpansionFull}
\end{align}
where
\begin{equation}
\mathcal E[\phib]=
-\partial^2\phib+m^2\phib+\frac{\lambda}{3!}\phib^3,
\label{eq:classicalEOM}
\end{equation}
\begin{equation}
\Kop_{\phib}(x,y)=
\left[-\partial_x^2+m^2+\frac{\lambda}{2}\phib(x)^2\right]
\delta^{(d)}(x-y),
\label{eq:backgroundK}
\end{equation}
\begin{equation}
S_3[h;\phib]=\int\dd^dx\,\frac{\lambda}{3!}\phib h^3,
\qquad
S_4[h]=\int\dd^dx\,\frac{\lambda}{4!}h^4.
\label{eq:S3S4}
\end{equation}
The quadratic background covariance is
\begin{equation}
G_{\phib}=\Kop_{\phib}^{-1}.
\label{eq:backgroundG}
\end{equation}
Writing
\begin{equation}
\Kop_{\phib}=\Kop_0+V_{\phib},
\qquad
\Kop_0=-\partial^2+m^2,
\qquad
V_{\phib}=\frac{\lambda}{2}\phib^2,
\end{equation}
shows that
\begin{align}
G_{\phib}
&=(\Kop_0+V_{\phib})^{-1}
\nonumber\\
&=G_0-G_0V_{\phib}G_0
+G_0V_{\phib}G_0V_{\phib}G_0-\cdots.
\label{eq:NeumannBackground}
\end{align}
Thus using $G_{\phib}$ exactly resums all numbers of quadratic background insertions $\lambda\phib^2h^2/4$. Ordinary Feynman theory may instead use $G_0$ and expand in all three fluctuation vertices
\begin{equation}
\frac{\lambda}{4}\phib^2h^2,
\qquad
\frac{\lambda}{3!}\phib h^3,
\qquad
\frac{\lambda}{4!}h^4.
\end{equation}
The two organisations are algebraically equivalent before a finite endpoint is assigned.

\section{Loop expansion through two loops}

\subsection{Rescaling and Gaussian measure}

Write
\begin{equation}
\Gamma[\phib]=S[\phib]+\hbar\Gamma_1[\phib]+\hbar^2\Gamma_2[\phib]+\order(\hbar^3)
\label{eq:loopExpansionGamma}
\end{equation}
and rescale the fluctuation in Eq.~\eqref{eq:exactEffectiveIdentity},
\begin{equation}
\eta=\sqrt\hbar\,h.
\end{equation}
Taylor expansion gives
\begin{align}
S[\phib+\sqrt\hbar h]
={}&S[\phib]
+\sqrt\hbar\,S_{,i}h_i
+\frac\hbar2h_i\Kop_{ij}h_j
+\frac{\hbar^{3/2}}{3!}S^{(3)}_{ijk}h_ih_jh_k
\nonumber\\
&+\frac{\hbar^2}{4!}S^{(4)}_{ijkl}h_ih_jh_kh_l.
\label{eq:actionHbarExpansion}
\end{align}
Because
\begin{equation}
\Gamma_{,i}=S_{,i}+\hbar\Gamma_{1,i}+\hbar^2\Gamma_{2,i}+\cdots,
\end{equation}
the linear combination in the exact identity is
\begin{align}
S[\phib+\sqrt\hbar h]-\sqrt\hbar\Gamma_{,i}h_i
={}&S[\phib]+\frac\hbar2h\Kop h
\nonumber\\
&+\hbar^{3/2}
\left(\frac1{3!}S^{(3)}h^3-\Gamma_{1,i}h_i\right)
+\frac{\hbar^2}{4!}S^{(4)}h^4
+\order(\hbar^{5/2}).
\label{eq:sourceCancelledExpansion}
\end{align}
Define the normalised Gaussian expectation with covariance $G=\Kop^{-1}$,
\begin{equation}
\vev{\mathcal O[h]}_0
=\frac{\int\calD h\,\mathcal O[h]\ee^{-h\Kop h/2}}
{\int\calD h\,\ee^{-h\Kop h/2}},
\qquad
\vev{h_ih_j}_0=G_{ij}.
\label{eq:GaussianExpectation}
\end{equation}
Introduce
\begin{equation}
A[h]=\frac1{3!}S^{(3)}_{ijk}h_ih_jh_k-\Gamma_{1,i}h_i,
\qquad
B[h]=\frac1{4!}S^{(4)}_{ijkl}h_ih_jh_kh_l.
\label{eq:ABDefinitions}
\end{equation}
Then
\begin{align}
\Gamma[\phib]
={}&S[\phib]+\frac\hbar2\Tr\ln\Kop
-\hbar\ln\vev{\exp[-\sqrt\hbar A-\hbar B+\order(\hbar^{3/2})]}_0
+\text{constant}.
\label{eq:GammaCumulantMaster}
\end{align}

\subsection{Cumulant expansion}

Since the centred Gaussian has vanishing odd moments,
\begin{equation}
\vev{A}_0=0.
\end{equation}
Expanding the logarithm through order $\hbar$ inside Eq.~\eqref{eq:GammaCumulantMaster},
\begin{align}
\ln\vev{\ee^{-\sqrt\hbar A-\hbar B}}_0
&=-\hbar\vev{B}_0
+\frac\hbar2\vev{A^2}_0+\order(\hbar^{3/2}).
\label{eq:cumulantExpansion}
\end{align}
It follows that
\begin{equation}
\Gamma_1=\frac12\Tr\ln\Kop,
\label{eq:Gamma1}
\end{equation}
\begin{equation}
\Gamma_2=\vev{B}_0-\frac12\vev{A^2}_0.
\label{eq:Gamma2BeforeWick}
\end{equation}

\subsection{Derivative of the one-loop functional}

Varying Eq.~\eqref{eq:Gamma1},
\begin{equation}
\Gamma_{1,i}
=\frac12\Tr(G\Kop_{,i})
=\frac12S^{(3)}_{iab}G_{ab}.
\label{eq:Gamma1Derivative}
\end{equation}
The second equality follows because $\Kop_{ab}=S^{(2)}_{ab}$.

\subsection{Quartic contraction: the figure-eight coefficient}

The Wick contraction formula gives
\begin{equation}
\vev{h_ih_jh_kh_l}_0
=G_{ij}G_{kl}+G_{ik}G_{jl}+G_{il}G_{jk}.
\end{equation}
Because $S^{(4)}_{ijkl}$ is fully symmetric,
\begin{align}
\vev{B}_0
&=\frac1{4!}S^{(4)}_{ijkl}
\left(G_{ij}G_{kl}+G_{ik}G_{jl}+G_{il}G_{jk}\right)
\nonumber\\
&=\frac18S^{(4)}_{ijkl}G_{ij}G_{kl}.
\label{eq:BExpectation}
\end{align}
For $\lambda\phi^4$,
\begin{equation}
S^{(4)}(x_1,x_2,x_3,x_4)
=\lambda\delta(x_1-x_2)\delta(x_1-x_3)\delta(x_1-x_4),
\end{equation}
so
\begin{equation}
\vev{B}_0=\frac\lambda8\int\dd^dx\,G(x,x)^2.
\label{eq:figureEight}
\end{equation}

\subsection{Two cubic vertices and exact cancellation of the one-particle-reducible (1PR) term}

Let
\begin{equation}
C[h]=\frac1{3!}S^{(3)}_{abc}h_ah_bh_c,
\qquad
T[h]=\Gamma_{1,i}h_i,
\end{equation}
so $A=C-T$. The six-field Gaussian moment has two classes of pairings. Six pairings connect all three lines from one cubic vertex to the other, while nine pairings contain one line between vertices and one tadpole at each vertex. Therefore
\begin{align}
\vev{C^2}_0
={}&\frac16S^{(3)}_{abc}S^{(3)}_{def}
G_{ad}G_{be}G_{cf}
+\frac14B_iG_{ij}B_j,
\label{eq:C2Pairings}
\end{align}
where
\begin{equation}
B_i=S^{(3)}_{iab}G_{ab}=2\Gamma_{1,i}.
\end{equation}
Hence the second term in Eq.~\eqref{eq:C2Pairings} is
\begin{equation}
\frac14B_iG_{ij}B_j=\Gamma_{1,i}G_{ij}\Gamma_{1,j}.
\label{eq:onePRpiece}
\end{equation}
The mixed contraction is
\begin{align}
\vev{CT}_0
&=\frac1{3!}S^{(3)}_{abc}\Gamma_{1,i}
\vev{h_ah_bh_ch_i}_0
\nonumber\\
&=\frac12S^{(3)}_{abc}\Gamma_{1,i}G_{ab}G_{ci}
=\Gamma_{1,c}G_{ci}\Gamma_{1,i}.
\label{eq:CT}
\end{align}
Finally,
\begin{equation}
\vev{T^2}_0=\Gamma_{1,i}G_{ij}\Gamma_{1,j}.
\label{eq:T2}
\end{equation}
Combining Eqs.~\eqref{eq:C2Pairings}--\eqref{eq:T2},
\begin{align}
\vev{A^2}_0
&=\vev{C^2}_0-2\vev{CT}_0+\vev{T^2}_0
\nonumber\\
&=\frac16S^{(3)}_{abc}S^{(3)}_{def}G_{ad}G_{be}G_{cf}.
\label{eq:A2ConnectedOnly}
\end{align}
The $1-2+1$ cancellation is the cancellation of the one-particle-reducible contraction by the source term required by the Legendre transform.

\begin{figure}[!t]
    \centering
    \includegraphics[width=0.72\linewidth]{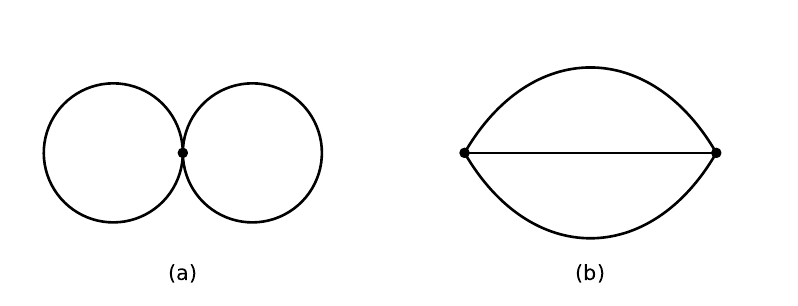}
    \caption{The two two-loop one-particle-irreducible topologies generated by the background-field cumulant expansion: (a) the figure-eight from one quartic fluctuation vertex and (b) the sunset from two cubic fluctuation vertices.}
    \label{fig:twoLoopEffectiveAction}
\end{figure}

Substituting into Eq.~\eqref{eq:Gamma2BeforeWick},
\begin{equation}
\boxed{
\Gamma_2[\phib]
=\frac18S^{(4)}_{ijkl}G_{ij}G_{kl}
-\frac1{12}S^{(3)}_{ijk}S^{(3)}_{lmn}G_{il}G_{jm}G_{kn}.}
\label{eq:Gamma2General}
\end{equation}
For scalar $\lambda\phi^4$,
\begin{equation}
\boxed{
\Gamma_2[\phib]
=\frac\lambda8\int\dd^dx\,G_{\phib}(x,x)^2
-\frac{\lambda^2}{12}\int\dd^dx\,\dd^dy\,
\phib(x)G_{\phib}(x,y)^3\phib(y).}
\label{eq:Gamma2Phi4}
\end{equation}
These are the complete two-loop one-particle-irreducible structures: the figure-eight and the sunset, shown in Fig.~\ref{fig:twoLoopEffectiveAction}.

\subsection{Why no other terms occur at two loops}

The loop order of a connected vacuum graph is $L=I-V+1$. A single quartic vertex has $V=1$, $I=2$, hence $L=2$. Two cubic vertices have $V=2$, $I=3$, hence $L=2$. Other cumulants begin later:
\begin{itemize}
\item $\vev{S_3}_0=0$ by Gaussian parity;
\item $\vev{S_3S_4}_0=0$ because it contains seven fields;
\item the connected part of $\vev{S_4^2}_0$ has $V=2$, $I=4$, hence $L=3$;
\item $\vev{S_3^3}_0=0$ by parity;
\item four cubic vertices first contribute beyond two loops.
\end{itemize}
Thus Eq.~\eqref{eq:Gamma2Phi4} is complete at order $\hbar^2$ while each exact $G_{\phib}$ still contains arbitrarily many quadratic background insertions through Eq.~\eqref{eq:NeumannBackground}.

\section{Gaussian differentiation, determinant, and trace-log equivalence}

\subsection{Finite-dimensional derivation}

Let $K$ be a real symmetric positive $N\times N$ matrix and
\begin{equation}
I[K]=\int_{\mathbb R^N}\dd^Nh\,\ee^{-h^TKh/2}.
\end{equation}
Completing the square gives
\begin{equation}
I[K]=(2\pi)^{N/2}(\det K)^{-1/2}.
\end{equation}
The normalised covariance is
\begin{equation}
\frac1{I[K]}\int\dd^Nh\,h_ih_j\ee^{-h^TKh/2}=(K^{-1})_{ij}.
\end{equation}
Differentiate the integral directly:
\begin{align}
\delta I[K]
&=-\frac12\int\dd^Nh\,(h^T\delta K h)\ee^{-h^TKh/2}
\nonumber\\
&=-\frac12I[K]\Tr(K^{-1}\delta K).
\end{align}
Therefore
\begin{equation}
\delta[-\ln I[K]]=\frac12\Tr(K^{-1}\delta K)
=\delta\left(\frac12\Tr\ln K\right).
\label{eq:gaussianTraceLogEquiv}
\end{equation}
The determinant is not required as an intermediate calculation, but it is implicit in the Gaussian normalisation.

\subsection{Functional derivatives in scalar theory}

For Eq.~\eqref{eq:backgroundK},
\begin{align}
\frac{\delta\Kop_{\phib}(x,y)}{\delta\phib(x_1)}
&=\lambda\phib(x)\delta(x-x_1)\delta(x-y),
\label{eq:Kfirst}\\
\frac{\delta^2\Kop_{\phib}(x,y)}{\delta\phib(x_1)\delta\phib(x_2)}
&=\lambda\delta(x-x_1)\delta(x-x_2)\delta(x-y),
\label{eq:Ksecond}\\
\frac{\delta^r\Kop_{\phib}}{\delta\phib^r}&=0,
\qquad r\ge3.
\label{eq:KhigherZero}
\end{align}
The first variation of the one-loop action is
\begin{equation}
\delta\Gamma_1=\frac12\Tr(G_{\phib}\delta\Kop_{\phib}).
\end{equation}
The second variation is
\begin{equation}
\delta_2\delta_1\Gamma_1
=\frac12\Tr\left[-G\Kop_{,2}G\Kop_{,1}+G\Kop_{,12}\right].
\label{eq:secondTraceLog}
\end{equation}
At $\phib=0$, $\Kop_{,1}=0$, so
\begin{equation}
\left.\frac{\delta^2\Gamma_1}{\delta\phib(x_1)\delta\phib(x_2)}\right|_{\phib=0}
=\frac\lambda2G_0(x_1,x_1)\delta(x_1-x_2).
\label{eq:tadpoleResult}
\end{equation}
This is the ordinary one-loop tadpole self-energy obtained without first writing either a determinant or a proper-time integral.

\section{All-order spectral operator calculus}

\subsection{Spectral functions and divided differences}

Let $\Kop$ be positive self-adjoint and let $f$ be analytic on a domain containing its spectrum. Let $\mathcal C$ be a positively oriented contour enclosing the relevant spectrum. The functional calculus is
\begin{equation}
f(\Kop)=\frac1{2\pi\ii}\oint_\mathcal C\dd z\,f(z)(z-\Kop)^{-1}.
\label{eq:contourFunctionalCalculus}
\end{equation}
The scalar divided differences are defined recursively by
\begin{equation}
f^{[0]}(x_0)=f(x_0),
\qquad
f^{[n]}(x_0,\ldots,x_n)
=\frac{f^{[n-1]}(x_1,\ldots,x_n)-f^{[n-1]}(x_0,\ldots,x_{n-1})}{x_n-x_0},
\label{eq:dividedDifferenceDef}
\end{equation}
with coincident arguments defined by continuity. In particular,
\begin{equation}
f^{[n]}(x,\ldots,x)=\frac{f^{(n)}(x)}{n!}.
\end{equation}
If $\Kop=\sum_a\kappa_aP_a$, where $\kappa_a$ are spectral values and $P_a$ the corresponding spectral projections, and where $V_1,\ldots,V_n$ are operator insertion directions, the $n$th Fr\'echet derivative is
\begin{align}
D^nf(\Kop)[V_1,\ldots,V_n]
={}&\sum_{\sigma\in S_n}\sum_{a_0,\ldots,a_n}
 f^{[n]}(\kappa_{a_0},\ldots,\kappa_{a_n})
\nonumber\\
&\times P_{a_0}V_{\sigma(1)}P_{a_1}\cdots
V_{\sigma(n)}P_{a_n}.
\label{eq:FrechetSpectral}
\end{align}
Equivalently, differentiating Eq.~\eqref{eq:contourFunctionalCalculus},
\begin{align}
D^nf(\Kop)[V_1,\ldots,V_n]
={}&\frac1{2\pi\ii}\oint_\mathcal C\dd z\,f(z)
\sum_{\sigma\in S_n}(z-\Kop)^{-1}V_{\sigma(1)}(z-\Kop)^{-1}
\nonumber\\
&\hspace{38mm}\cdots V_{\sigma(n)}(z-\Kop)^{-1}.
\label{eq:FrechetContour}
\end{align}
Equations~\eqref{eq:FrechetSpectral} and \eqref{eq:FrechetContour} are valid for noncommuting insertions. Replacing them by ordinary scalar derivatives is generally incorrect.

\subsection{Ordered insertion formula (Duhamel form)}

For $f_T(z)=\ee^{-Tz}$, let $S_n$ denote the permutation group of $n$ objects. The operator derivative can be written as the following ordered integral, usually called the Duhamel formula
\begin{align}
D^n\ee^{-T\Kop}[V_1,\ldots,V_n]
={}&(-1)^n\sum_{\sigma\in S_n}
\int_{u_j\ge0}\left(\prod_{j=0}^n\dd u_j\right)
\delta\left(T-\sum_{j=0}^{n}u_j\right)
\nonumber\\
&\times\ee^{-u_0\Kop}V_{\sigma(1)}\ee^{-u_1\Kop}
\cdots V_{\sigma(n)}\ee^{-u_n\Kop}.
\label{eq:DuhamelAllOrder}
\end{align}
The variables $u_j$ are non-negative segments of one original history and satisfy
\begin{equation}
\sum_{j=0}^{n}u_j=T.
\label{eq:DuhamelSum}
\end{equation}
Functional differentiation partitions a history; it does not create $n+1$ independently bounded complete histories.

\subsection{\texorpdfstring{All-order functional derivatives in $\lambda\phi^4$}{All-order functional derivatives in lambda phi4}}

Let
\begin{equation}
\mathcal A_f[\phib]=\frac12\Tr f(\Kop_{\phib}).
\end{equation}
Because only the first and second derivatives of $\Kop_{\phib}$ are nonzero, the $n$th functional derivative is a sum over set partitions whose blocks have size one or two:
\begin{equation}
\delta_1\cdots\delta_n\mathcal A_f
=\frac12\sum_{\pi\in\calP_{1,2}(n)}
\Tr D^{|\pi|}f(\Kop)
\left[\Kop_{,B_1},\ldots,\Kop_{,B_{|\pi|}}\right].
\label{eq:partitionFormula}
\end{equation}
Here $\calP_{1,2}(n)$ is the set of partitions of $\{1,\ldots,n\}$ into singleton and pair blocks, and $\Kop_{,B}$ denotes the corresponding first or second background derivative. Singletons generate ordinary insertions, while pairs generate contact vertices. At $\phib=0$, all singleton blocks vanish and only pair partitions survive; odd derivatives are therefore zero by $\mathbb Z_2$ symmetry.

\section{Schwinger, heat-kernel, worldline, and Feynman equivalence}

\subsection{Free covariance}

For the free Euclidean operator $\Kop_0=-\partial^2+m^2$,
\begin{equation}
G_0=\Kop_0^{-1}=\int_0^\infty\dd s\,\ee^{-s\Kop_0}.
\label{eq:GSchwinger}
\end{equation}
In momentum space,
\begin{equation}
G_0(p)=\frac1{p^2+m^2}.
\end{equation}
The heat kernel is
\begin{equation}
K_s(x,y)=\bra{x}\ee^{-s\Kop_0}\ket{y}
=\frac{\ee^{-m^2s-(x-y)^2/(4s)}}{(4\pi s)^{d/2}}.
\label{eq:freeHeatKernel}
\end{equation}
It satisfies the semigroup law
\begin{equation}
\int\dd^dz\,K_{s_1}(x,z)K_{s_2}(z,y)=K_{s_1+s_2}(x,y).
\label{eq:semigroup}
\end{equation}
The same kernel has the worldline representation
\begin{equation}
K_s(x,y)=\ee^{-m^2s}
\int_{x(0)=y}^{x(s)=x}\calD x(\tau)
\exp\left[-\int_0^s\dd\tau\,\frac{\dot x^2}{4}\right].
\label{eq:worldlineKernel}
\end{equation}
Thus the momentum propagator, Schwinger integral, heat kernel, and worldline path integral are representations of the same inverse operator.

\subsection{Source generator and ordinary diagrams}

Let $Z_0[J]$ denote the free generating functional. Then
\begin{equation}
Z_0[J]=Z_0[0]\exp\left(\frac1{2\hbar}J_iG_{0,ij}J_j\right).
\end{equation}
The interacting scalar functional may be written formally as
\begin{equation}
Z[J]=\exp\left[-\frac\lambda{4!}\hbar^3\int\dd^dx\,
\frac{\delta^4}{\delta J(x)^4}\right]Z_0[J].
\label{eq:sourceGeneratorFeynman}
\end{equation}
Each pair of source derivatives acting on the Gaussian produces one covariance $G_0$. Expanding the interaction exponential therefore produces Wick contractions and ordinary Feynman graphs. Replacing each covariance by Eq.~\eqref{eq:GSchwinger} assigns a Schwinger parameter to each Gaussian contraction. Performing the momentum integrals gives the heat-kernel or worldline network representation. Before imposing the finite endpoint,
\begin{equation}
\boxed{\begin{aligned}
\text{source generator}
&=\text{Wick/Feynman expansion}\\
&=\text{Schwinger-parametric expansion}\\
&=\text{heat-kernel/worldline networks}.
\end{aligned}}
\label{eq:equivalenceUndeformed}
\end{equation}

\subsection{Background covariance as a resummed Feynman series}

The Schwinger representation of the quadratic background covariance is
\begin{equation}
G_{\phib}=\int_0^\infty\dd T\,\ee^{-T(\Kop_0+V_{\phib})}.
\end{equation}
Expanding the exponential with Eq.~\eqref{eq:DuhamelAllOrder} gives
\begin{align}
G_{\phib}
={}&\sum_{n=0}^\infty(-1)^n
\int_0^\infty\dd T
\int_{u_j\ge0}\prod_{j=0}^n\dd u_j\,
\delta\left(T-\sum_{j=0}^{n}u_j\right)
\nonumber\\
&\times\ee^{-u_0\Kop_0}V_{\phib}\ee^{-u_1\Kop_0}
\cdots V_{\phib}\ee^{-u_n\Kop_0}.
\label{eq:backgroundDuhamelExpansion}
\end{align}
Integrating over $T$ removes the delta function and reproduces Eq.~\eqref{eq:NeumannBackground}. The individual $u_j$ become the Schwinger parameters of the free propagator segments between quadratic background insertions.

\section{Finite proper time as operator functions}
\label{sec:operatorFunctions}

\subsection{One history, two topologies: interval and circle}

{\color{papergreen}The open and closed objects below are not two unrelated choices of spectral function. They are the interval and circle forms of the same proper-time history. The interval has fixed endpoints and measure $\dd T$. Closing it gives a trace. A circle has no preferred origin, so the common translation of the origin is divided out and the measure becomes $\dd T/T$. Choosing one reference point on the circle removes this cyclic freedom and restores the interval measure.}

For a positive operator $\Kop$, define
\begin{equation}
\boxed{
\Rop_{s_0}(\Kop)
=\int_{s_0}^\infty\dd s\,\ee^{-s\Kop}
=\ee^{-s_0\Kop}\Kop^{-1}.}
\label{eq:Roperator}
\end{equation}
Define also
\begin{equation}
\boxed{
\Fop_{s_0}(\Kop)
=-\Eone(s_0\Kop),
\qquad
\Eone(z)=\int_z^\infty\frac{\ee^{-t}}t\dd t.}
\label{eq:Foperator}
\end{equation}
The closed one-loop functional is
\begin{equation}
\boxed{
\Gamma_{1,s_0}[\Kop]
=\frac12\Tr\Fop_{s_0}(\Kop)
=-\frac12\int_{s_0}^\infty\frac{\dd s}{s}\Tr\ee^{-s\Kop}.}
\label{eq:closedFPT}
\end{equation}
Since
\begin{equation}
\frac{\dd}{\dd z}\left[-\Eone(s_0z)\right]
=\frac{\ee^{-s_0z}}z,
\end{equation}
the trace variation is
\begin{equation}
\boxed{
\delta\Gamma_{1,s_0}
=\frac12\Tr\left[\Rop_{s_0}(\Kop)\delta\Kop\right].}
\label{eq:sewingVariation}
\end{equation}
In particular, if $\partial\Kop/\partial m^2=\id$, with $\id$ the identity operator,
\begin{equation}
\frac{\partial\Gamma_{1,s_0}}{\partial m^2}
=\frac12\Tr\Rop_{s_0}(\Kop).
\label{eq:massSewing}
\end{equation}
Equation~\eqref{eq:massSewing} is the operator expression of the interval--circle relation: differentiating with respect to $m^2$ marks a point on the closed history and removes the cyclic $1/T$ factor. The relation follows from the common topology; it is not the origin of that topology. The finite-$s_0$ construction can therefore be stated entirely through spectral operator functions. Proper-time integrals are useful representations of those functions.

\subsection{\texorpdfstring{\textcolor{papergreen}{Complete histories after interaction sewing}}{Complete histories after interaction sewing}}
\label{sec:operatorCalculus}

{\color{papergreen}
The one-loop spectral functions fix the interval and circle sectors before any graph expansion is chosen. At higher loop order, local interactions join propagating segments into a network. The remaining question is which combinations of those segments form complete closed virtual histories. This can be determined from local momentum conservation.

Let $G=(V,E)$ be a connected Feynman multigraph with oriented internal edges. Let $B$ be its vertex--edge incidence matrix. If $q\in(\mathbb R^d)^E$ denotes the internal momenta and $p$ the external momentum injected at the vertices, local conservation is
\begin{equation}
Bq+p=0.
\label{eq:incidenceMomentum}
\end{equation}
Choose one solution $q^{(0)}$. Any other internal assignment with the same external data has
\begin{equation}
q=q^{(0)}+\delta q,
\qquad
B\,\delta q=0.
\label{eq:circulationCondition}
\end{equation}
Thus the unconstrained internal virtual momentum directions are the circulations in $\ker B$. For a connected graph,
\begin{equation}
\dim\ker B=|E|-|V|+1=L,
\end{equation}
the usual loop number.

{\color{paperred}\paragraph{Primitive closed history.}
A non-zero internal circulation $z\in\ker B$ is called primitive when no non-zero circulation has support strictly contained in $\operatorname{supp}z$. Its support is exactly a circuit of the Feynman multigraph. A circuit may be a self-loop, a pair of parallel edges, or an ordinary simple closed cycle.
\label{res:primitiveCirculation}

\paragraph{Derivation.}
Take $z\neq0$ with $Bz=0$ and start on an edge with $z_e\neq0$. At every vertex touched by the support, conservation prevents the non-zero flow from terminating there. Because the graph is finite, following the support eventually returns to a previously visited vertex and therefore contains a circuit. If $z$ has support-minimality, that circuit cannot be a proper subset of its support, because an oriented circuit itself defines a non-zero element of $\ker B$. Hence the support is exactly a circuit. Conversely, orient any circuit consistently and assign equal magnitude to its edges. The resulting vector lies in $\ker B$. Removing any edge opens the circuit into a path and destroys the non-zero conserved circulation, so the support is minimal.}

Introduce a Schwinger parameter $s_e\ge0$ on each internal edge. For a primitive circuit $c$, choose the unit circulation $z_c$ with $(z_c)_e=\pm1$ on $c$ and zero elsewhere. If the corresponding loop deformation is $\ell z_c$, the coefficient of $\ell^2$ in the Euclidean exponent is
\begin{equation}
\sum_{e\in E}s_e(z_c)_e^2
=
\sum_{e\in c}s_e.
\end{equation}
The proper-time length conjugate to that primitive closed virtual circulation is therefore
\begin{equation}
\boxed{
T_c=\sum_{e\in c}s_e.}
\label{eq:circuitProperTime}
\end{equation}
Applying the retained complete-history boundary to the primitive closed histories gives the graph domain
\begin{equation}
\boxed{
\mathcal D_G(s_0)
=
\left\{
s_e\ge0:
\sum_{e\in c}s_e\ge s_0
\quad\text{for every }c\in\mathcal C(G)
\right\},}
\label{eq:circuitDomain}
\end{equation}
where $\mathcal C(G)$ is the circuit set of $G$.

Equation~\eqref{eq:circuitDomain} is basis independent. A choice of $L$ loop momenta is only a basis of $\ker B$, whereas the circuits are the support-minimal conserved flows themselves. The same graph may contain more than $L$ circuits, and all of them are required. No individual edge receives a lower bound merely because it is an internal propagator. A bridge lies on no circuit and therefore appears in no condition in Eq.~\eqref{eq:circuitDomain}.

The one-loop operator calculus is recovered as a special case. A tadpole is a one-edge circuit and gives $s\ge s_0$. A bubble obtained by partitioning one closed loop into two propagating segments is a two-edge circuit and gives $s_1+s_2\ge s_0$. More generally an $n$-segment one-loop history has the single condition $\sum_{j=1}^n s_j\ge s_0$, exactly as obtained by the Duhamel hierarchy.

The two-loop topologies derived in Eq.~\eqref{eq:Gamma2Phi4} keep their ordinary vertex coefficients and symmetry factors. Their finite-$s_0$ form is obtained by integrating their Schwinger parameters over Eq.~\eqref{eq:circuitDomain}:
\begin{equation}
\boxed{
\Gamma_{2,s_0}[\phib]
=
\frac{\lambda}{8}\,\mathcal H_{\mathrm{fig8}}[\phib;s_0]
-
\frac{\lambda^2}{12}\,\mathcal H_{\mathrm{sun}}[\phib;s_0],}
\label{eq:FPTTwoLoop}
\end{equation}
where $\mathcal H_{\mathrm{fig8}}$ and $\mathcal H_{\mathrm{sun}}$ denote the same figure-eight and sunset kernels as in the ordinary loop expansion, with their proper-time integrations restricted by the circuit domain. Their explicit translation-invariant forms are worked out in Section~\ref{sec:twoLoopParametric}.

\begin{figure}[H]
    \centering
    \includegraphics[width=0.72\linewidth]{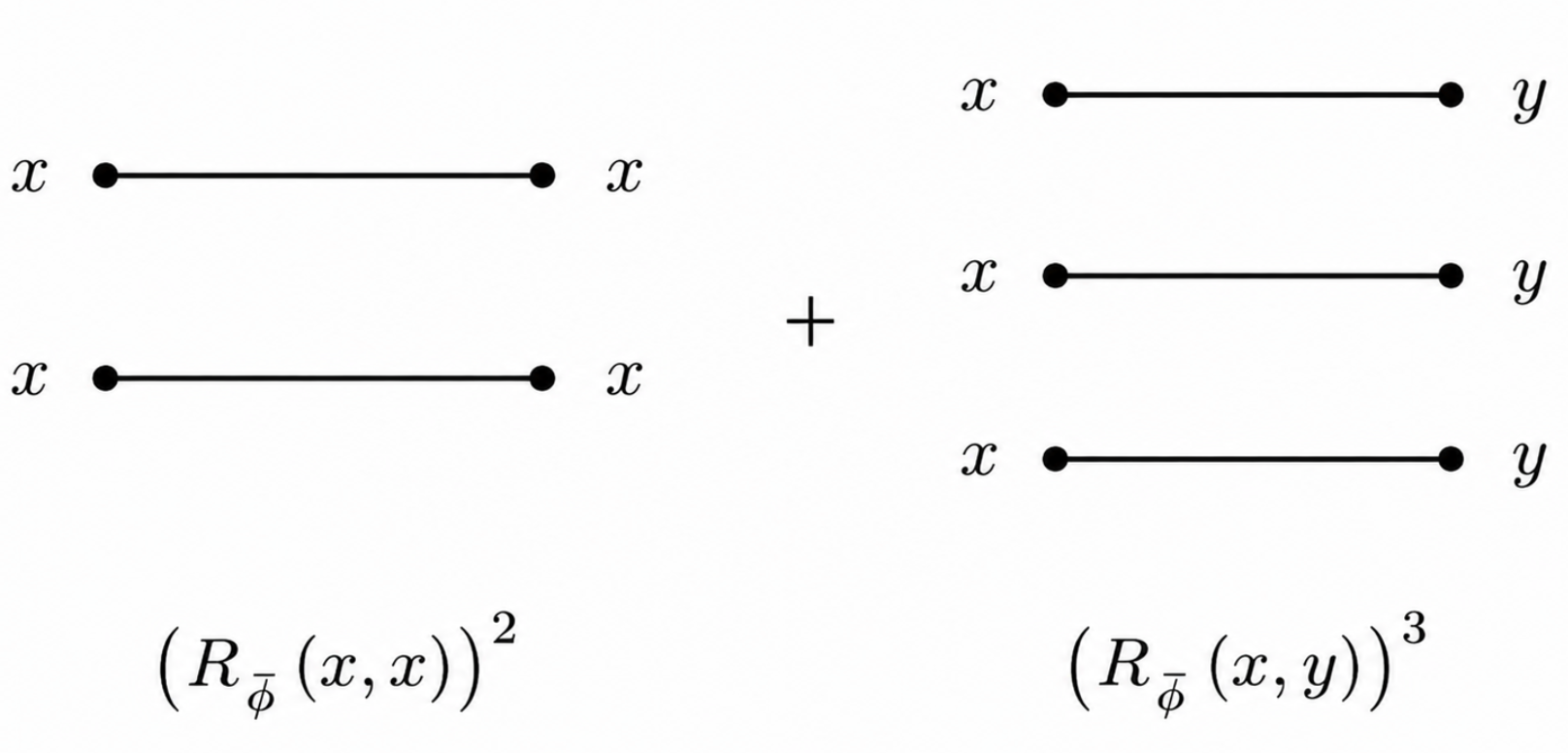}
    \caption{\textcolor{papergreen}{The two two-loop topologies. In the figure-eight each self-loop is itself a primitive circuit. In the sunset, the three propagating edges form three two-edge circuits. The retained endpoint acts on the total proper time of these closed circulations rather than independently on every edge. Duhamel insertions partition an edge total without creating a new circuit endpoint.}}
    \label{fig:schematicworldlineness}
\end{figure}

{\color{paperred}\paragraph{Why this is not an ordinary Gaussian determinant?}
A Gaussian measure with covariance $\Rop_{s_0}(\Kop_{\phib})$ would assign that kernel independently to every Wick contraction and would have normalisation proportional to $\det[\Rop_{s_0}(\Kop_{\phib})^{-1}]^{-1/2}$. This is not the closed complete-history functional in Eq.~\eqref{eq:closedFPT}. In the circuit construction, the one-loop open interval $\Rop_{s_0}$ remains the derivative of the closed spectral function, while a multi-loop graph is constrained collectively through Eq.~\eqref{eq:circuitDomain}. An independently damped Gaussian covariance is therefore a different finite-$s_0$ theory.}
}

\subsection{\texorpdfstring{\textcolor{papergreen}{Why complete histories are not independently modified propagators}}{Why complete histories are not independently modified propagators}}

{\color{papergreen}
Equations~\eqref{eq:Roperator} and \eqref{eq:closedFPT} are related, but they are not the covariance and normalisation of one ordinary Gaussian measure. If one chooses
\begin{equation}
C_{s_0}=\Rop_{s_0}(\Kop_0)
\label{eq:ordinaryCovModel}
\end{equation}
as a free Gaussian covariance, then every Wick contraction carries $C_{s_0}$ independently. Its Gaussian normalisation contains
\begin{equation}
\frac12\Tr\ln C_{s_0}^{-1}
=
\frac12\Tr\ln\Kop_0+\frac{s_0}{2}\Tr\Kop_0,
\label{eq:ordinaryCovDet}
\end{equation}
which is not the closed complete-history functional
\begin{equation}
-\frac12\Tr\Eone(s_0\Kop_0).
\end{equation}
The independently damped Gaussian model and the complete-history spectral construction therefore differ at finite $s_0$, even though their formal boundary-removal correspondence limits $s_0\to0^+$ reproduce the local theory after the appropriate local subtractions. The value $s_0=0$ is outside the defined fixed-$s_0$ FPT theory.

The interval--circle relation remains exact. An open history has measure $\dd T$, while a closed trace has measure $\dd T/T$ because no point on the closed circle is preferred. Differentiating the closed functional with respect to $m^2$ chooses one reference point on the circle and restores the interval measure:
\begin{equation}
\frac{\partial}{\partial m^2}
\left[
-\frac12\int_{s_0}^{\infty}\frac{\dd T}{T}\Tr\ee^{-T\Kop}
\right]
=
\frac12\int_{s_0}^{\infty}\dd T\,\Tr\ee^{-T\Kop}
=
\frac12\Tr\Rop_{s_0}(\Kop).
\label{eq:circleIntervalRelation}
\end{equation}
Thus $\Rop_{s_0}$ is the open-interval sector of the same spectral history. It is not thereby promoted to the covariance of every edge in an arbitrary graph.

Functional differentiation inserts operators on an interval or circle that already exists and partitions its total $T$. Local interaction vertices then join propagating segments into a graph. Once that sewing is performed, the primitive complete closed histories are determined by the circuit result in Sec.~\ref{sec:operatorCalculus}: they are the graph circuits, and the lower boundary is imposed on the circuit totals in Eq.~\eqref{eq:circuitDomain}. An edge may belong to several circuits, one circuit, or none. In particular a bridge belongs to no closed circulation and carries no independent circuit endpoint.

This distinction is also required by the exact Legendre structure. The connected two-point kernel and the 1PI two-point kernel obey
\begin{equation}
W^{(2)}=(\Gamma^{(2)})^{-1}.
\label{eq:LegendreTwoPointInverse}
\end{equation}
Consequently one-particle-reducible connected amplitudes are sewn by the appropriate full connected propagator generated by this inverse relation. At tree level that bridge is the ordinary $K_0^{-1}$ propagator. The finite endpoint enters through the 1PI closed-circuit subgraphs rather than through an additional independent factor placed on the bridge.
}

\subsection{How insertions divide one complete history}

Differentiate Eq.~\eqref{eq:Roperator}. The $n$th derivative contains
\begin{align}
D^n\Rop_{s_0}(\Kop)[V_1,\ldots,V_n]
={}&(-1)^n\sum_{\sigma\in S_n}
\int_{T\ge s_0}\dd T
\int_{u_j\ge0}\prod_{j=0}^n\dd u_j\,
\delta\left(T-\sum_{j=0}^{n}u_j\right)
\nonumber\\
&\times\ee^{-u_0\Kop}V_{\sigma(1)}\ee^{-u_1\Kop}\cdots
V_{\sigma(n)}\ee^{-u_n\Kop}.
\label{eq:FPTDuhamel}
\end{align}
The finite domain is
\begin{equation}
\boxed{u_j\ge0,
\qquad
\sum_{j=0}^{n}u_j\ge s_0,}
\label{eq:globalHistoryDomain}
\end{equation}
not $u_j\ge s_0$ for every segment. This distinction is the scalar version of the complete-history versus per-segment distinction.

\begin{figure}[!t]
\centering
\includegraphics[width=0.72\linewidth]{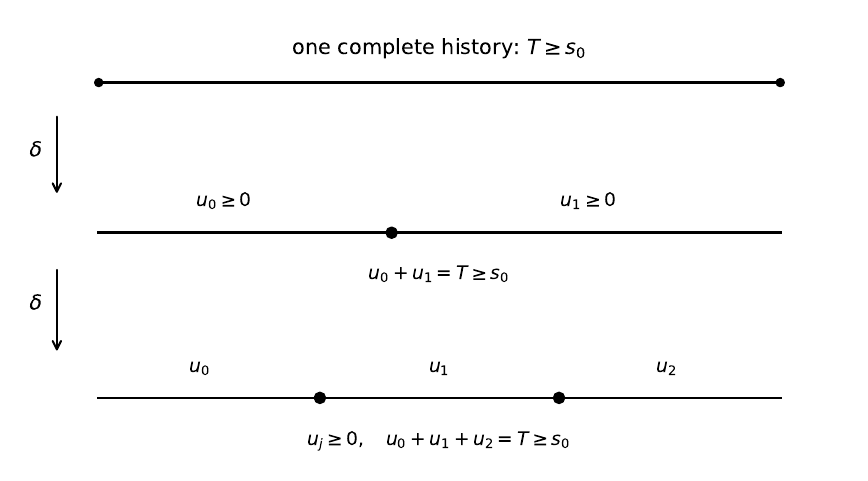}
\caption{Insertion rule for one complete open history. A functional derivative inserts an operator and partitions the same total proper time $T$ into non-negative segments. Repeated differentiation produces further segments, but the only lower-endpoint condition is on their sum, $\sum_j u_j=T\ge s_0$.}
\label{fig:operatorAncestry}
\end{figure}

Figure~\ref{fig:operatorAncestry} summarizes this insertion rule. Functional differentiation inserts operators along one pre-existing complete history and partitions its total proper time; it does not create new complete histories.

\section{One-loop examples}

\subsection{FPT tadpole in arbitrary dimension}

For the free operator,
\begin{align}
\Rop_{s_0}(x,x)
&=\int\frac{\dd^dp}{(2\pi)^d}
\frac{\ee^{-s_0(p^2+m^2)}}{p^2+m^2}
\nonumber\\
&=\frac1{(4\pi)^{d/2}}
\int_{s_0}^\infty\dd s\,s^{-d/2}\ee^{-m^2s}
\nonumber\\
&=\frac{(m^2)^{d/2-1}}{(4\pi)^{d/2}}
\Gamma\left(1-\frac d2,m^2s_0\right).
\label{eq:tadpoleGeneralD}
\end{align}
Here $\Gamma(\nu,z)$ is the upper incomplete gamma function.

\begin{figure}[!t]
    \centering
    \includegraphics[width=0.72\linewidth]{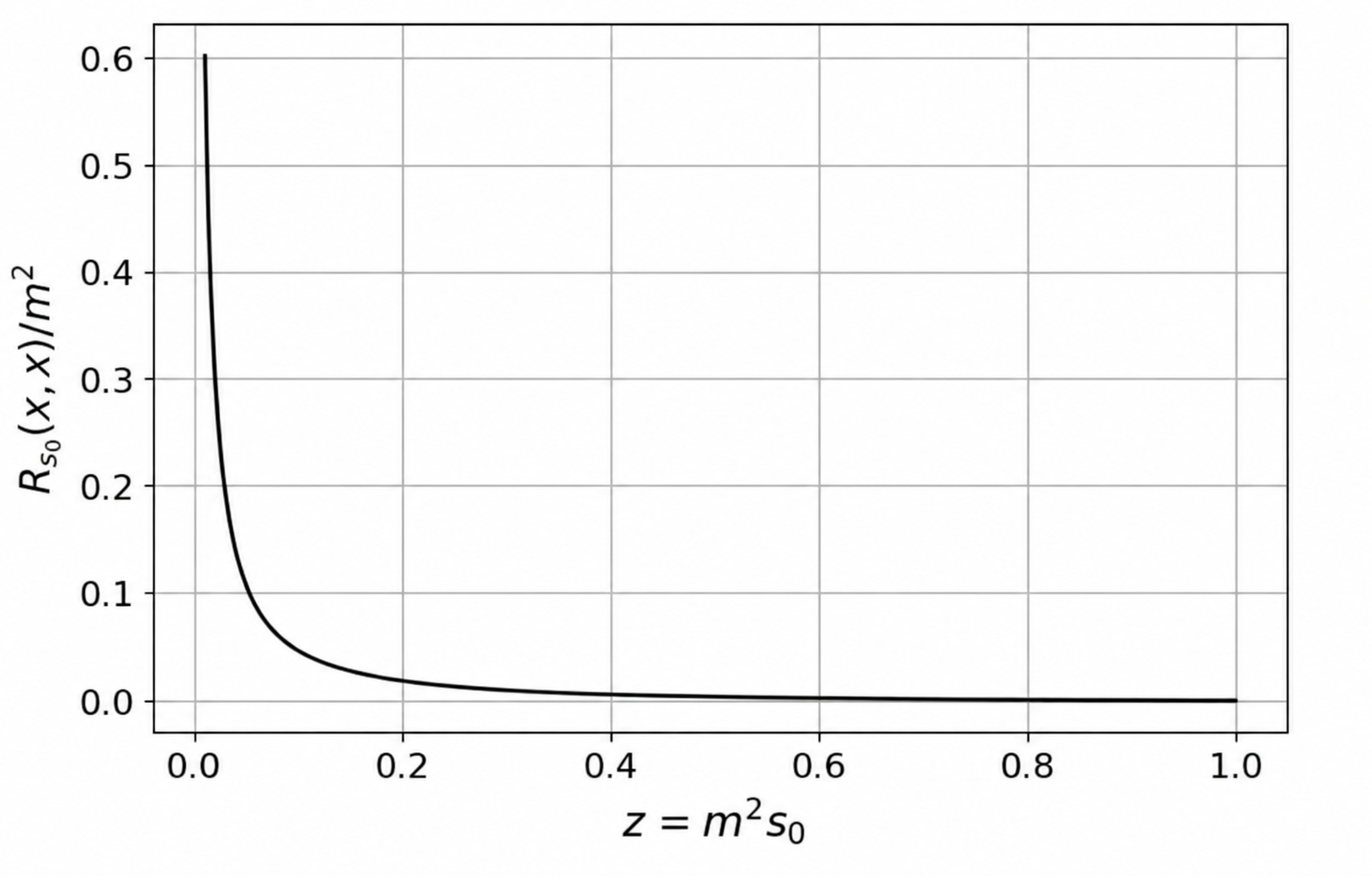}
    \caption{Dimensionless coincident finite-endpoint covariance in four dimensions, $\Rop_{s_0}(x,x)/m^2=\left[e^{-z}/z-E_1(z)\right]/(16\pi^2)$ with $z=m^2s_0=m^2/\Lambda^2$. The quantity is finite for every fixed $s_0>0$ and diverges as $\Rop_{s_0}(x,x)/m^2\sim 1/(16\pi^2z)$ only in the formal boundary-removal comparison $s_0\to0^+$, which lies outside the defined fixed-$s_0$ theory.
    }
    \label{fig:gamma_minus_one}
\end{figure}

The dimensionless four-dimensional result is plotted in Fig.~\ref{fig:gamma_minus_one}; for every fixed $s_0>0$ it is finite, while the formal boundary-removal comparison $s_0\to0^+$ reproduces the ordinary ultraviolet divergence.

In four dimensions,
\begin{equation}
\Rop_{s_0}(x,x)
=\frac1{16\pi^2}
\left[
\frac{\ee^{-m^2s_0}}{s_0}
-m^2\Eone(m^2s_0)
\right].
\label{eq:tadpoleD4}
\end{equation}
The tree plus one-loop 1PI two-point kernel at $\phib=0$ is
\begin{equation}
\Gamma^{(2)}_{s_0}(x_1,x_2)
=\left[-\partial^2+m^2+\frac\lambda2\Rop_{s_0}(x_1,x_1)\right]
\delta(x_1-x_2)+\order(\lambda^2),
\end{equation}
Here $\Gamma_{s_0}^{(2)}=\Gamma_0^{(2)}+\hbar\Gamma_{1,s_0}^{(2)}+\cdots$ in the convention of Eq.~\eqref{eq:GammaNotationConvention}; the displayed equation keeps the tree and one-loop terms.

\begin{figure}[!t]
    \centering
    \includegraphics[width=0.72\linewidth]{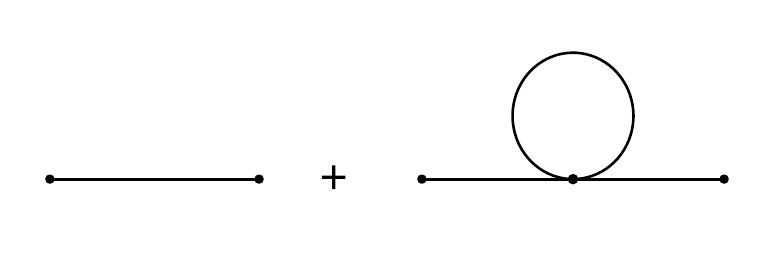}
    \caption{Schematic one-loop one-particle-irreducible two-point kernel: the tree inverse kernel plus the scalar tadpole self-energy.}
    \label{fig:oneLoopTwoPointKernel}
\end{figure}

The tree inverse kernel and tadpole self-energy represented above are shown schematically in Fig.~\ref{fig:oneLoopTwoPointKernel}.

\subsection{One-loop bubble as one closed history}

\begin{figure}[H]
    \centering
    \includegraphics[width=0.72\linewidth]{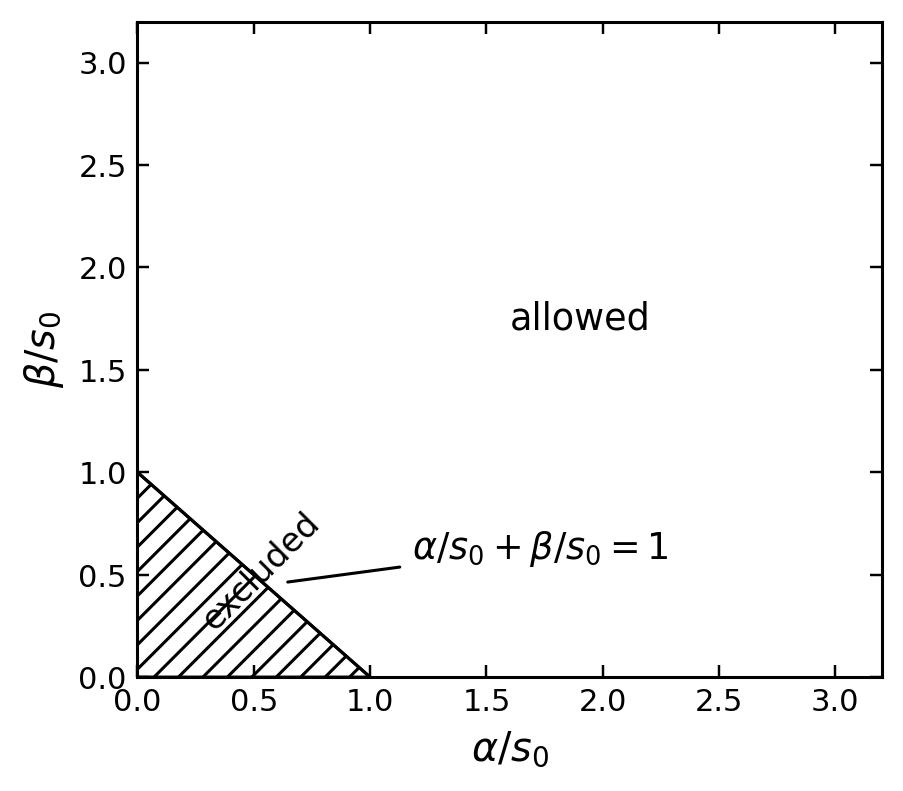}
    \caption{Schwinger parameter domain for two insertions on one closed history. The excluded triangle is $\alpha+\beta<s_0$; the individual segments may approach zero provided the total history satisfies $\alpha+\beta\ge s_0$.}
    \label{fig:completeHistoryBubble}
\end{figure}

The one-loop bubble diagram is obtained from four derivatives of the closed one-loop functional Eq.~\eqref{eq:closedFPT}. Taking four derivatives, bearing in mind that single functional derivatives on $\mathcal{K}$ vanish at $\phi=0$ and applying twice the Duhamel formula, we get
\begingroup\small
\begin{equation}
\begin{aligned}
\label{eq:bubbleFourthDerivative}
\left.
\frac{\delta^4 \Gamma_{1,s_0}[\phi]}
{\delta \phi(x_1)\delta\phi(x_2)\delta \phi(x_3)\delta \phi(x_4)}
\right|_{\phi=0}
&=
-\frac{1}{2}
\int_{s_0}^{\infty} \dd s
\int_0^s \dd\alpha\,
\mathrm{Tr}
\left(
e^{-\alpha \mathcal{K}}
\frac{\delta^2\mathcal{K}}
{\delta\phi(x_3)\delta\phi(x_4)}
e^{-(s-\alpha)\mathcal{K}}
\frac{\delta^2\mathcal{K}}
{\delta\phi(x_1)\delta\phi(x_2)}
\right)_{\phi=0}
\\[1em]
&\quad
-\frac{1}{2}
\int_{s_0}^{\infty} \dd s
\int_0^s \dd \alpha\,
\mathrm{Tr}
\left(
e^{-\alpha \mathcal{K}}
\frac{\delta^2\mathcal{K}}
{\delta\phi(x_2)\delta\phi(x_3)}
e^{-(s-\alpha)\mathcal{K}}
\frac{\delta^2\mathcal{K}}
{\delta\phi(x_1)\delta\phi(x_4)}
\right)_{\phi=0}
\\[1em]
&\quad
-\frac{1}{2}
\int_{s_0}^{\infty} \dd s
\int_0^s \dd \alpha\,
\mathrm{Tr}
\left(
e^{-\alpha \mathcal{K}}
\frac{\delta^2\mathcal{K}}
{\delta\phi(x_2)\delta\phi(x_4)}
e^{-(s-\alpha)\mathcal{K}}
\frac{\delta^2\mathcal{K}}
{\delta\phi(x_1)\delta\phi(x_3)}
\right)_{\phi=0}.
\end{aligned}
\end{equation}
\endgroup
Defining $\beta=s-\alpha$, we can change variables to obtain

\begin{equation}
\int_{s_0}^{\infty} \dd s
\int_0^s \dd \alpha\,
\mathrm{Tr}
\left(
e^{-\alpha \mathcal{K}}
\frac{\delta^2\mathcal{K}}
{\delta\phi(x_3)\delta\phi(x_4)}
e^{-(s-\alpha)\mathcal{K}}
\frac{\delta^2\mathcal{K}}
{\delta\phi(x_1)\delta\phi(x_2)}
\right)_{\phi=0}
\end{equation}

\begin{equation}
=
\int_{\substack{\alpha,\beta\geq 0\\ \alpha+\beta\geq s_0}}
\dd\alpha\,\dd\beta\,
\mathrm{Tr}
\left(
e^{-\alpha\mathcal{K}}
\frac{\delta^2\mathcal{K}}
{\delta\phi(x_1)\delta\phi(x_2)}
e^{-\beta\mathcal{K}}
\frac{\delta^2\mathcal{K}}
{\delta\phi(x_3)\delta\phi(x_4)}
\right)_{\phi=0}.
\end{equation}
\begin{equation}
=
\lambda^2
\delta(x_1-x_2)
\delta(x_3-x_4)
\int_{\substack{\alpha,\beta\geq 0\\ \alpha+\beta\geq s_0}}
\dd\alpha\,\dd\beta\,K_{\alpha}(x_1,x_3)K_\beta(x_3,x_1).
\end{equation}
The corresponding Schwinger domain is shown in Fig.~\ref{fig:completeHistoryBubble}; it is a single-history constraint rather than two independent line lower bounds. The other two terms in Eq.~\eqref{eq:bubbleFourthDerivative} behave similarly.

{\color{papergreen}This calculation shows why the endpoint must be assigned before the differentiated expression is read as a set of propagator factors. The two heat-kernel segments descend from one closed history, so their allowed region is $\alpha,\beta\ge0$ with $\alpha+\beta\ge s_0$. Assigning $s_0$ separately to $\alpha$ and $\beta$ would define a different theory.}

For a constant background define
\begin{equation}
M^2=m^2+\frac\lambda2\phib^2.
\end{equation}
With $p_E$ the external Euclidean momentum, the ordinary scalar bubble is
\begin{equation}
B(p_E^2)=\int\frac{\dd^dk}{(2\pi)^d}
\frac1{(k^2+M^2)((k+p)^2+M^2)}.
\label{eq:bubbleMomentum}
\end{equation}

Introduce two Schwinger segments $\alpha,\beta\ge0$ and integrate over $k$:
\begin{align}
B(p_E^2)
&=\frac1{(4\pi)^{d/2}}
\int_0^\infty\dd\alpha\int_0^\infty\dd\beta\,
(\alpha+\beta)^{-d/2}
\nonumber\\
&\quad\times\exp\left[-M^2(\alpha+\beta)
-\frac{\alpha\beta}{\alpha+\beta}p_E^2\right].
\label{eq:bubbleAlphaBeta}
\end{align}
Set
\begin{equation}
T=\alpha+\beta,
\qquad
x=\frac\beta T,
\qquad
\dd\alpha\,\dd\beta=T\,\dd T\,\dd x.
\end{equation}
Then
\begin{equation}
B(p_E^2)=\frac1{(4\pi)^{d/2}}
\int_0^1\dd x\int_0^\infty\dd T\,
T^{1-d/2}\ee^{-T[M^2+x(1-x)p_E^2]}.
\label{eq:bubbleTx}
\end{equation}
If the two denominator factors arise from two insertions on one closed heat-kernel history, the complete-history endpoint is
\begin{equation}
T\ge s_0,
\qquad
0\le x\le1.
\end{equation}
Thus, using the superscript $\mathrm{CH}$ to denote the complete-history prescription,
\begin{equation}
B^{\mathrm{CH}}_{s_0}(p_E^2)
=\frac1{(4\pi)^{d/2}}
\int_0^1\dd x\int_{s_0}^\infty\dd T\,
T^{1-d/2}\ee^{-TQ(x,p_E^2)},
\label{eq:bubbleCompleteHistory}
\end{equation}
where
\begin{equation}
Q(x,p_E^2)=M^2+x(1-x)p_E^2.
\end{equation}
In $d=4$,
\begin{equation}
\boxed{
B^{\mathrm{CH}}_{s_0}(p_E^2)
=\frac1{16\pi^2}\int_0^1\dd x\,
\Eone\left(s_0Q(x,p_E^2)\right).}
\label{eq:bubbleE1}
\end{equation}

\subsection{Higher derivatives of the finite-endpoint one-loop functional}

The general result follows directly from the partition formula in Eq.~\eqref{eq:partitionFormula}. At $\phib=0$, every first derivative $\Kop_{,i}$ vanishes and only pair blocks survive. Hence all odd derivatives of the one-loop functional vanish, while for $2n$ derivatives one obtains
\begin{equation}
\left.
\delta_1\cdots\delta_{2n}\Gamma_{1,s_0}
\right|_{\phib=0}
=
\frac12
\sum_{\pi\in\calP_2(2n)}
\Tr\,
D^n\Fop_{s_0}(\Kop_0)
\left[
\Kop_{,B_1},\ldots,\Kop_{,B_n}
\right],
\label{eq:closedEvenDerivative}
\end{equation}
where $\calP_2(2n)$ is the set of pairings of the $2n$ external derivatives and $\Kop_{,B_j}$ denotes the corresponding second background derivative. Using Eq.~\eqref{eq:closedFPT} and the Duhamel formula gives
\begin{align}
\left.
\delta_1\cdots\delta_{2n}\Gamma_{1,s_0}
\right|_{\phib=0}
={}&
-\frac12(-1)^n
\sum_{\pi\in\calP_2(2n)}
\sum_{\sigma\in S_n}
\int_{T\ge s_0}\frac{\dd T}{T}
\int_{u_j\ge0}
\left(\prod_{j=0}^{n}\dd u_j\right)
\delta\!\left(T-\sum_{j=0}^{n}u_j\right)
\nonumber\\
&\times
\Tr\!\left[
\ee^{-u_0\Kop_0}
\Kop_{,B_{\sigma(1)}}
\ee^{-u_1\Kop_0}\cdots
\Kop_{,B_{\sigma(n)}}
\ee^{-u_n\Kop_0}
\right].
\label{eq:closedEvenDerivativeDuhamel}
\end{align}
Thus an arbitrary even derivative of one closed one-loop functional contains one complete proper time $T\ge s_0$. Its $n+1$ Duhamel segments are non-negative and sum to that same $T$; they do not acquire independent lower bounds. The four-point bubble above is the case $n=2$.

\section{\texorpdfstring{\textcolor{papergreen}{Complete two-loop Schwinger-parametric treatment}}{Complete two-loop Schwinger-parametric treatment}}
\label{sec:twoLoopParametric}

{\color{papergreen}The two-loop topologies provide the first non-trivial test of the circuit domain. Their ordinary combinatorial coefficients are unchanged. Only the allowed Schwinger-parameter region is altered, and that region is determined by the primitive circuits of each graph.}

\subsection{\texorpdfstring{\textcolor{papergreen}{Figure-eight}}{Figure-eight}}

{\color{papergreen}
For a translation-invariant constant background, let $\mathcal V_d$ denote the Euclidean spacetime volume. The ordinary figure-eight density is
\begin{equation}
\frac{\Gamma_{2,\mathrm{fig8}}}{\mathcal V_d}
=\frac\lambda8
\left[\int\frac{\dd^dp}{(2\pi)^d}\frac1{p^2+M^2}\right]^2.
\end{equation}
Its Schwinger form is
\begin{equation}
\frac{\Gamma_{2,\mathrm{fig8}}}{\mathcal V_d}
=\frac\lambda8\frac1{(4\pi)^d}
\int_0^\infty\dd s_1\int_0^\infty\dd s_2\,
(s_1s_2)^{-d/2}\ee^{-M^2(s_1+s_2)}.
\label{eq:fig8Schwinger}
\end{equation}
Each internal edge is a self-loop attached to the quartic vertex. A self-loop is a one-edge graph circuit, so Eq.~\eqref{eq:circuitDomain} gives
\begin{equation}
\boxed{
s_1\ge s_0,
\qquad
s_2\ge s_0.}
\label{eq:fig8FPTDomain}
\end{equation}
Thus the figure-eight contains two independent primitive closed circulations. Its finite-$s_0$ value may be written
\begin{equation}
\frac{Gamma_{2,s_0,\mathrm{fig8}}}{\mathcal V_d}
=
\frac\lambda8\Rop_{s_0}(x,x)^2,
\end{equation}
because in this topology each one-edge loop is itself a complete circuit.

\begin{figure}[H]
    \centering
    \includegraphics[width=0.72\linewidth]{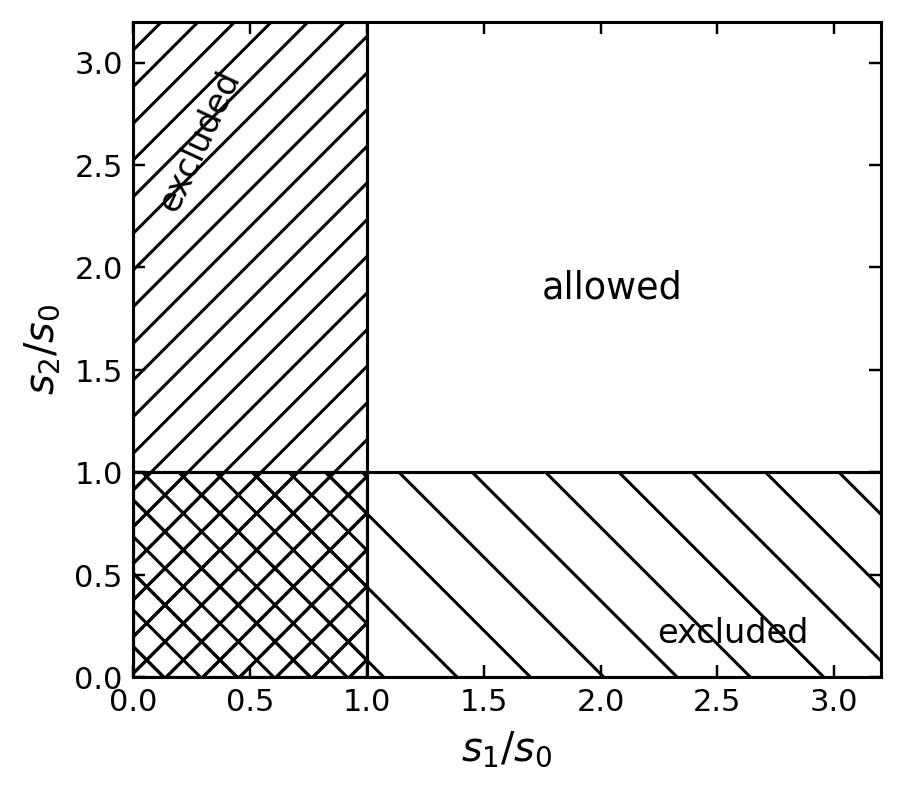}
    \caption{\textcolor{papergreen}{Schwinger-parameter domain for the figure-eight. Each self-loop is a primitive circuit, so the two independent circuit conditions are $s_1\ge s_0$ and $s_2\ge s_0$.}}
    \label{fig:figureEightDomain}
\end{figure}

Introduce a graph radial scale and projective coordinate,
\begin{equation}
T=s_1+s_2,
\qquad
x=\frac{s_1}{T}.
\end{equation}
Then
\begin{equation}
\dd s_1\dd s_2=T\dd T\dd x,
\qquad
(s_1s_2)^{-d/2}=T^{-d}[x(1-x)]^{-d/2}.
\end{equation}
A graph-global condition $T\ge s_0$ alone leaves $x\to0$ and $x\to1$ accessible and therefore does not control either one-loop subgraph. In $d=4$ the projective integral contains
\begin{equation}
\int_0^1\dd x\,[x(1-x)]^{-2},
\end{equation}
which diverges at both endpoints. The two self-loop circuit conditions in Eq.~\eqref{eq:fig8FPTDomain} remove both ultraviolet regions.
}

\subsection{\texorpdfstring{\textcolor{papergreen}{Sunset vacuum graph at constant background}}{Sunset vacuum graph at constant background}}

{\color{papergreen}
The sunset part of the constant-background effective potential is
\begin{equation}
\frac{\Gamma_{2,\mathrm{sun}}}{\mathcal V_d}
=-\frac{\lambda^2\phib^2}{12}
\int\frac{\dd^dp}{(2\pi)^d}\frac{\dd^dq}{(2\pi)^d}
\frac1{(p^2+M^2)(q^2+M^2)((p+q)^2+M^2)}.
\label{eq:sunsetVacuumMomentum}
\end{equation}
Introduce $s_1,s_2,s_3\ge0$. The momentum exponent is
\begin{equation}
s_1p^2+s_2q^2+s_3(p+q)^2
=\begin{pmatrix}p&q\end{pmatrix}
A(s)
\begin{pmatrix}p\\q\end{pmatrix},
\qquad
A(s)\equiv
\begin{pmatrix}s_1+s_3&s_3\\s_3&s_2+s_3\end{pmatrix}.
\label{eq:sunsetQuadraticMatrix}
\end{equation}
The determinant is the first Symanzik polynomial
\begin{equation}
\boxed{
U(s_1,s_2,s_3)=s_1s_2+s_2s_3+s_3s_1.}
\label{eq:Usunset}
\end{equation}
The $2d$-dimensional Gaussian integral is
\begin{equation}
\int\frac{\dd^dp\,\dd^dq}{(2\pi)^{2d}}
\ee^{-s_1p^2-s_2q^2-s_3(p+q)^2}
=\frac1{(4\pi)^d}U^{-d/2}.
\label{eq:sunsetGaussian}
\end{equation}
Hence
\begin{equation}
\frac{\Gamma_{2,\mathrm{sun}}}{\mathcal V_d}
=-\frac{\lambda^2\phib^2}{12(4\pi)^d}
\int_0^\infty\prod_{i=1}^3\dd s_i\,
U^{-d/2}\ee^{-M^2(s_1+s_2+s_3)}.
\label{eq:sunsetVacuumParametric}
\end{equation}

The sunset is a two-vertex multigraph with three parallel internal edges. Its loop number is $L=2$, but it has three support-minimal circulations: the two-edge circuits
\[
c_{12}=\{e_1,e_2\},
\qquad
c_{13}=\{e_1,e_3\},
\qquad
c_{23}=\{e_2,e_3\}.
\]
The routing-independent complete-history domain is therefore
\begin{equation}
\boxed{
s_1+s_2\ge s_0,
\qquad
s_1+s_3\ge s_0,
\qquad
s_2+s_3\ge s_0.}
\label{eq:sunsetPerLineDomain}
\end{equation}
In particular an individual edge may have $s_i=0$ while the graph remains inside the allowed domain. What is forbidden is the simultaneous collapse of the two edges that support any primitive loop circulation.

The same matrix in Eq.~\eqref{eq:sunsetQuadraticMatrix} already exhibits the ultraviolet mechanism. If it had a zero direction, a non-zero linear combination of $p$ and $q$ would have support only on edges with zero Schwinger parameter. That support would contain a two-edge circuit, contradicting Eq.~\eqref{eq:sunsetPerLineDomain}. Thus the loop Gaussian is positive throughout the circuit domain. Appendix~\ref{app:circuitDerivations} gives the corresponding result for an arbitrary graph.
}

\subsection{\texorpdfstring{\textcolor{papergreen}{Sunset two-point function with external momentum}}{Sunset two-point function with external momentum}}

{\color{papergreen}
For external Euclidean momentum $P$, choose denominators
\begin{equation}
A_1=k^2+M^2,
\qquad
A_2=q^2+M^2,
\qquad
A_3=(P-k-q)^2+M^2.
\end{equation}
The exponent is
\begin{align}
&s_1k^2+s_2q^2+s_3(P-k-q)^2
\nonumber\\
&=\begin{pmatrix}k&q\end{pmatrix}A
\begin{pmatrix}k\\q\end{pmatrix}
-2s_3P\cdot(k+q)+s_3P^2,
\end{align}
with the same positive matrix $A(s)$ as in Eq.~\eqref{eq:sunsetQuadraticMatrix}. Completing the square gives
\begin{equation}
s_3P^2-b^TA(s)^{-1}b
=P^2\frac{s_1s_2s_3}{U},
\qquad
b=s_3P\begin{pmatrix}1\\1\end{pmatrix}.
\end{equation}
The finite-$s_0$ sunset integral is therefore
\begin{equation}
\boxed{
I_{\mathrm{sun}}^{s_0}(P_E^2)
=\frac1{(4\pi)^d}
\int_{\mathcal D_{\mathrm{sun}}(s_0)}
\prod_{i=1}^3\dd s_i\,
U^{-d/2}
\exp\left[-M^2T-P_E^2\frac{s_1s_2s_3}{U}\right],}
\label{eq:sunsetExternalParametric}
\end{equation}
where
\begin{equation}
T=s_1+s_2+s_3
\end{equation}
and $\mathcal D_{\mathrm{sun}}(s_0)$ is the pairwise circuit domain in Eq.~\eqref{eq:sunsetPerLineDomain}. The second contribution is $P_E^2s_1s_2s_3$. For real Euclidean $P$, the completed-square remainder is non-negative, so the external momentum does not weaken the ultraviolet Gaussian bound.
}

\subsection{\texorpdfstring{\textcolor{papergreen}{Projective radial variables}}{Projective radial variables}}

{\color{papergreen}
Set
\begin{equation}
s_i=\rho x_i,
\qquad
x_i\ge0,
\qquad
x_1+x_2+x_3=1.
\label{eq:projectiveVars}
\end{equation}
The Jacobian is
\begin{equation}
\dd s_1\dd s_2\dd s_3
=\rho^2\dd\rho\,\dd x_1\dd x_2,
\qquad
x_3=1-x_1-x_2.
\label{eq:projectiveJacobian}
\end{equation}
Define
\begin{equation}
u(x)=x_1x_2+x_2x_3+x_3x_1.
\end{equation}
Then
\begin{equation}
U=\rho^2u(x),
\qquad
\frac{s_1s_2s_3}{U}
=\rho\frac{x_1x_2x_3}{u(x)}.
\end{equation}
Let $\Delta_2=\{(x_1,x_2,x_3):x_i\ge0,\ x_1+x_2+x_3=1\}$ denote the projective two-simplex. Equation~\eqref{eq:sunsetExternalParametric} becomes
\begin{align}
I_{\mathrm{sun}}^{s_0}(P_E^2)
={}&\frac1{(4\pi)^d}
\int_{\Delta_2}\dd x_1\dd x_2\,u(x)^{-d/2}
\nonumber\\
&\times
\int_{\rho_{\min}(x)}^\infty
\dd\rho\,\rho^{2-d}
\exp\left[
-\rho\left(
M^2+
P_E^2\frac{x_1x_2x_3}{u(x)}
\right)
\right].
\label{eq:sunsetProjective}
\end{align}
The three circuit conditions are
\[
\rho(x_1+x_2)\ge s_0,
\qquad
\rho(x_1+x_3)\ge s_0,
\qquad
\rho(x_2+x_3)\ge s_0.
\]
Hence
\begin{equation}
\boxed{
\rho_{\min}(x)
=
\frac{s_0}
{\min(x_1+x_2,x_1+x_3,x_2+x_3)}
=
\frac{s_0}{1-\max(x_1,x_2,x_3)}.}
\label{eq:rhoMin}
\end{equation}

This geometry removes precisely the projective boundaries on which a complete loop collapses. For example, $x_1\to1$ means that $s_2$ and $s_3$ are simultaneously short relative to the graph scale. The circuit $c_{23}$ then collapses and $\rho_{\min}\to\infty$. By contrast, taking only $x_3\to0$ with $x_1,x_2>0$ leaves the circuit $c_{12}$ at finite size and need not be excluded. The circuit rule therefore removes the ultraviolet subgraph boundaries without forbidding an isolated short propagator that does not by itself carry an independent loop momentum.

\begin{figure}
    \centering
    
    \includegraphics[width=0.5\linewidth]{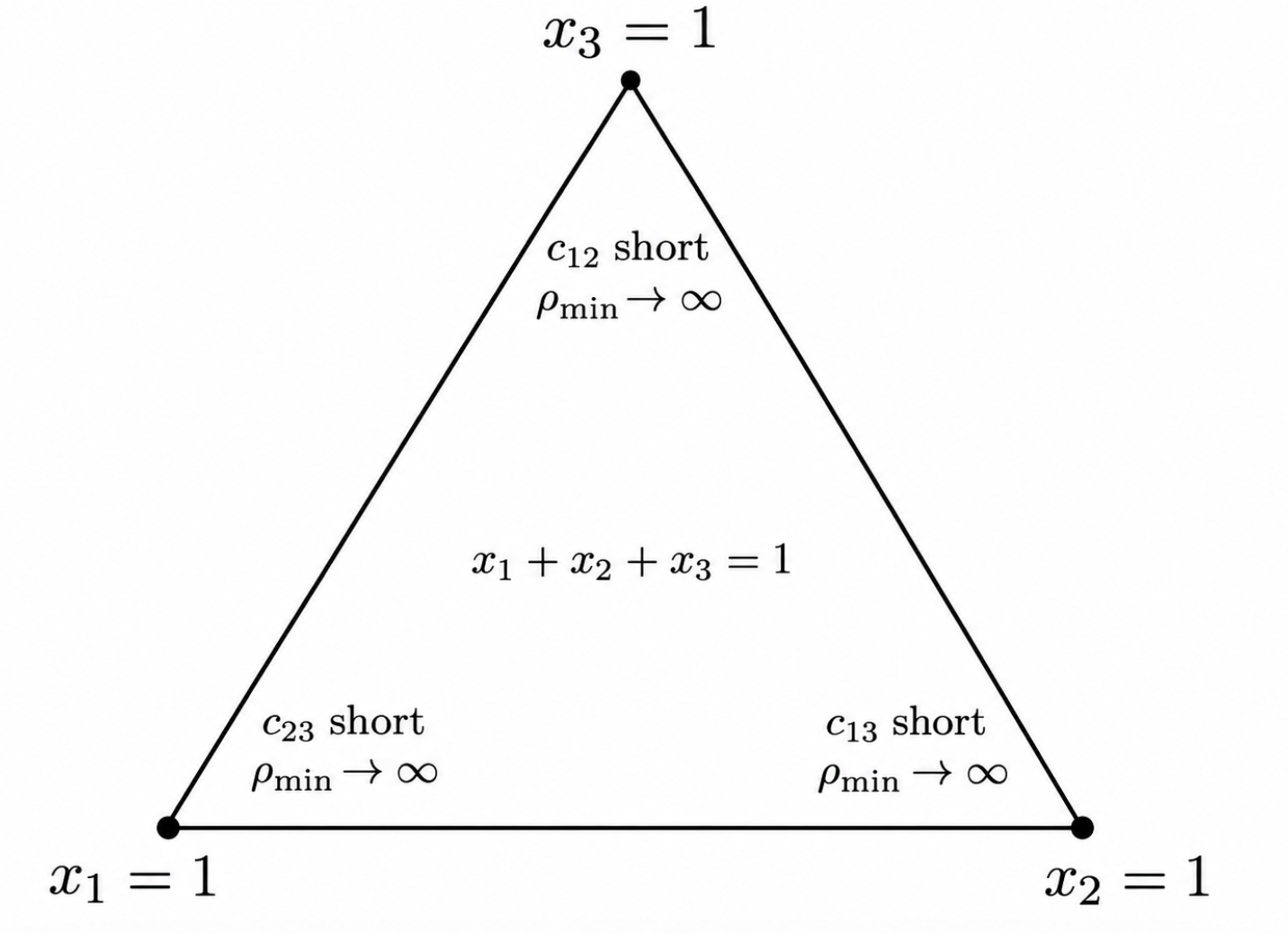}
\caption{\textcolor{papergreen}{Projective simplex for the sunset. Divergence of $\rho_{\min}$ occurs at the three vertices, where one of the two-edge primitive circuits collapses. A generic point on an edge corresponds to only one short propagator and is not by itself a loop ultraviolet limit.}}
    \label{fig:sunsetSimplex}
\end{figure}

\subsection{\texorpdfstring{\textcolor{papergreen}{Arbitrarily many quadratic background insertions inside the two-loop graphs}}{Arbitrarily many quadratic background insertions inside the two-loop graphs}}

{\color{papergreen}
Replace each exact line by Eq.~\eqref{eq:NeumannBackground}. The figure-eight becomes
\begin{equation}
\Gamma_{2,\mathrm{fig8}}
=\frac\lambda8\int\dd^dx
\sum_{n,m\ge0}(-1)^{n+m}
\left[G_0(V_{\phib}G_0)^n\right](x,x)
\left[G_0(V_{\phib}G_0)^m\right](x,x).
\label{eq:fig8AllInsertions}
\end{equation}
The sunset becomes
\begin{align}
\Gamma_{2,\mathrm{sun}}
={}&-\frac{\lambda^2}{12}\int\dd^dx\dd^dy\,\phib(x)\phib(y)
\nonumber\\
&\times\sum_{n_1,n_2,n_3\ge0}(-1)^{n_1+n_2+n_3}
\prod_{r=1}^3\left[G_0(V_{\phib}G_0)^{n_r}\right](x,y).
\label{eq:sunsetAllInsertions}
\end{align}
Thus Eq.~\eqref{eq:Gamma2Phi4} contains all ordinary two-loop diagrams with arbitrary numbers of two-leg background insertions.

For an edge $e$, let
\begin{equation}
S_e=\sum_{j=0}^{m_e}u_{e,j}
\end{equation}
be the sum of all Duhamel segments descending from that edge. Functional differentiation changes the partition of $S_e$ but does not create a new primitive circulation. The graph-level finite endpoint therefore acts on the circuit totals
\begin{equation}
\sum_{e\in c}S_e\ge s_0.
\label{eq:insertedCircuitDomain}
\end{equation}
For the figure-eight this reduces to $S_1\ge s_0$ and $S_2\ge s_0$. For the sunset it gives
\begin{equation}
S_1+S_2\ge s_0,
\qquad
S_1+S_3\ge s_0,
\qquad
S_2+S_3\ge s_0.
\end{equation}
Individual insertion segments remain non-negative and may approach zero. Only a complete primitive closed circulation is prevented from collapsing below $s_0$.
}

\section{\texorpdfstring{\textcolor{papergreen}{Comparison of finite endpoint prescriptions}}{Explicit comparison of finite endpoint prescriptions}}

{\color{papergreen}
\subsection{Two-segment example}

For positive numbers $A,B$, ordinary Schwinger factorisation gives
\begin{equation}
\frac1{AB}
=
\int_0^\infty\dd\alpha
\int_0^\infty\dd\beta\,
\ee^{-A\alpha-B\beta}.
\end{equation}
If the two segments descend from one complete history, the finite domain is
\begin{equation}
\alpha,\beta\ge0,
\qquad
\alpha+\beta\ge s_0.
\label{eq:twoGlobalDomain}
\end{equation}
By contrast, an independent-segment prescription would impose
\begin{equation}
\alpha\ge s_0,
\qquad
\beta\ge s_0.
\label{eq:twoIndependentDomain}
\end{equation}
The independent result is
\begin{equation}
I_{\mathrm{ind}}(A,B)
=
\frac{\ee^{-s_0(A+B)}}{AB},
\label{eq:independentAB}
\end{equation}
whereas the complete-history (CH) result is
\begin{equation}
I_{\mathrm{CH}}(A,B)
=
\int_{s_0}^\infty\dd T\,T
\int_0^1\dd x\,
\ee^{-T[(1-x)A+xB]}.
\label{eq:globalAB}
\end{equation}
The two expressions are unequal for generic $A,B,s_0$. The complete-history domain removes only the triangle $\alpha+\beta<s_0$; it does not remove the two strips independently.

\subsection{Five distinct graph domains}

The following possibilities should be kept separate.
\begin{enumerate}
\item \textbf{One complete operator history.} A single interval or trace history has one total $T\ge s_0$. Duhamel segments are non-negative and sum to that same $T$.
\item \textbf{Independent Duhamel segments.} Every segment created by differentiation is assigned $u_j\ge s_0$. This is a different theory because it removes strips of parameter space that are present in the operator history.
\item \textbf{Independent propagator endpoints.} Every internal edge is assigned $s_e\ge s_0$ merely because it is a propagator. This reproduces an independently damped covariance model in generic graphs and is not the complete-history construction.
\item \textbf{Circuit-complete domain.} Each edge has only $s_e\ge0$, while every primitive closed circulation obeys
\begin{equation}
\sum_{e\in c}s_e\ge s_0,
\qquad
c\in\mathcal C(G).
\label{eq:circuitRuleComparison}
\end{equation}
This is the graph-level completion of the operator-history rule obtained from the circuit result in Sec.~\ref{sec:operatorCalculus}.
\item \textbf{Single graph-wide radial (total-proper-time) condition.} Write $s_e=\rho x_e$ with $x_e\ge0$ and $\sum_{e\in E(G)}x_e=1$, so that $\rho=\sum_{e\in E(G)}s_e$. Imposing only $\rho\ge s_0$ places a constant lower bound on the overall radial Schwinger coordinate, irrespective of how many primitive circuits the graph contains, but leaves the projective directions $x_e$ unconstrained. It excludes simultaneous collapse of the full graph while still allowing a proper loop subgraph, and hence one of its circuit sums, to collapse below $s_0$.
\end{enumerate}

The differences are already visible at two loops. For the figure-eight, the two edges are self-loop circuits, so the circuit-complete domain happens to coincide with independent edge bounds:
\[
s_1\ge s_0,
\qquad
s_2\ge s_0.
\]
For the sunset, the circuit-complete domain is instead
\[
s_1+s_2\ge s_0,
\qquad
s_1+s_3\ge s_0,
\qquad
s_2+s_3\ge s_0.
\]
It suppresses exactly the collapse of a closed loop while allowing one isolated edge parameter to approach zero. A single graph-wide radial condition $\rho\ge s_0$ would not prevent, for example, $s_2,s_3\to0$ with $s_1$ fixed, which is the ultraviolet contraction of the circuit $c_{23}$.
}

\section{\texorpdfstring{\textcolor{paperred}{Repeated derivatives on an open history}}{Repeated derivatives on an open history}}

{\color{papergreen}
{\color{paperred}The kernel $\Rop_{s_0}(\Kop)$ is the open interval associated with a complete spectral history. Functional derivatives insert operators on this interval and divide its original total proper time into non-negative pieces. They do not assign $\Rop_{s_0}$ independently to every edge of a multi-loop graph. After the interaction vertices are joined, the endpoint conditions are fixed by the graph circuits in Eq.~\eqref{eq:circuitDomain}.}
}
\subsection{Repeated derivatives of one open history}

Equation~\eqref{eq:sewingVariation} produces the open-history operator $\Rop_{s_0}(\Kop)$. Repeated functional derivatives of this operator keep one total proper-time condition. The first derivative is
\begin{equation}
    \frac{\delta\mathcal{R}_{s_0}(\mathcal{K})}{\delta \phi(x_i)}=-\int_{s_0}^\infty \dd s\int_0^s \dd u\,e^{-u\mathcal{K}}\frac{\delta \mathcal{K}}{\delta \phi(x_i)}e^{-(s-u)\mathcal{K}}=-\int_{\substack{s_1,s_2\geq 0\\ s_1+s_2\geq s_0}}
\dd s_1\,\dd s_2\,
\left(\,e^{-s_1\mathcal{K}}\frac{\delta \mathcal{K}}{\delta \phi(x_i)}e^{-s_2\mathcal{K}}
\right)
\label{eq:RfirstDerivative}
\end{equation}
Further derivatives will act on this object by the product rule. Suppose at a stage of the calculation where $q$ derivatives have been applied, a term where a total of $m \leq q$ derivatives have been applied on exponential factors is considered. Objects ${I}^{(m)}$ generated in this way take the form
\begin{equation}
    {I}^{(m)}=(-1)^m\int_{\substack{s_1,s_2,\cdots s_{m+1}\geq 0\\ \sum^{m+1}_{i=1} s_i\geq s_0}}
\dd s_1\,\dd s_2\,\cdots\,\dd s_{m+1}
\left(\,e^{-s_1\mathcal{K}}{\tilde{\delta}} \mathcal{K} e^{-s_2\mathcal{K}}{\tilde{\delta}} \mathcal{K} e^{-s_3\mathcal{K}}\cdots {\tilde{\delta}} \mathcal{K} e^{-s_{m+1}\mathcal{K}}
\right)
\label{eq:ImGeneral}
\end{equation}
where $\tilde{\delta}\mathcal{K}$ takes either the form $\delta \mathcal{K}/\delta \phi(x_{\xi_1})$ or $\delta^2 \mathcal{K}/\delta \phi(x_{\xi_1})\delta \phi(x_{\xi_2})$ with $\xi_1,\xi_2 \in \{i_1,i_2,\ldots,i_n\}$. These come from the $q-m$ derivatives which do not act on exponential factors. Equation~\eqref{eq:RfirstDerivative} gives the case $m=1$. At the next step a further derivative simply divides one of the existing heat-kernel intervals. Taking the $b$th interval in $I^{(m)}$, let a further functional derivative $\delta/\delta \phi(x_{i_{q+1}})$ act on the $b$th exponential factor in ${I}^{(m)}$, where $1\leq b \leq m+1$. We obtain
\begin{equation}
\begin{aligned}
{I}^{(m+1)}
&=(-1)^m
\int_{\substack{s_1,s_2,\cdots,s_{m+1}\geq 0\\
\sum_{i=1}^{m+1}s_i\geq s_0}}
\dd s_1\,\dd s_2\cdots \dd s_{m+1}\,
\Bigg(
e^{-s_1\mathcal{K}}\tilde{\delta}\mathcal{K}
e^{-s_2\mathcal{K}}\tilde{\delta}\mathcal{K}
\cdots
e^{-s_{b-1}\mathcal{K}}\tilde{\delta}\mathcal{K}
\\
&\qquad\times
\left[
-\int_0^{s_b}d\alpha\,
e^{-\alpha\mathcal{K}}
\left(\frac{\delta\mathcal{K}}{\delta \phi{(x_{i_{q+1}})}}\right)
e^{-(s_b-\alpha)\mathcal{K}}
\right]
\tilde{\delta}\mathcal{K}
e^{-s_{b+1}\mathcal{K}}
\cdots
\tilde{\delta}\mathcal{K}
e^{-s_{m+1}\mathcal{K}}
\Bigg).
\end{aligned}
\end{equation}
We now make the coordinate change $s_b \rightarrow \beta=s_b-\alpha$. The Jacobian factor has magnitude one. We now obtain
\begingroup\footnotesize
\begin{equation}
\begin{aligned}
{I}^{(m+1)}
&=
(-1)^{(m+1)}
\int_{\substack{\alpha,\beta,s_1,s_2,\ldots,s_{b-1},s_{b+1},\ldots,s_{m+1}\geq 0\\
\alpha+\beta+\sum_{i\neq b}s_i\geq s_0}}
\dd\alpha\,\dd\beta\,\dd s_1\,\dd s_2\cdots \dd s_{m+1}
\\
&\quad\times
\left(
e^{-s_1\mathcal{K}}\tilde{\delta}\mathcal{K}
e^{-s_2\mathcal{K}}\tilde{\delta}\mathcal{K}
e^{-s_3\mathcal{K}}\cdots
e^{-s_{b-1}\mathcal{K}}\tilde{\delta}\mathcal{K}
e^{-\alpha\mathcal{K}}
\frac{\delta\mathcal{K}}{\delta\phi(x_{i_{q+1}})}
e^{-\beta\mathcal{K}}
\tilde{\delta}\mathcal{K}
e^{-s_{b+1}\mathcal{K}}
\tilde{\delta}\mathcal{K}
\cdots
e^{-s_{m}\mathcal{K}}
\tilde{\delta}\mathcal{K}
e^{-s_{m+1}\mathcal{K}}
\right)
\end{aligned}
\end{equation}
\endgroup
After relabelling the integration variables this has exactly the form of Eq.~\eqref{eq:ImGeneral}, so the same structure is preserved at every derivative order.

We now evaluate at $\phi=0$, where the first background derivative of $\mathcal K$ vanishes. Odd derivatives are therefore zero. Writing the even order as $n=2r$, let $\calP_2(2r)$ denote the pairings of the external functional derivatives and let $\mathcal K_{,B_j}$ denote the corresponding second background derivative. The result can then be written without additional combinatorial notation as
\begin{equation}
\begin{aligned}
\left.
\delta_1\cdots\delta_{2r}\,
\mathcal R_{s_0}(\mathcal K)
\right|_{\phi=0}
={}&(-1)^r
\sum_{\pi\in\calP_2(2r)}
\sum_{\sigma\in S_r}
\int_{\substack{u_0,\ldots,u_r\ge0\\
\sum_{j=0}^{r}u_j\ge s_0}}
\prod_{j=0}^{r}\dd u_j
\nonumber\\
&\times
\left.
 e^{-u_0\mathcal K}\mathcal K_{,B_{\sigma(1)}}
 e^{-u_1\mathcal K}\cdots
 \mathcal K_{,B_{\sigma(r)}}e^{-u_r\mathcal K}
\right|_{\phi=0}.
\end{aligned}
\label{eq:derivativesonR}
\end{equation}
With $(2r-1)!!$ denoting the odd double factorial, there are $(2r-1)!!$ pairings and $r!$ orderings of the noncommuting insertions, giving $(2r)!/2^r$ ordered terms. Equation~\eqref{eq:derivativesonR} makes the geometric point explicit: repeated differentiation partitions one complete history into $r+1$ non-negative segments whose \emph{sum}, rather than each segment separately, is bounded below by $s_0$.

\subsection{Diagrammatic bookkeeping}
\label{sec:diagrammatic}

As shown in Fig.~\ref{fig:schematicworldlineness}, we express the Schwinger-parameter representation of the open history $\Rop_{s_0}(x,y)$ as a line.

\[
\mathcal{R}_{s_0}(x,y)
=
\simpleLineDiagram
\]

As shown in Fig.~\ref{fig:operatorAncestry}, we represent insertions with dots partitioning the line. We label the derivatives used to produce the insertion by brackets beneath the line. For example,

\[
\begin{aligned}
\insertedLineDiagram
&=(-1)^2
\int_{\substack{s_1,s_2,s_3 \geq 0\\
s_1+s_2+s_3\geq s_0}}
\dd s_1\,\dd s_2\,\dd s_3
\\
&\quad\times
\left.
\left[
e^{-s_1\mathcal{K}}
\mathcal{K}_{,12}
e^{-s_2\mathcal{K}}
\mathcal{K}_{,34}
e^{-s_3\mathcal{K}}
\right](x,y)
\right|_{\phi=0},
\end{aligned}
\]
If an insertion is produced by two derivatives but we do not wish to specify which pair of points the derivatives act on, we replace the label $(x_i,x_j)$ by $(2)$; more generally $(r)$ denotes the derivative order of an insertion. At $\phi=0$ in $\lambda\phi^4$ theory, only second-background-derivative insertions survive. At a general background, first-derivative insertions are also present, and more general theories may admit higher derivative orders. Diagrammatically, the left-hand side of Eq.~\eqref{eq:derivativesonR} scales as
\[
   \left.
\frac{\delta}{\delta \phi(x_{i_n})}\cdots
\frac{\delta}{\delta \phi(x_{i_2})}
\frac{\delta}{\delta\phi(x_{i_1})}
\mathcal{R}_{s_0}(x,y)
\right|_{\phi=0} \sim \sum \longLineDiagram
\]
where we have $n/2$ insertions. We must sum over all different combinations of derivatives, noting that different orderings of derivatives in each individual insertion are considered the same. There are $n!$ ways to order the $n$ derivatives, but we have overcounted by a factor of $2^{n/2}$. This leads to the $n!/2^{n/2}$ terms as follows from the pairing count. Thus, we can write
\begingroup\scriptsize
\[
   \left.
\frac{\delta}{\delta \phi(x_{i_n})}\cdots
\frac{\delta}{\delta \phi(x_{i_2})}
\frac{\delta}{\delta\phi(x_{i_1})}
\mathcal{R}_{s_0}(x,y)
\right|_{\phi=0} = \frac{1}{2^{n/2}} \sum_{\text{all permutations}} \longLineDiagram.
\]
\endgroup
{\color{papergreen}
If an operator expression genuinely contains a product of open-history kernels, the external derivatives are distributed among those factors in the usual Leibniz fashion. For example,
\[
\begin{gathered}
\left.
\frac{\delta}{\delta \phi(x_{1})}
\frac{\delta}{\delta \phi(x_{2})}
\frac{\delta}{\delta \phi(x_{3})}
\frac{\delta}{\delta\phi(x_{4})}
\mathcal{R}(x,y)_{s_0}
\mathcal{R}(w,v)_{s_0}
\right|_{\phi=0}
\sim \\[1ex] \sum
\diagramOne
+
\diagramTwo
+
\diagramThree.
\end{gathered}
\]
This identity concerns differentiation of the open-interval operators themselves. In a multi-loop graph, it must not be read as a rule that every internal edge is independently replaced by $\mathcal R_{s_0}$. The graph topology determines which edge totals enter a complete closed history.

The bookkeeping can be stated as follows.
\begin{itemize}
\item For each open-interval factor, distribute the non-vanishing functional derivatives over the allowed ordered insertions. At $\phi=0$ in $\lambda\phi^4$ theory, only the second background derivative of $K$ survives.
\item If an edge $e$ contains $m_e$ Duhamel insertions, write its descendant segments as $u_{e,0},\ldots,u_{e,m_e}\ge0$ and define the edge total
\begin{equation}
S_e=\sum_{j=0}^{m_e}u_{e,j}.
\label{eq:edgeDescendantTotal}
\end{equation}
The insertions partition $S_e$ but do not create independent endpoint conditions on the individual $u_{e,j}$.
\item Preserve the operator ordering along every open interval and the ordinary local vertex combinatorics.
\item Once the propagating segments are sewn into a graph $G$, identify its primitive circuits and impose
\begin{equation}
\sum_{e\in c}S_e\ge s_0,
\qquad
c\in\mathcal C(G).
\label{eq:descendantCircuitRule}
\end{equation}
For a one-edge self-loop this reduces to the one-edge circuit condition $S_e\ge s_0$. For a multi-edge circuit it is the sum over the participating edge totals that is bounded.
\item Sum over all derivative assignments and operator orderings with the same permutation multiplicities as in the ordinary functional calculus.
\end{itemize}
Thus Duhamel ancestry and graph topology play different roles. Functional differentiation determines how an edge total is partitioned; local momentum conservation determines which collections of edge totals form complete closed virtual histories.
}

\subsection{\texorpdfstring{\textcolor{papergreen}{Complete-history Feynman rules from graph circuits}}{Complete-history Feynman rules from graph circuits}}

{\color{papergreen}
The effective action generates one-particle-irreducible vertices. The tree term is $S[\phib]$, the one-loop term is the closed spectral trace, and higher-loop topologies come from the connected cumulants in Eq.~\eqref{eq:GammaCumulantMaster}. In $\lambda\phi^4$ theory the only bare fluctuation vertices are
\begin{equation}
S^{(3)}(x_1,x_2,x_3)
=
\lambda\phib(x_1)
\delta(x_1-x_2)
\delta(x_1-x_3)
\end{equation}
and
\begin{equation}
S^{(4)}(x_1,x_2,x_3,x_4)
=
\lambda
\delta(x_1-x_2)
\delta(x_1-x_3)
\delta(x_1-x_4).
\end{equation}
A derivative acting on the explicit $\phib$ in a cubic vertex attaches an external leg at that vertex. A derivative acting on an open-history kernel inserts $\delta\Kop$ and partitions that interval according to the Duhamel formula. Neither operation changes the ordinary local vertex combinatorics.

The endpoint assignment is topological and does not depend on a chosen momentum routing. For a connected tree,
\begin{equation}
L=|E|-|V|+1=0,
\qquad
\ker B=\{0\},
\end{equation}
so the vertex delta functions fix every internal momentum in terms of the external momenta. For $L>0$, the general solution is
\begin{equation}
q=q^{(0)}(p)+Z\ell,
\qquad
BZ=0,
\label{eq:routingIndependentLoopSolution}
\end{equation}
where the columns of $Z$ are any basis of the cycle space and $\ell=(\ell_1,\ldots,\ell_L)$ are loop coordinates. Replacing $Z$ by $ZA$ with $A\in GL(L,\mathbb R)$, the general linear group of invertible real $L\times L$ matrices, merely changes the loop-momentum coordinates. It can move the explicit appearance of a loop variable from one propagator denominator to another, but it cannot change the graph circuits. The finite endpoint is therefore not assigned according to whether a chosen routing makes a particular propagator carry an explicit loop variable. It belongs to the routing-independent closed circulation that supports that momentum freedom.

A Duhamel insertion does not alter this conclusion. For example, let one edge of a two-edge circuit contain a single insertion. If that edge is partitioned into $u_{1,0},u_{1,1}\ge0$ and the second edge has parameter $s_2\ge0$, then
\begin{equation}
S_1=u_{1,0}+u_{1,1},
\qquad
S_2=s_2,
\qquad
\boxed{u_{1,0}+u_{1,1}+s_2\ge s_0}.
\label{eq:insertedTwoEdgeCircuitExample}
\end{equation}
There is no condition $u_{1,0}\ge s_0$, $u_{1,1}\ge s_0$, or $s_2\ge s_0$ separately. The insertion marks and partitions propagation already belonging to the circuit; it does not create another complete closed history. By contrast, a bridge belongs to no circuit and receives no closed-history endpoint. Figure~\ref{fig:circuitInsertionBridge} summarises these two cases.

\begin{figure}[H]
\centering
\includegraphics[width=0.8\linewidth]{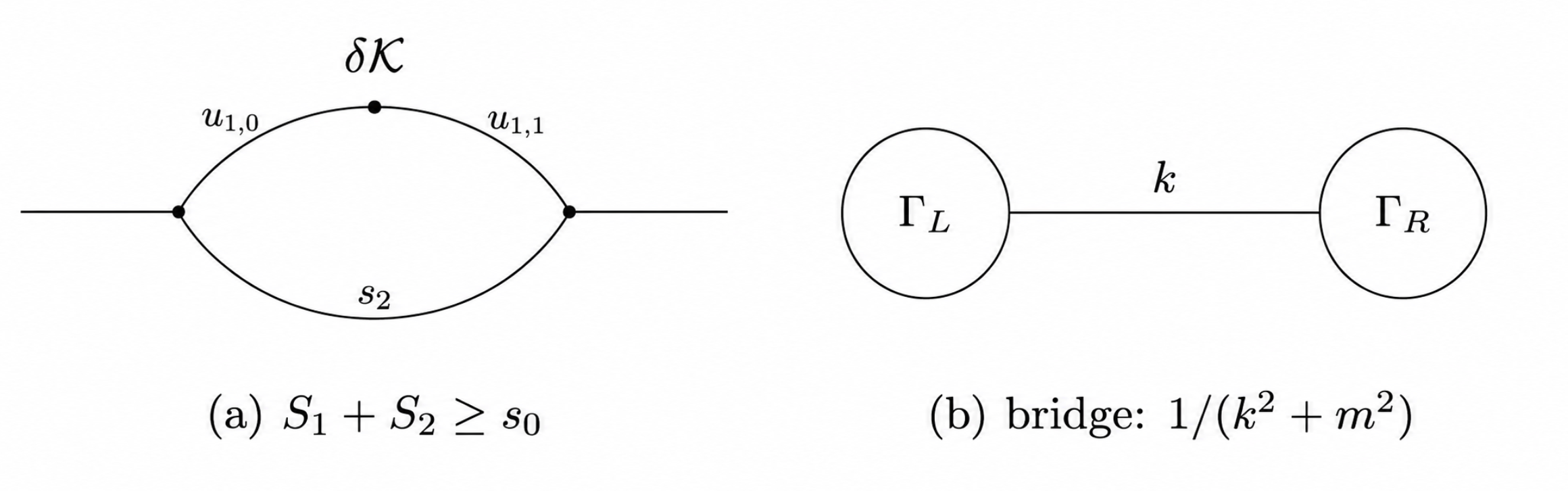}
\caption{\textcolor{papergreen}{Circuit and insertion bookkeeping. (a) A two-edge closed virtual circuit with one Duhamel insertion on the upper edge. The insertion partitions $S_1=u_{1,0}+u_{1,1}$ but creates no independent endpoint; the single complete-history condition is $S_1+S_2\ge s_0$. (b) A bridge between two 1PI blocks lies on no circuit. Its proper-time integral is unrestricted and gives the ordinary propagator. Momentum labels are routing variables; the circuit/bridge distinction is graph invariant.}}
\label{fig:circuitInsertionBridge}
\end{figure}

For a fixed graph $G$, the momentum-space rules are:
\begin{enumerate}
\item Draw the ordinary 1PI topology and keep its ordinary vertex factor, momentum-conservation delta functions, symmetry factor, and loop-momentum measure.
\item Assign one non-negative Schwinger parameter $s_e$ to each internal edge. If functional differentiation has partitioned edge $e$, replace $s_e$ by the sum $S_e$ of its descendant Duhamel segments as in Eq.~\eqref{eq:edgeDescendantTotal}.
\item Solve the local momentum-conservation equations or, equivalently, construct the incidence matrix $B$. The primitive complete closed virtual histories are the support-minimal non-zero solutions of $Bz=0$, hence the circuits $c\in\mathcal C(G)$.
\item Impose the single universal boundary
\begin{equation}
\sum_{e\in c}S_e\ge s_0
\qquad
\text{for every }c\in\mathcal C(G).
\label{eq:feynmanCircuitRule}
\end{equation}
No internal edge is assigned an independent lower bound merely because it is a propagator. Only the circuit sums in Eq.~\eqref{eq:feynmanCircuitRule} are constrained; a one-edge self-loop is the special case in which a circuit sum consists of a single edge total.
\item Perform the Schwinger and loop-momentum integrations and sum all derivative assignments and operator orderings. {\color{paperred}As a separate formal correspondence check, removing the boundary through $s_0\to0^+$ makes the circuit inequalities inactive and recovers the ordinary graph. The value $s_0=0$ is not part of the defined FPT spectral theory and this comparison is not used in its definition.}
\end{enumerate}

The bridge rule follows rather than being added separately. An internal edge $b$ is a bridge if and only if it belongs to no circuit. Its Schwinger parameter therefore occurs in no endpoint inequality, so
\begin{equation}
\int_0^\infty\dd s_b\,\ee^{-s_b(p_b^2+m^2)}
=
\frac1{p_b^2+m^2}.
\label{eq:bridgeOrdinaryPropagator}
\end{equation}
Its momentum is fixed by the external and, when appropriate, cut momenta on one side of the bridge. It is not an independent loop variable.

Connected one-particle-reducible amplitudes are then assembled consistently with the exact Legendre transform. The 1PI vertices are supplied by $\Gamma$, while the connecting full two-point function is
\begin{equation}
W^{(2)}=(\Gamma^{(2)})^{-1}.
\end{equation}
At tree level this gives the ordinary bridge in Eq.~\eqref{eq:bridgeOrdinaryPropagator}.

{\color{paperred}The open-history kernel $\Rop_{s_0}(\Kop)$ remains an essential operator object because repeated insertions divide one pre-existing proper-time interval. It must be distinguished from a generic internal bridge of a connected amplitude. The open history retains the endpoint of the closed history from which the interval relation was obtained; a bridge lies on no circuit and is fixed by the connected/1PI composition law. This preserves both the interval--circle identity and the exact Legendre relation.}
}

\textcolor{orange}{\subsection{Failure of universal self-energy resummation}}
For a fixed connected graph $G$, let $\mathcal N_G$ collect its ordinary local coupling and symmetry prefactors, and let $m_e$ denote the mass associated with internal edge $e$. The Euclidean scattering amplitude is
\begin{equation}
\mathcal{M}_{G}(p)
=
\mathcal{N}_{G}
\int
\prod_{r=1}^{L}
\frac{\dd^{d}\ell_{r}}{(2\pi)^{d}}\,
I_{G}\!\left(\{q_{e}(\ell,p)\};s_{0}\right),
\end{equation}
where \(L=|E|-|V|+1\) is the loop number, \(\ell_{r}\) are the
independent loop momenta, and $q_{e}$ is the momentum carried
by internal edge \(e\). The proper-time factor is
\begin{equation}
I_{G}\!\left(\{q_{e}\};s_{0}\right)
=
\int_{\mathcal D_{G}(s_{0})}
\left(\prod_{e\in E} \dd s_{e}\right)
\exp\!\left[
-\sum_{e\in E}
s_{e}\left(q_{e}^{2}+m_{e}^{2}\right)
\right],
\end{equation}
with graph domain
\begin{equation}
\mathcal D_{G}(s_{0})
=
\left\{
(s_{e})_{e\in E}\in\mathbb{R}_{\geq 0}^{|E|}
\;:\;
\sum_{e\in c}s_{e}\geq s_{0}
\quad
\text{for every }
c\in\mathcal{C}(G)
\right\},
\end{equation}
where \(\mathcal{C}(G)\) denotes the set of primitive circuits of
\(G\). Equivalently, defining the edge-denominator variables
\begin{equation}
a_{e}
=
q_{e}^{2}+m_{e}^{2},
\end{equation}
and writing $\mathbf a=(a_e)_{e\in E}$, the proper-time factor may be written as
\begin{equation}
I_{G}(\mathbf{a};s_{0})
=
\int_{\mathcal D_{G}(s_{0})}
\left(\prod_{e\in E} \dd s_{e}\right)
\exp\!\left(
-\sum_{e\in E}a_{e}s_{e}
\right).
\end{equation}
Define
\begin{equation}
    c_i(s_{E\setminus i})=\max\left[0,\max_{\substack{c\in\mathcal C(G)\\ i\in c}}\left(s_0-\sum_{\substack{e\in c\\ e\neq i}}s_e\right)\right].
\end{equation}
We can isolate the contribution from propagator $i$ by
\begin{equation}
I_G=\int_{\mathcal D_{G\setminus i}^*}\prod_{e\neq i}\dd s_e\,e^{-\sum_{e\neq i}a_es_e}\int_{c_i(s_{E\setminus i})}^{\infty}\dd s_i\,e^{-a_is_i}.
\end{equation}
Here
\begin{equation}
\mathcal D_{G\setminus i}^*
=
\left\{
(s_e)_{e\neq i}\in\mathbb R_{\ge0}^{|E|-1}:
\sum_{e\in c}s_e\ge s_0
\;\text{for every }c\in\mathcal C(G)\text{ with }i\notin c
\right\}
\end{equation}
is the domain imposed by all parent circuits that do not contain edge $i$.
If propagator $i$ is dressed by one or more insertion subgraphs, the total backbone proper time that participates in parent circuits has the lower endpoint $c_i(s_{E\setminus i})$, while primitive circuits contained entirely within the inserted 1PI subgraphs retain their own endpoint $s_0$. After integrating over loop momenta internal to the insertion, the general proper-time factor takes the form
\begin{equation}
\int_{\mathcal D_{G\setminus i}^*}\prod_{e\neq i}\dd s_e\,e^{-\sum_{e\neq i}a_es_e}\int_{c_i(s_{E\setminus i})}^{\infty}\dd s_i\,e^{-a_is_i}
\to
\int_{\mathcal D_{G\setminus i}^*}\prod_{e\neq i}\dd s_e\,e^{-\sum_{e\neq i}a_es_e}f\!\left(c_i(s_{E\setminus i}),a_i\right).
\end{equation}
The dependence of \(f(c_i(s_{E\setminus i}),a_i)\) on both \(a_i\) and the circuit-determined endpoint \(c_i(s_{E\setminus i})\) prevents insertions from being represented by a momentum-dependent factor alone. For example, the one-loop tadpole correction gives
\begin{equation}
f\!\left(c_i(s_{E\setminus i}),a_i\right)=\frac{\lambda}{2}\frac{e^{-a_ic_i(s_{E\setminus i})}\left[1+a_ic_i(s_{E\setminus i})\right]}{a_i^2}\Rop_{s_0}(x,x)
\end{equation}
compared with
\begin{equation}
f\!\left(c_i(s_{E\setminus i}),a_i\right)=\frac{e^{-a_ic_i(s_{E\setminus i})}}{a_i}
\end{equation}
for the tree-level propagator. Hence, self-energy corrections cannot in general be resummed into a universal momentum-dependent dressed propagator: different insertion topologies inherit different dependences on $c_i(s_{E\setminus i})$, leaving irreducible dependence on the surrounding circuit parameters. This does not invalidate the standalone identity $W^{(2)}=(\Gamma^{(2)})^{-1}$ or the full two-point function used as a bridge between 1PI blocks. It states instead that an off-shell two-point insertion embedded inside a larger closed circuit generally cannot be replaced by a parent-independent momentum-only dressed line.

\section{\texorpdfstring{\textcolor{paperred}{Euclidean bounds and the all-order local construction}}{Euclidean bounds and the all-order local construction}}
\label{sec:scalarNonperturbative}

{\color{papergreen}
The scalar model has a useful property that is absent in spinor QED. For a real background and $\lambda\ge0$,
\begin{equation}
\Kop_{\phib}
=
-\partial^2+m^2+\frac{\lambda}{2}\phib^2
=
\Kop_0+V_{\phib},
\qquad
V_{\phib}(x)\ge0.
\label{eq:scalarPositivePotential}
\end{equation}
Hence
\begin{equation}
\boxed{
\Kop_{\phib}\ge\Kop_0\ge m^2.
}
\label{eq:scalarOperatorOrder}
\end{equation}
This simple order relation gives nonperturbative control of the real Euclidean background.

\subsection{Background-uniform heat-kernel domination}

{\color{paperred}For a non-negative local potential, the heat kernel can be written as the free massive heat kernel multiplied by an average of
$\exp[-\int_0^T V_{\phib}(x(\tau))\,\dd\tau]$
over paths joining $y$ to $x$. The exponential weight lies between zero and one because $V_{\phib}\ge0$. Therefore}
\begin{equation}
\boxed{
0\le
\ee^{-T\Kop_{\phib}}(x,y)
\le
\frac{\exp[-m^2T-|x-y|^2/(4T)]}{(4\pi T)^{d/2}}.
}
\label{eq:scalarFKbound}
\end{equation}
No small-background assumption is used.

Integrating Eq.~\eqref{eq:scalarFKbound} gives the same bound for the open history,
\begin{equation}
0\le
\Rop_{s_0}(\Kop_{\phib})(x,y)
\le
\int_{s_0}^{\infty}
\frac{\dd T}{(4\pi T)^{d/2}}
\exp\left[-m^2T-\frac{|x-y|^2}{4T}\right].
\label{eq:scalarOpenDomination}
\end{equation}
On a finite region $X$, let $\Tr_X$ denote the trace restricted to $X$. Then
\begin{equation}
0\le
\Tr_X\ee^{-T\Kop_{\phib}}
\le
|X|(4\pi T)^{-d/2}\ee^{-m^2T}.
\label{eq:scalarLocalHeatTrace}
\end{equation}
These inequalities are uniform in the value and shape of the real background field.

\subsection{Proper-time shells and locality}

Choose $M>1$ and
\begin{equation}
a_j=s_0M^{2j},
\qquad
\kappa_j=a_j^{-1/2},
\qquad
j=0,1,\ldots.
\end{equation}
Define the positive shell operator
\begin{equation}
Q_j(\Kop)
=
\int_{a_j}^{a_{j+1}}
\frac{\dd T}{T}\,\ee^{-T\Kop}.
\label{eq:scalarShellDef}
\end{equation}
Then
\begin{equation}
-\frac12\int_{s_0}^{\infty}\frac{\dd T}{T}\Tr\ee^{-T\Kop_{\phib}}
=
-\frac12\sum_{j=0}^{\infty}\Tr Q_j(\Kop_{\phib}).
\label{eq:scalarShellSum}
\end{equation}
For $d=4$, Eq.~\eqref{eq:scalarLocalHeatTrace} gives
\begin{equation}
\boxed{
0\le
\Tr_X Q_j(\Kop_{\phib})
\le
\frac{|X|}{16\pi^2}
\int_{a_j}^{a_{j+1}}
\frac{\dd T}{T^3}\,\ee^{-m^2T}.
}
\label{eq:scalarShellTraceBound}
\end{equation}
The mass makes the sum over large $j$ absolutely convergent. The retained $s_0$ sets the lower spectral boundary of the short-history domain.

The same estimate gives spatial locality. Since $T\le a_{j+1}$ inside one shell,
\begin{align}
0\le Q_j(\Kop_{\phib})(x,y)
&\le
\frac{\exp[-|x-y|^2/(4a_{j+1})]}{(4\pi)^2}
\int_{a_j}^{a_{j+1}}
\frac{\dd T}{T^3}\,\ee^{-m^2T}
\nonumber\\
&=
\boxed{
A_j\,
\exp\left[-\frac{\kappa_j^2|x-y|^2}{4M^2}\right],
}
\label{eq:scalarShellLocality}
\end{align}
where $A_j$ is the finite positive coefficient displayed on the first line. Thus each complete shell is local on the shell length scale $\kappa_j^{-1}$, uniformly in the real background.

{\color{paperred}\paragraph{Background-uniform shell bound.}
For $m^2>0$, $\lambda\ge0$, and $s_0>0$, every proper-time shell of the scalar complete-history one-loop functional is finite on a compact Euclidean volume and obeys Eqs.~\eqref{eq:scalarShellTraceBound} and \eqref{eq:scalarShellLocality} uniformly for all real backgrounds $\phib$. Equation~\eqref{eq:scalarOperatorOrder} and the positive path weight described above give Eq.~\eqref{eq:scalarFKbound}; integrating it over one finite proper-time interval gives the trace and spatial bounds directly.}

\subsection{\texorpdfstring{\textcolor{paperred}{All-order local construction at fixed $s_0$}}{All-order local construction at fixed s0}}

{\color{paperred}
The heat-kernel bounds above control the virtual spectral part of the theory. The local interaction still has to be multiplied and sewn at coincident spacetime points. We do this without treating a point source as a physical state. Let
\begin{equation}
\phi(f)=\int\dd^4x\,f(x)\phi(x),
\qquad
f\in C_c^\infty(\mathbb R^4),
\label{eq:smearedField}
\end{equation}
so every field appearing in a physical source history is first smeared with a smooth compact function. Products are defined when the insertion points are distinct and are then extended to coincident points as distributions. This is the standard way to define local quantum fields without multiplying singular point values directly \cite{Scharf1995}.

For a graph $G$, let $\Theta$ denote the Heaviside step function and define the history factor
\begin{equation}
\chi_{G,s_0}(s)
=
\prod_{c\in\mathcal C(G)}
\Theta\!\left(\sum_{e\in c}s_e-s_0\right).
\label{eq:historyFactor}
\end{equation}
If $G_1$ is a subgraph of $G_2$, every circuit already present in $G_1$ remains a circuit of $G_2$. Therefore the final history factor automatically includes every earlier circuit condition,
\begin{equation}
\boxed{\chi_{G_2,s_0}\chi_{G_1,s_0}=\chi_{G_2,s_0}.}
\label{eq:historyAbsorption}
\end{equation}
This elementary identity is what makes the global circuit rule compatible with step-by-step interaction sewing.

A few small graphs make the rule transparent. Joining two previously disconnected pieces by a single bridge creates no new closed circulation, so the bridge parameter still runs from zero to infinity. Closing an open path of total proper time $u_1+\cdots+u_r$ with one new edge of proper time $v$ creates one circuit and adds only
\begin{equation}
 u_1+\cdots+u_r+v\ge s_0.
\label{eq:sewingSingleCircuitExample}
\end{equation}
If three parallel edges are sewn between the same two vertices, the resulting theta graph has three primitive circuits and therefore the three conditions $s_1+s_2\ge s_0$, $s_1+s_3\ge s_0$, and $s_2+s_3\ge s_0$. These examples show why the endpoint follows the completed circulation rather than the individual line.

Let $F_{G_1}$ and $H_{G_2}$ be two partially constructed histories. For every allowed contraction $\sigma$ that joins them into a graph $G$, define
\begin{equation}
(F\star_{s_0}H)_G
=
\sum_{\sigma:G_1\#_\sigma G_2=G}
\chi_{G,s_0}\,
\operatorname{Contr}_{\sigma}(F_{G_1}\otimes H_{G_2}).
\label{eq:historyStarProduct}
\end{equation}
The ordinary contraction operation is associative. Equation~\eqref{eq:historyAbsorption} then gives
\begin{equation}
\boxed{
(F\star_{s_0}H)\star_{s_0}Q
=
F\star_{s_0}(H\star_{s_0}Q).}
\label{eq:historyAssociativity}
\end{equation}
Indeed, both orders of sewing generate the same final graph $K$. The two intermediate history factors reduce to $\chi_{K,s_0}$ by Eq.~\eqref{eq:historyAbsorption}. The circuit prescription is therefore independent of the order in which local vertices are assembled.

The same argument shows that the retained history boundary does not select a preferred spacetime point or direction. Proper time and every circuit sum $\sum_{e\in c}s_e$ are scalars. A translation or rotation of all insertion points changes the spacetime arguments of the fields but leaves $\chi_{G,s_0}$ unchanged. The continuation to physical time introduces no preferred direction because the same proper-time sums remain scalar. Thus the local spacetime transformation laws of the scalar field are preserved by the history restriction.

At coincident points the FPT factor satisfies
\begin{equation}
0\le \chi_{G,s_0}\le1.
\label{eq:historyFactorBound}
\end{equation}
It can remove part of the Schwinger domain but cannot create a stronger short-distance singularity than the corresponding ordinary scalar graph. This can be checked directly by power counting. For a connected subgraph $H$ with $V_H$ quartic vertices, $E_H$ internal edges, $L_H$ independent loop momenta, and $N_H$ external legs,
\begin{equation}
4V_H=2E_H+N_H,
\qquad
L_H=E_H-V_H+1.
\end{equation}
At large common loop momentum $Q$, the integration measure contributes $Q^{4L_H}$ and the scalar denominators contribute $Q^{-2E_H}$. The short-distance degree is therefore
\begin{equation}
4L_H-2E_H=4-N_H.
\label{eq:plainPowerCounting}
\end{equation}
Multiplication by $\chi_{G,s_0}$ cannot increase this degree, and the circuit inequalities additionally suppress every independent loop direction. Thus only the vacuum, two-point, and four-point local structures can require a finite matching choice. In four-dimensional $\lambda\phi^4$ theory the corresponding local terms are
\begin{equation}
1,
\qquad
\phi^2,
\qquad
(\partial\phi)^2,
\qquad
\phi^4.
\label{eq:localMatchingSet}
\end{equation}
Note the order of reasoning: we first assert the only interacting vertices of the theory are quartic, then conclude that all the independent local terms are specified by Eq.~\eqref{eq:localMatchingSet}. Their coefficients are fixed by the vacuum convention, the measured mass, the field normalisation, and the measured quartic coupling. The higher-derivative $s_0$ dependence is then predicted by the complete-history rule rather than fixed by a new local parameter at every order. Physically, two allowed ways of defining the product exactly at a coincidence can differ only by a term supported at that coincidence. Equation~\eqref{eq:plainPowerCounting} limits those local terms to the four structures in Eq.~\eqref{eq:localMatchingSet}; once their measured coefficients are fixed there is no additional independent short-distance choice.

These observations give an all-order causal source construction at fixed $s_0$. Let $T_n^{s_0}(x_1,\ldots,x_n)$ denote the $n$-vertex time-ordered coefficient. If the insertion points split into two sets $I$ and $J$ such that every point in $I$ lies later than every point in $J$, the separated-point coefficient is defined by
\begin{equation}
\boxed{
T_n^{s_0}(X_I,X_J)
=
T_{|I|}^{s_0}(X_I)\star_{s_0}T_{|J|}^{s_0}(X_J).}
\label{eq:causalHistoryFactorization}
\end{equation}
Different allowed splittings agree on their overlaps because the sewing product is associative, Eq.~\eqref{eq:historyAssociativity}. The coincident-point values are then fixed by the local conditions in Eq.~\eqref{eq:localMatchingSet}. This recursively determines $T_n^{s_0}$ from lower orders. The factor $\lambda^n$ is kept outside $T_n^{s_0}$ so that the coupling expansion is explicit. For a smooth compact switching function $g$ define
\begin{equation}
\boxed{
\mathcal S_{s_0}[g;\lambda]
=
\id+
\sum_{n=1}^{\infty}
\frac{\lambda^n}{n!}
\int\prod_{j=1}^{n}\dd^4x_j\,
T_n^{s_0}(x_1,\ldots,x_n)
\prod_{j=1}^{n}g(x_j).}
\label{eq:causalSourceSeries}
\end{equation}
At each finite order this construction is finite, respects causal ordering of separated insertions, preserves the local spacetime transformation laws, and obeys the same circuit domain as the graph rules derived above. The series in Eq.~\eqref{eq:causalSourceSeries} is an all-order expansion. Whether it defines one numerical function for a measured non-zero $\lambda$ is a separate question about convergence or summation in the coupling variable. We return to that question after constructing the physical boundary.
}
}

\section{\texorpdfstring{\textcolor{paperred}{The open Euclidean history and the physical particle state}}{The open Euclidean history and the physical particle state}}

{\color{paperred}
The open-history kernel
\begin{equation}
\Rop_{s_0}(\Kop_0)=\ee^{-s_0\Kop_0}\Kop_0^{-1}
\label{eq:openKernelAgain}
\end{equation}
is positive as a Euclidean operator because $\ee^{-s_0\kappa}/\kappa>0$ for every spectral value $\kappa>0$. Its position-space kernel is also positive,
\begin{equation}
\mathcal R_{s_0}(x-y)
=
\int_{s_0}^{\infty}\dd s\,
\frac{\ee^{-m^2s-(x-y)^2/(4s)}}{(4\pi s)^{d/2}}>0.
\label{eq:pointwisePositive}
\end{equation}
This positivity is useful for Euclidean estimates. It does not mean that $\mathcal R_{s_0}$ is the physical particle two-point function.

A Euclidean two-point kernel intended to generate physical positive-energy states must satisfy a stronger positivity condition under reflection in Euclidean time; equivalently, its fixed-spatial-momentum dependence must admit a positive energy-spectral decomposition \cite{OsterwalderSchrader1,OsterwalderSchrader2}. At fixed spatial momentum $\mathbf p$, define $\omega^2=\mathbf p^2+m^2$ and $z=p_0^2\ge0$ for Euclidean energy $p_0$. A physical positive-energy spectral representation would have the form
\begin{equation}
F(z)=\int_0^\infty\frac{\dd\rho(\mu)}{z+\mu},
\qquad
\dd\rho\ge0.
\label{eq:positiveSpectralRep}
\end{equation}
For the open Euclidean history,
\begin{equation}
F_{s_0}(z)=\frac{\ee^{-s_0(z+\omega^2)}}{z+\omega^2}.
\end{equation}
If Eq.~\eqref{eq:positiveSpectralRep} held with non-zero positive weight, then $zF(z)$ would approach the total positive spectral weight as $z\to\infty$. Instead
\begin{equation}
\lim_{z\to\infty}zF_{s_0}(z)=0.
\end{equation}
Thus the open-history kernel cannot itself be the positive physical state kernel. A direct two-point test reaches the same conclusion. Define the fixed-spatial-momentum Euclidean-time kernel by $C(\tau)=\int_{-\infty}^{\infty}\frac{\dd p_0}{2\pi}\,\ee^{\ii p_0\tau}F_{s_0}(p_0^2)$. For $\omega=1$, $s_0=1$, $t_1=0.1$, and $t_2=0.5$,
\begin{equation}
\left[C(t_i+t_j)\right]
=
\begin{pmatrix}
0.07814884&0.07426795\\
0.07426795&0.06713350
\end{pmatrix},
\end{equation}
whose determinant is $-2.69324\times10^{-4}$.

This result fixes the physical interpretation rather than obstructing the theory. $\Rop_{s_0}$ is a Euclidean spectral response for a complete virtual history, not a physical state kernel. The ordinary physical mass-shell pole is kept in the asymptotic particle sector, while $s_0$ enters the complete interaction amplitudes. The physical inner product is therefore imposed only after the complete Euclidean amplitudes have been continued to their physical boundary. This separation is also consistent with the bridge result: an ordinary bridge is not replaced by $\Rop_{s_0}$ merely because it is an internal line.

The theory is defined at an admissible non-zero spectral boundary $s_0>0$; $s_0=0$ is not part of this spectral theory. A finite spectral-mode cutoff used in a calculation is only a numerical approximation to the fixed-$s_0$ operator. For example, if $P_N$ projects onto the first $N$ eigenmodes of $\Kop_0$, then
\begin{equation}
\|(1-P_N)\Rop_{s_0}(\Kop_0)\|
\le
\frac{\ee^{-s_0\kappa_{N+1}}}{\kappa_{N+1}},
\label{eq:spectralApproxConvergence}
\end{equation}
where $\kappa_{N+1}$ is the first omitted spectral value. Thus increasing $N$ improves the representation without changing the spectral boundary $s_0$. Infinite physical volume is a different limit. It restores exact translation invariance of scattering states and is controlled by the mass and the spatial decay established in Section~\ref{sec:scalarNonperturbative}.
}

\section{\texorpdfstring{\textcolor{papergreen}{Lorentzian boundary, cutting, and the physical state space}}{Lorentzian boundary, cutting, and the physical state space}}
\label{sec:LorentzianContinuation}

{\color{paperred}
The open Euclidean history kernel is not used as a physical state kernel. The physical question is instead posed for a complete amplitude after all virtual histories have been assembled. This order of operations is essential because an isolated finite-$s_0$ factor can grow after a timelike continuation even though the complete amplitude has the ordinary pole and cut structure. We therefore form the full Euclidean graph first and only then approach its upper or lower physical boundary \cite{KoshelevTokareva2021}.
}

{\color{papergreen}

\subsection{Pole plus entire remainder}

The Euclidean open history satisfies
\begin{equation}
\frac{\ee^{-s_0A}}A
=
\frac1A+
\frac{\ee^{-s_0A}-1}{A},
\qquad
A=p_E^2+m^2.
\label{eq:poleEntireEuclidean}
\end{equation}
The second term is entire in $A$. Its apparent pole at $A=0$ is removable. Formal continuation gives
\begin{equation}
D_{F,s_0}(p)
=
\frac{\ii\ee^{s_0(p^2-m^2)}}{p^2-m^2+\ii0}.
\label{eq:LorentzianPropagator}
\end{equation}
The pole is still at $p^2=m^2$ and
\begin{equation}
\operatorname*{Res}_{p^2=m^2}
\left[
\frac{\ee^{s_0(p^2-m^2)}}{p^2-m^2}
\right]
=1.
\label{eq:unitResidue}
\end{equation}
There are no additional finite poles. The separate entire factor grows in timelike directions, so Eq.~\eqref{eq:LorentzianPropagator} must not be used as a stand-alone tempered propagator. The continuation is defined from the complete Euclidean amplitude.

\subsection{One-loop bubble discontinuity}

For the complete-history bubble in four dimensions, continue
\begin{equation}
p_E^2\longrightarrow-(s+\ii0).
\end{equation}
For $s>4M^2$, the argument of $\Eone$ crosses its negative-axis branch cut for
\begin{equation}
x\in[x_-,x_+],
\qquad
x_\pm=\frac12(1\pm\beta),
\qquad
\beta=\sqrt{1-\frac{4M^2}{s}}.
\end{equation}
With $\Disc F=F(s+\ii0)-F(s-\ii0)$, the continuation of the argument gives
\begin{equation}
\Eone(-y-\ii0)-\Eone(-y+\ii0)=2\pi\ii,
\qquad y>0.
\label{eq:E1BranchDisc}
\end{equation}
Therefore
\begin{align}
\Disc B^{\mathrm{CH}}_{s_0}(s)
&=
\frac{2\pi\ii}{16\pi^2}
\int_{x_-}^{x_+}\dd x
\nonumber\\
&=
\boxed{
\frac{\ii}{8\pi}
\sqrt{1-\frac{4M^2}{s}}
}.
\label{eq:bubbleDisc}
\end{align}
The endpoint does not change the two-particle phase-space weight.

\subsection{\texorpdfstring{\textcolor{papergreen}{Fixed-order cuts and deletion consistency}}{Fixed-order cuts and deletion consistency}}

{\color{papergreen}
The one-history result provides the local pole model. Appendix E gives, for one complete history partitioned into $n$ segments,
\begin{equation}
\boxed{
\mathcal I_n(a_1,\ldots,a_n)
=
\prod_{r=1}^{n}\frac1{a_r}
-
\mathcal Q_{n,s_0}(a_1,\ldots,a_n),}
\label{eq:scalarHistoryPoleEntire}
\end{equation}
where
\begin{equation}
\mathcal Q_{n,s_0}
=
\int_{\substack{u_r\ge0\\ \sum_r u_r<s_0}}
\prod_{r=1}^{n}\dd u_r\,
\exp\left[-\sum_{r=1}^{n}a_ru_r\right].
\label{eq:scalarHistoryEntireRemainder}
\end{equation}
The excluded region is compact, so $\mathcal Q_{n,s_0}$ is entire in the variables $a_r$. In particular
\begin{equation}
\Rop_{s_0}(A)
=
\frac1A+
\frac{\ee^{-s_0A}-1}{A},
\label{eq:scalarPoleEntireSplit}
\end{equation}
and the pole residue is unity.

For a general graph the endpoint restrictions overlap because an edge can belong to more than one circuit. The appropriate statement is therefore a deletion identity for the complete circuit domain. Define
\begin{equation}
\chi_G(s)
=
\prod_{c\in\mathcal C(G)}
\Theta\left(\sum_{e\in c}s_e-s_0\right)
\label{eq:graphCircuitIndicator}
\end{equation}
and, for $\operatorname{Re}a_e>0$,
\begin{equation}
J_G(a)
=
\int_{\mathbb R_+^{|E|}}
\left(\prod_{e\in E}\dd s_e\right)
\exp\left[-\sum_{e\in E}a_es_e\right]\chi_G(s).
\label{eq:graphCircuitLaplace}
\end{equation}

{\color{paperred}\paragraph{Fixed-order cut factorisation.}
Let $C\subset E(G)$ be a non-empty set of internal edges placed on shell by an admissible finite physical pinch. Assume a fixed infrared regulator when needed and that the pinch is reached by an admissible continuation from the Euclidean domain without an independent pinch on the proper-time boundary. Then the endpoint part of the pole residue obeys
\begin{equation}
\boxed{
\lim_{\{a_r\to0^+\}_{r\in C}}
\left(\prod_{r\in C}a_r\right)J_G(a)
=
J_{G\setminus C}(a_{E\setminus C}).}
\label{eq:graphDeletionResidue}
\end{equation}
If deletion of $C$ separates the graph into $G_L\sqcup G_R$, then
\begin{equation}
J_{G\setminus C}=J_{G_L}J_{G_R}.
\label{eq:cutGraphFactorisation}
\end{equation}
Consequently the two factors appearing in the physical cut are the same finite-$s_0$ subamplitudes obtained by constructing $G_L$ and $G_R$ independently.
\label{prop:scalarCutting}

\paragraph{Derivation.}
For every $r\in C$ set $t_r=a_rs_r$. Then
\begin{align}
\left(\prod_{r\in C}a_r\right)J_G
={}&
\int\prod_{u\notin C}\dd s_u
\prod_{r\in C}\dd t_r\,
\exp\left[-\sum_{u\notin C}a_us_u-\sum_{r\in C}t_r\right]
\nonumber\\
&\times
\chi_G\left(s_{E\setminus C},\{t_r/a_r\}_{r\in C}\right).
\label{eq:cutRescaledIntegral}
\end{align}
For almost every $t_r>0$, $t_r/a_r\to\infty$ as $a_r\to0^+$. If a parent circuit $c$ contains at least one cut edge, then
\[
\sum_{e\in c}s_e\longrightarrow\infty,
\]
so its step factor tends to one. {\color{paperred}In the limit $a_r\to0^+$, every circuit that contains a cut edge has step factor one for almost every $t_r>0$. Every circuit disjoint from the cut keeps its original step factor. Hence the pointwise limiting indicator is exactly $\chi_{G\setminus C}(s_{E\setminus C})$. The integrations over the rescaled cut parameters then separate and give}
\begin{equation}
{\color{paperred}
\prod_{r\in C}\int_0^\infty\dd t_r\,\ee^{-t_r}=1.}
\label{eq:cutOnShellUnitIntegral}
\end{equation}
If $c\cap C=\varnothing$, its step factor is unchanged. Graph deletion gives the exact identity
\begin{equation}
\mathcal C(G\setminus C)
=
\left\{
c\in\mathcal C(G):c\cap C=\varnothing
\right\}.
\label{eq:circuitDeletionIdentity}
\end{equation}
Since $0\le\chi_G\le1$, the exponential in Eq.~\eqref{eq:cutRescaledIntegral} supplies an integrable dominating function. Dominated convergence therefore gives Eq.~\eqref{eq:graphDeletionResidue}. If the deleted graph is disconnected, every surviving circuit lies wholly in one connected component, so both the indicator and the remaining Schwinger integrations factor, giving Eq.~\eqref{eq:cutGraphFactorisation}.
}

Equation~\eqref{eq:graphDeletionResidue} is stronger than the statement that a cut propagator has unit residue. It also determines the endpoint status of every uncut edge. If an edge becomes a bridge after the cut, it belongs to no circuit of $G\setminus C$ and its remaining proper-time integration is unrestricted. If an uncut loop survives wholly on one side, that loop is a circuit of the cut graph and retains exactly the endpoint condition it would have in the standalone subamplitude.

After continuation to the physical Lorentzian boundary, the standard local contour pinch replaces each cut scalar pole by
\begin{equation}
2\pi\,\delta_+(q^2-m^2),
\end{equation}
where $\delta_+$ denotes the positive-energy mass-shell delta distribution.
The local vertices and symmetry factors are unchanged, and Eq.~\eqref{eq:graphDeletionResidue} supplies no additional cut residue. Thus, with the usual convention for the discontinuity,
\begin{equation}
\boxed{
\operatorname{Disc}_C\mathcal M_G^{s_0}
=
\ii\int\dd\Phi_C\,
\mathcal M_{G_L}^{s_0}\,
\mathcal M_{G_R}^{s_0\,*},}
\label{eq:fixedOrderCircuitCut}
\end{equation}
where the cut phase-space measure is
\begin{equation}
\dd\Phi_C(P)
=
(2\pi)^4\delta^{(4)}
\left(P-\sum_{\ell\in C}q_\ell\right)
\prod_{\ell\in C}
\frac{\dd^3\bm q_\ell}{(2\pi)^3\,2E_\ell}
\ge0.
\label{eq:ordinaryCutPhaseSpace}
\end{equation}
Here $\bm q_\ell$ is the three-momentum of cut line $\ell$ and $E_\ell=\sqrt{\bm q_\ell^2+m^2}$ its positive on-shell energy. The graph identities and ultraviolet derivations, together with the one-loop, sunset, and $K_4$ examples, are given in Appendix~\ref{app:circuitDerivations}. Equation~\eqref{eq:fixedOrderCircuitCut} is a fixed-order statement; passing it through an infinite perturbative sum requires the uniform bounds stated below.
}

\subsection{\texorpdfstring{\textcolor{paperred}{The all-order source series and one numerical coupling}}{The all-order source series and one numerical coupling}}

{\color{paperred}
Equation~\eqref{eq:causalSourceSeries} constructs every coefficient of the interacting source series at fixed $s_0$. This fixes every perturbative order by one common rule; no finite order is given a different definition. A further question remains if $\lambda$ is assigned one numerical value: does the complete series determine one analytic function of that value?

A coefficient bound
\begin{equation}
\|\mathcal G_n\|\le C A^n n!
\label{eq:coefficientGevrey}
\end{equation}
controls the large-order coefficients but does not by itself determine a unique numerical function. Let $\mathcal G_{N,V}(\lambda;\mathcal K)$ denote an exact connected source kernel, where $N$ labels the finite spectral approximation and $V$ the finite Euclidean volume. If these kernels obey the uniform remainder estimate
\begin{equation}
\boxed{
\left\|
\mathcal G_{N,V}(\lambda;\mathcal K)
-
\sum_{n=0}^{r-1}\lambda^n\mathcal G_n(\mathcal K)
\right\|
\le
C_{\mathcal K}A_{\mathcal K}^{r}r!|\lambda|^r,}
\label{eq:scalarUniformFactorial}
\end{equation}
with constants independent of $N$ and $V$ on a common Nevanlinna--Sokal complex-coupling domain satisfying the analyticity hypotheses of the summability theorem, the resulting numerical sum is unique \cite{Sokal1980}. This estimate concerns the exact remainder rather than the individual coefficients. Tree reorganisations provide a natural way to test such uniform connected bounds \cite{BrydgesKennedy1987,AbdesselamRivasseau1995}.

Define the factorially rescaled series
\begin{equation}
\widehat{\mathcal G}(t;\mathcal K)
=
\sum_{n=0}^{\infty}
\frac{\mathcal G_n(\mathcal K)}{n!}t^n.
\label{eq:scalarBorelTransform}
\end{equation}
When the continuation of $\widehat{\mathcal G}$ along $t\ge0$ obeys
\begin{equation}
\|\widehat{\mathcal G}(t;\mathcal K)\|
\le
C'_{\mathcal K}\ee^{t/R_{\mathcal K}},
\label{eq:scalarBorelExpBound}
\end{equation}
the direct Borel--Laplace sum is defined in the sector where $\operatorname{Re}(1/\lambda)>1/R_{\mathcal K}$ by
\begin{equation}
\boxed{
\mathcal G(\lambda;\mathcal K)
=
\frac1\lambda
\int_0^\infty\dd t\,
\ee^{-t/\lambda}\widehat{\mathcal G}(t;\mathcal K).}
\label{eq:scalarBorelSum}
\end{equation}
The same exponential bound is required uniformly while the collection $E$ of external invariants approaches the physical boundary, with $\epsilon>0$ denoting the boundary prescription,
\begin{equation}
\sup_{0<\epsilon\le\epsilon_0}
\|\widehat{\mathcal G}(t;E\pm\ii\epsilon)\|
\le
C_E\ee^{t/R_E}.
\label{eq:scalarBoundaryBorelBound}
\end{equation}
Then the physical limit and the $t$ integral can be interchanged by dominated convergence. This gives one boundary value of the complete summed amplitude rather than a separate continuation of each finite-endpoint factor.

The fixed-order cut relation is polynomial in the amplitudes. With the normalization in Eq.~\eqref{eq:scalarBorelTransform}, if $C(\lambda)=A(\lambda)B(\lambda)$ then $\widehat C(t)=\frac{\dd}{\dd t}\int_0^t\dd u\,\widehat A(u)\widehat B(t-u)$, with the constant terms understood in the usual way. Thus multiplication is represented by the derivative of the Borel convolution, and the stated uniform exponential bounds make the Borel-summable class an algebra. The exact cut identity can therefore pass through the Borel--Laplace integral in Eq.~\eqref{eq:scalarBorelSum}. With $S=\id+\ii T$, upper and lower physical boundaries related by complex conjugation, and $\rho$ the positive on-shell channel phase-space operator,
\begin{equation}
\boxed{
T^+-T^-
=
\ii\,T^-\rho T^+,
\qquad
\rho\ge0.}
\label{eq:scalarExactCut}
\end{equation}
This is the all-order physical unitarity relation whenever Eqs.~\eqref{eq:scalarUniformFactorial} and \eqref{eq:scalarBoundaryBorelBound} hold for the complete source kernels.
}

\subsection{\texorpdfstring{\textcolor{paperred}{Positive physical states from complete amplitudes}}{Positive physical states from complete amplitudes}}

{\color{paperred}
Positivity is now imposed on complete physical channels, not on the virtual kernel $\Rop_{s_0}$. Let $\alpha$ label an on-shell scalar channel and let $\dd\Phi_\alpha\ge0$ be its ordinary phase-space measure. A compact source history $J$ defines the physical amplitude vector
\begin{equation}
\Psi_J=\{\mathcal A_\alpha[J]\}_{\alpha}.
\end{equation}
Define
\begin{equation}
\boxed{
\langle\Psi_{J_i},\Psi_{J_j}\rangle
=
\sum_\alpha\int\dd\Phi_\alpha\,
\mathcal A_\alpha[J_i]^*\mathcal A_\alpha[J_j].}
\label{eq:scalarHistoryKernel}
\end{equation}
For arbitrary coefficients $c_i$,
\begin{equation}
\sum_{ij}c_i^*c_j
\langle\Psi_{J_i},\Psi_{J_j}\rangle
=
\sum_\alpha\int\dd\Phi_\alpha
\left|\sum_i c_i\mathcal A_\alpha[J_i]\right|^2
\ge0.
\label{eq:scalarHistoryPositive}
\end{equation}
Thus the physical state space is obtained by taking the span of the $\Psi_J$, removing zero-norm combinations, and completing the resulting positive inner-product space. No positivity property of the virtual-history kernel is needed. Denote by $|J\rangle$ the equivalence class represented by the source amplitude vector $\Psi_J$ in this quotient-and-completion construction.

The same source vectors define interacting observables. For a smooth test function $f$, set on the dense source-generated domain
\begin{equation}
\Phi(f)|J\rangle
=
\left.
\frac1{\ii}\frac{\dd}{\dd\tau}
|J+\tau f\rangle
\right|_{\tau=0}.
\label{eq:observableSourceDerivative}
\end{equation}
Whenever the physical source kernel is differentiable at $J=0$, every finite matrix of source-derived states has the form
\begin{equation}
M_{ab}=\langle\Psi_a,\Psi_b\rangle\succeq0.
\label{eq:scalarAllNPositive}
\end{equation}
Translation invariance of the complete amplitudes makes translations preserve this inner product. Strong continuity then gives self-adjoint energy and momentum generators. Because the channel measure in Eq.~\eqref{eq:scalarHistoryKernel} is supported on ordinary positive-energy scalar states, the physical spectrum lies in the positive-energy region.
}

\subsection{\texorpdfstring{\textcolor{paperred}{Massive propagation and observable locality}}{Massive propagation and observable locality}}

{\color{paperred}
The mass $m>0$ suppresses the large-proper-time part of every Euclidean shell and gives exponential spatial decay. Finite-source observables are defined before any scattering limit is taken. Their locality is tested through commutators of the reconstructed smeared observables, not through $\Rop_{s_0}(x-y)$. For smooth compactly supported test functions $f$ and $g$, define $f_a(x)\equiv f(x-a)$. For vectors $\Psi,\Xi$ in the source-generated dense domain, the desired spacelike-separation condition is
\begin{equation}
\langle\Psi,[\Phi(f_a),\Phi(g)]\Xi\rangle
\longrightarrow0
\qquad
\text{as the spacelike separation }|a|\to\infty,
\label{eq:asymptoticLocalityPlain}
\end{equation}
with decay fast enough for separated wave packets to define asymptotic particle states. The shell estimate~\eqref{eq:scalarShellLocality} gives the Euclidean input for this decay, while the uniform physical-boundary bound~\eqref{eq:scalarBoundaryBorelBound} is what transfers it to the physical amplitudes.

Let $\Sigma(z)$ denote the renormalised two-point self-energy as a function of the invariant variable $z$. After the measured mass is fixed, the one-particle pole is simple as long as
\begin{equation}
\sup_{z\in B_r}|\partial_z\Sigma(z)|<1
\label{eq:massShellDerivativeBound}
\end{equation}
on a convex subthreshold neighbourhood $B_r$ of the pole. Define $q\equiv\sup_{z\in B_r}|\Sigma'(z)|<1$. Then
\begin{equation}
D(z)=z-m_{\rm phys}^2-[\Sigma(z)-\Sigma(m_{\rm phys}^2)]
\end{equation}
has only one zero in $B_r$. Indeed, if $D(z_1)=D(z_2)=0$, then
\begin{equation}
|z_1-z_2|
=|\Sigma(z_1)-\Sigma(z_2)|
\le q|z_1-z_2|,
\qquad q<1,
\end{equation}
so $z_1=z_2$. Below threshold $\Sigma'(m_{\rm phys}^2)$ is real, and the pole residue
\begin{equation}
Z_{\rm pole}=\frac1{1-\Sigma'(m_{\rm phys}^2)}
\end{equation}
is positive. A finite gap to the first multiparticle threshold then separates the one-particle state from the continuum. This statement belongs to the particle sector and is not required to define observables with finite source support. When asymptotic wave-packet scattering is considered in addition, Eq.~\eqref{eq:asymptoticLocalityPlain} gives the usual massive scattering construction, with the same physical boundary amplitudes that enter Eq.~\eqref{eq:scalarExactCut}.

The cycle domain controls the closed virtual histories, smooth source smearing defines local field products, and the background heat bound controls their Euclidean size and spatial range. Complete boundary amplitudes then define the positive physical state space. If the uniform remainder and boundary bounds above hold, the same all-order expansion has one numerical sum at the chosen coupling.
}
}

\section{Relation to renormalisation and running scales}

\subsection{Running cutoffs in renormalisation-group calculations}

In a Wilsonian renormalisation-group construction, the running cutoff is an auxiliary coarse-graining parameter and the effective action changes with it while the same long-distance theory is represented; Polchinski made this explicit for four-dimensional $\lambda\phi^4$ with a momentum cutoff and a separation into relevant and irrelevant parts \cite{Polchinski1984}. Proper-time kernels can also organise RG flows. Litim and Pawlowski showed that commonly used proper-time flows need not be exact functional renormalisation-group (FRG) equations and analysed their relation to exact flows \cite{LitimPawlowski2002}; explicitly Wilsonian or exact Schwinger proper-time formulations have subsequently been constructed \cite{Bonanno2020,AbelHeurtier2025}. Thus ``proper-time regularisation'' by itself does not specify whether a proper-time scale is a removable regulator, an RG coarse-graining scale, or a retained parameter of a different theory.

Recent calculations reinforce this distinction. Proper-time flows reproduce universal perturbative beta-function information in scalar and Yang--Mills theories even though the standard proper-time flow is not itself an exact FRG equation \cite{GiacomettiRizzoZappala2026}; gauge and parametrisation dependence can remain in truncated gravitational applications \cite{BonannoOglialoroZappala2025}. None of these results identifies the running RG cutoff with a physical retained endpoint.

The exact-RG distinction is especially important here. Latorre and Morris showed that changes of the cutoff scheme in an exact RG can be represented by field redefinitions \cite{LatorreMorris2000}. Consequently, changing an off-shell propagator or the shape of a cutoff function is not sufficient to establish physically inequivalent dynamics. This gives a direct physical-distinction test for the present construction.

\subsection{\texorpdfstring{Traditional renormalisation: separating $\Lambda_{\rm UV}$, $\mu$, and $s_0$}{Traditional renormalisation: separating LambdaUV, mu, and s0}}

The distinction can be made explicitly in the one-loop four-point function. Let $\Lambda_{\rm UV}$ denote any temporary ultraviolet regulator used in a conventional calculation and let $\mu$ denote the subtraction scale at which the renormalised quartic coupling is defined. Neither symbol is identified with the retained endpoint scale $\Lambda=s_0^{-1/2}$. Around the zero background, define the Euclidean channel momenta
\begin{equation}
q_s=p_1+p_2,
\qquad
q_t=p_1-p_3,
\qquad
q_u=p_1-p_4.
\end{equation}
Before imposing the finite endpoint, the one-loop 1PI four-point vertex contains the three bubble channels
\begin{equation}
\Gamma_1^{(4)}
=
-\frac{\lambda^2}{2}
\left[
B(q_s^2)+B(q_t^2)+B(q_u^2)
\right],
\label{eq:threeChannelBubble}
\end{equation}
with the usual local counterterm fixing one renormalisation condition. In the complete-history finite-endpoint theory the corresponding bubble is Eq.~\eqref{eq:bubbleE1},
\begin{equation}
B_{s_0}(q^2)
=
\frac1{16\pi^2}
\int_0^1\dd x\,
\Eone\!\left(
s_0 Q_q(x)
\right),
\qquad
Q_q(x)=m^2+x(1-x)q^2.
\label{eq:Bs0Comparison}
\end{equation}

Define $\lambda_R(\mu)$ by a symmetric Euclidean subtraction condition in which all three channel invariants equal $\mu^2$. To one loop, the renormalised finite-endpoint vertex can then be written without reference to the temporary regulator $\Lambda_{\rm UV}$ as
\begin{equation}
\boxed{
\Gamma^{(4)}_{R,s_0}
=
\lambda_R(\mu)
-\frac{\lambda_R(\mu)^2}{2}
\sum_{c=s,t,u}
\left[
B_{s_0}(q_c^2)-B_{s_0}(\mu^2)
\right]
+\order(\lambda_R^3).}
\label{eq:renormalisedFPT4pt}
\end{equation}
The subtraction removes the local normalisation ambiguity at the chosen reference point, but it does not remove the endpoint from the momentum dependence. Indeed
\begin{equation}
\frac{\partial}{\partial q^2}
\left[
B_{s_0}(q^2)-B_{s_0}(\mu^2)
\right]
=
-\frac1{16\pi^2}
\int_0^1\dd x\,
\frac{x(1-x)\ee^{-s_0Q_q(x)}}{Q_q(x)}.
\label{eq:FPTBubbleSlope}
\end{equation}
The corresponding local result is obtained by setting the exponential factor to unity. A single local $\phi^4$ counterterm changes the value of the four-point function at the subtraction point; it cannot remove this entire momentum-dependent difference.

The same statement is visible by separating the local limit from the finite remainder. Define
\begin{align}
\Delta B_{s_0}(q^2;\mu^2)
={}&
\left[B_{s_0}(q^2)-B_{s_0}(\mu^2)\right]
+\frac1{16\pi^2}
\int_0^1\dd x\,
\ln\frac{Q_q(x)}{Q_\mu(x)}
\nonumber\\
={}&
\frac1{16\pi^2}
\int_0^1\dd x
\left[
\Eone(s_0Q_q)-\Eone(s_0Q_\mu)
+\ln\frac{Q_q}{Q_\mu}
\right].
\label{eq:FPTFiniteRemainder}
\end{align}
This quantity is finite, vanishes in the formal boundary-removal comparison $s_0\to0^+$, and is zero at the renormalisation point by construction. Using, with $\gamma_E$ the Euler--Mascheroni constant,
\begin{equation}
\Eone(z)=-\gamma_E-\ln z+z-\frac{z^2}{4}+\order(z^3),
\qquad z\to0^+,
\end{equation}
gives, channel by channel,
\begin{equation}
\Delta B_{s_0}(q^2;\mu^2)
=
\frac{s_0(q^2-\mu^2)}{96\pi^2}
+\order\!\left(s_0^2Q^2\right).
\label{eq:FPTLowEnergyExpansion}
\end{equation}
At low momentum the exact finite-$s_0$ amplitude can be written as a local derivative series in $s_0p^2$. This series does not define the theory; it is obtained by expanding the exact spectral function. A general low-energy description could fit a separate coefficient at each derivative order. The present construction instead fixes the whole sequence from the same $s_0$. At one loop the full momentum dependence is fixed by this one endpoint. At two loops Eq.~\eqref{eq:FPTTwoLoop} introduces no new endpoint parameter, so after the same low-energy matching conditions its derivative coefficients are calculable proper-time moments of $m$, $\lambda_R$, and $s_0$. A complete two-loop on-shell $2\to2$ evaluation is not carried out here.

\subsubsection{Fixed-endpoint momentum-subtraction flow}

The subtraction scale $\mu$ remains arbitrary, but the retained endpoint is held fixed. This is a momentum-subtraction (MOM) flow of the renormalised coupling, not a Wilsonian flow of $s_0$. Impose the symmetric Euclidean condition
\begin{equation}
\left.\Gamma^{(4)}_{R,s_0}\right|_{q_s^2=q_t^2=q_u^2=\mu^2}
=
\lambda_R(\mu).
\label{eq:MOMconditionFPT}
\end{equation}
For fixed underlying parameters and fixed $s_0$, differentiating Eq.~\eqref{eq:renormalisedFPT4pt} gives
\begin{equation}
\beta^{\mathrm{MOM}}_{s_0}(\lambda_R,\mu)
\equiv
\mu\frac{\dd\lambda_R}{\dd\mu}
=
-\frac{3\lambda_R^2}{2}\,
\mu\frac{\partial B_{s_0}(\mu^2)}{\partial\mu}
+\order(\lambda_R^3).
\label{eq:FPTbetaDefinition}
\end{equation}
Using Eq.~\eqref{eq:FPTBubbleSlope},
\begin{equation}
\boxed{
\beta^{\mathrm{MOM}}_{s_0}(\lambda_R,\mu)
=
\frac{3\lambda_R^2}{16\pi^2}
\int_0^1\dd x\,
\frac{\mu^2x(1-x)}
{m^2+\mu^2x(1-x)}
\exp\!\left[-s_0\!\left(m^2+\mu^2x(1-x)\right)\right]
+\order(\lambda_R^3).}
\label{eq:FPTbetaMassive}
\end{equation}
No additional regulator is introduced in this derivation: the scale differentiated is the arbitrary subtraction point, while $s_0$ remains part of the definition of the spectral functions.

For $m=0$, let $a=\mu^2s_0=\mu^2/\Lambda^2$. Then
\begin{equation}
\boxed{
\beta^{\mathrm{MOM}}_{s_0}
=
\frac{3\lambda_R^2}{16\pi^2}\,
\mathcal T(a)
+\order(\lambda_R^3),
\qquad
\mathcal T(a)
=
\int_0^1\dd x\,\ee^{-a x(1-x)}.}
\label{eq:FPTbetaMassless}
\end{equation}
Completing the square in $x(1-x)=1/4-(x-1/2)^2$ gives
\begin{equation}
\boxed{
\mathcal T(a)
=
\sqrt{\frac{\pi}{a}}\,
\ee^{-a/4}
\operatorname{erfi}\!\left(\frac{\sqrt a}{2}\right),\qquad \operatorname{erfi}(z)\equiv-\ii\operatorname{erf}(\ii z).}
\label{eq:FPTthresholdClosed}
\end{equation}
Here $\operatorname{erf}$ is the error function and $\operatorname{erfi}$ its imaginary counterpart. The limiting expansions are
\begin{align}
\mathcal T(a)
&=
1-\frac{a}{6}+\frac{a^2}{60}-\frac{a^3}{840}
+\order(a^4),
\qquad a\ll1,
\\
\mathcal T(a)
&=
\frac{2}{a}
\left(1+\frac{2}{a}+\frac{12}{a^2}
+\order(a^{-3})\right),
\qquad a\gg1.
\end{align}
Thus the ordinary one-loop coefficient $\beta_{\rm local}=3\lambda_R^2/(16\pi^2)$ is recovered for $\mu\ll\Lambda$, while the fixed-endpoint MOM running is suppressed when the subtraction momentum resolves the retained scale. Figure~\ref{fig:FPTMOMthreshold} shows this threshold function. This is not an asymptotic-safety claim: the beta function is explicitly non-autonomous through $\mu/\Lambda$, and the one-loop running simply accumulates a finite amount once $\mu$ is taken far above the retained scale.

\begin{figure}[!t]
\centering
\includegraphics[width=0.72\linewidth]{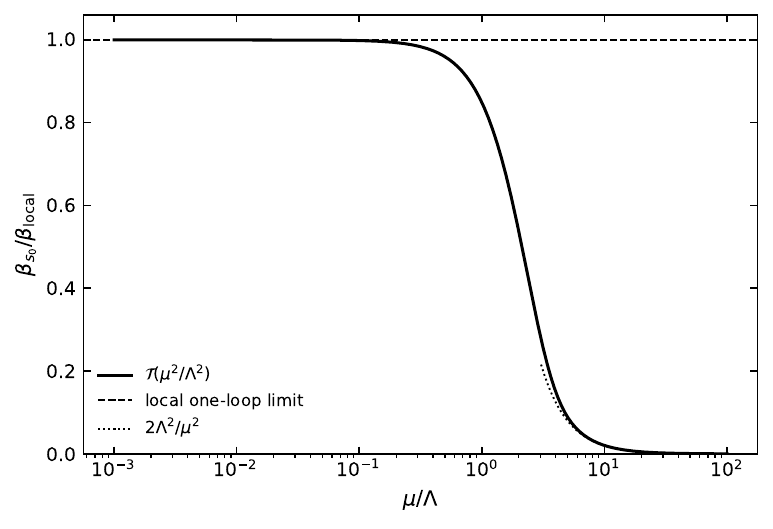}
\caption{One-loop fixed-endpoint momentum-subtraction flow in the massless scalar theory. The ratio $\beta^{\rm MOM}_{s_0}/\beta_{\rm local}=\mathcal T(\mu^2/\Lambda^2)$ approaches unity below the retained scale and falls as $2\Lambda^2/\mu^2$ for $\mu\gg\Lambda$. The endpoint $s_0$ is fixed throughout.}
\label{fig:FPTMOMthreshold}
\end{figure}

Integrating Eq.~\eqref{eq:FPTbetaDefinition} between two subtraction points gives
\begin{equation}
\frac{1}{\lambda_R(\mu)}
=
\frac{1}{\lambda_R(\mu_0)}
+\frac32\left[B_{s_0}(\mu^2)-B_{s_0}(\mu_0^2)\right]
+\order(\lambda_R).
\label{eq:FPTintegratedRunning}
\end{equation}
For the massless bubble, $B_{s_0}(\mu^2)\to0$ as $\mu\to\infty$. Hence the one-loop logarithmic growth saturates at a finite accumulated shift, provided perturbation theory does not encounter a pole before that limit. This result is a property of the fixed-endpoint theory, not of an auxiliary proper-time flow.

\subsubsection{Exact matched derivative tower}

It is useful to isolate the finite endpoint from the ordinary logarithmic amplitude in closed form. Define
\begin{equation}
\operatorname{Ein}(z)
\equiv
\gamma_E+\ln z+\Eone(z)
=
\sum_{n=1}^{\infty}
\frac{(-1)^{n+1}z^n}{n\,n!}.
\label{eq:EinDefinition}
\end{equation}
For arbitrary mass,
\begin{equation}
\boxed{
\Delta B_{s_0}(q^2;\mu^2)
=
\frac{1}{16\pi^2}
\sum_{n=1}^{\infty}
\frac{(-1)^{n+1}s_0^n}{n\,n!}
\int_0^1\dd x\,
\left[Q_q(x)^n-Q_\mu(x)^n\right].}
\label{eq:FPTExactTowerMassive}
\end{equation}
The series is the Taylor expansion of an entire function and converges for every finite $Q_q$ and $Q_\mu$. In the massless case,
\begin{equation}
\int_0^1\dd x\,[x(1-x)]^n
=
\frac{(n!)^2}{(2n+1)!},
\end{equation}
so
\begin{equation}
\boxed{
\Delta B_{s_0}(q^2;\mu^2)
=
\frac{1}{16\pi^2}
\sum_{n=1}^{\infty}
(-1)^{n+1}
\frac{(n-1)!}{(2n+1)!}
s_0^n
\left(q^{2n}-\mu^{2n}\right).}
\label{eq:FPTExactTower}
\end{equation}
Figure~\ref{fig:FPTMatchedRemainder} compares the exact matched function with its leading low-momentum term. A general local derivative expansion could fit any finite number of coefficients independently. The FPT construction instead predicts the complete correlated sequence from one common $s_0$.

\begin{figure}[!t]
\centering
\includegraphics[width=0.72\linewidth]{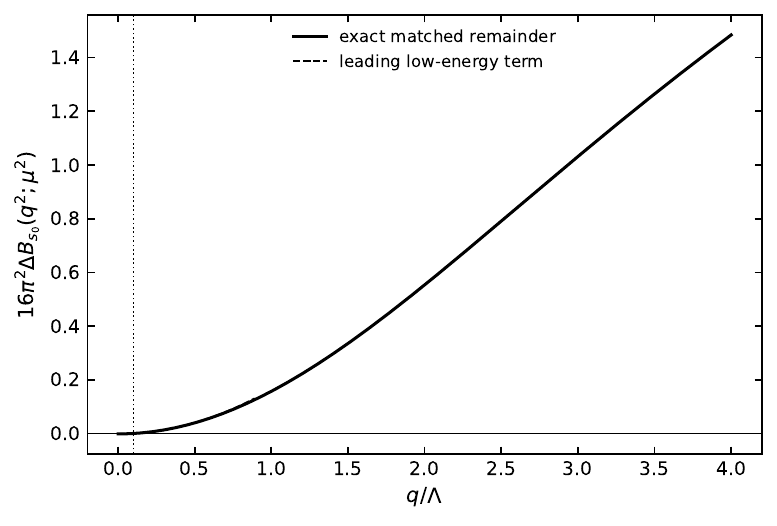}
\caption{Exact massless matched bubble remainder after the local quartic coupling has been fixed at $\mu/\Lambda=0.1$. The dashed line is the leading low-energy term. The nonzero function away from the subtraction point is finite and is fixed by the retained endpoint rather than by an additional local counterterm.}
\label{fig:FPTMatchedRemainder}
\end{figure}

\subsubsection{On-shell physical distinction of the retained endpoint}

Off-shell differences are not sufficient to establish physical inequivalence. Exact-RG scheme changes are field-redefinition redundancies \cite{LatorreMorris2000}, and the equivalence result states that admissible local or almost-local invertible field redefinitions leave the $S$ matrix unchanged \cite{Chisholm1961,KamefuchiORaifeartaighSalam1961,CriadoPerezVictoria2019}. We therefore test the matched amplitude on shell.

Under the fixed-order Euclidean-first continuation used in Sec.~\ref{sec:LorentzianContinuation}, $q_s^2\to-(s+\ii0)$, $q_t^2\to-(t+\ii0)$, and $q_u^2\to-(u+\ii0)$. The FPT--local difference after the same quartic matching is
\begin{equation}
\delta\Gamma^{(4)}_{\rm FPT-local}
=
-\frac{\lambda_R^2}{2}
\sum_{c=s,t,u}
\Delta B_{s_0}(q_c^2;\mu^2)
+\order(\lambda_R^3).
\label{eq:FPTOnShellDifference}
\end{equation}
The $n=1$ term of Eq.~\eqref{eq:FPTExactTowerMassive} is proportional, apart from the subtraction-point constant, to
\begin{equation}
q_s^2+q_t^2+q_u^2
\longrightarrow
-(s+t+u)=-4m^2
\end{equation}
for identical on-shell particles. It is therefore kinematically constant and can be absorbed into the matched quartic coupling. In the massless case it vanishes identically. This explicitly identifies the leading two-derivative four-field correction as on-shell redundant.

At $n=2$, all terms proportional to $m^2(s+t+u)$ are again constant, but the quadratic channel invariant survives:
\begin{equation}
\boxed{
\left.
\delta\Gamma^{(4)}_{\rm FPT-local}
\right|_{\rm nonconstant}
=
\frac{\lambda_R^2s_0^2}{3840\pi^2}
\left(s^2+t^2+u^2\right)
+\order(\lambda_R^2s_0^3E^6,\lambda_R^3).}
\label{eq:FPTOnShellLeadingPhysical}
\end{equation}
For massless centre-of-mass scattering,
\begin{equation}
t=-\frac{s}{2}(1-\cos\theta),
\qquad
u=-\frac{s}{2}(1+\cos\theta),
\end{equation}
and hence
\begin{equation}
\boxed{
\left.
\delta\Gamma^{(4)}_{\rm FPT-local}
\right|_{\rm nonconstant}
=
\frac{\lambda_R^2}{7680\pi^2}
\frac{s^2}{\Lambda^4}
\left(3+\cos^2\theta\right)
+\cdots.}
\label{eq:FPTAngularSignature}
\end{equation}
The characteristic angular factor is shown in Fig.~\ref{fig:FPTOnShellSignature}.

The massless formulas in this subsection are used only to display the coefficient relations in a simple closed form. The nonperturbative Lorentzian reconstruction below assumes a positive mass gap.

The massless exact tower gives a stronger on-shell statement than the leading term alone. After dropping subtraction-point constants and using $q_c^2\to-c$ for $c=s,t,u$, Eq.~\eqref{eq:FPTExactTower} yields the complete analytic one-loop remainder
\begin{equation}
\boxed{
\Delta\textcolor{revisionolive}{\mathcal M_{1,\rm an}}(s,t,u)
=
\frac{\lambda_R^2}{32\pi^2}
\sum_{n=2}^{\infty}
\frac{(n-1)!}{(2n+1)!}
 s_0^n\left(s^n+t^n+u^n\right)
+\order(\lambda_R^3).}
\label{eq:FPTOnShellExactTower}
\end{equation}
The sum starts at $n=2$ because the $n=1$ contribution vanishes for massless on-shell scattering. Define the matched on-shell coefficient $g_n$ by
\begin{equation}
\Delta\textcolor{revisionolive}{\mathcal M_{1,\rm an}}
=\sum_{n\ge2}g_n\left(s^n+t^n+u^n\right),
\qquad
 g_n=
\frac{\lambda_R^2}{32\pi^2}
\frac{(n-1)!}{(2n+1)!}s_0^n.
\label{eq:FPTgnDefinition}
\end{equation}
Then successive physical coefficients obey
\begin{equation}
\boxed{
\frac{g_{n+1}}{g_n}
=
\frac{n}{(2n+2)(2n+3)}s_0,
\qquad
s_0=
\frac{(2n+2)(2n+3)}{n}\frac{g_{n+1}}{g_n}.}
\label{eq:FPTs0Identifiable}
\end{equation}
In particular, $s_0=21g_3/g_2$. Eliminating $s_0$ gives parameter-free consistency relations,
\begin{equation}
\boxed{
\frac{g_{n+2}g_n}{g_{n+1}^2}
=
\frac{(n+1)^2(2n+3)}{n(n+2)(2n+5)},
\qquad
\frac{g_4g_2}{g_3^2}=\frac78.}
\label{eq:FPTParameterFreeRatios}
\end{equation}
Thus a general local low-momentum expansion can reproduce the FPT amplitude only when its higher-derivative coefficients lie on this correlated one-parameter trajectory. In that expansion $s_0$ is encoded in physical on-shell coefficients; it has not been removed as an arbitrary regulator.

\begin{figure}[!t]
\centering
\includegraphics[width=0.72\linewidth]{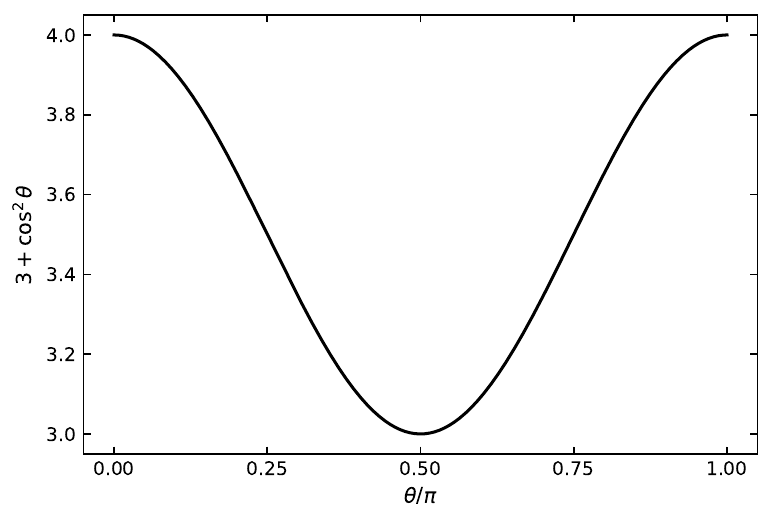}
\caption{Angular dependence of the leading non-constant finite-endpoint correction to massless $2\to2$ scalar scattering after quartic matching. The overall prefactor is $\lambda_R^2s^2/(7680\pi^2\Lambda^4)$; only the predicted shape $3+\cos^2\theta$ is shown.}
\label{fig:FPTOnShellSignature}
\end{figure}

{\color{paperred}\paragraph{Physical distinction after matching.}
Fix the physical mass, field normalisation, and quartic coupling by the same low-energy conditions in the local and finite-endpoint theories. Equation~\eqref{eq:FPTOnShellLeadingPhysical} then leaves a non-zero contribution proportional to $s^2+t^2+u^2$. This term varies over the physical $2\to2$ scattering surface and cannot be absorbed into the matched quartic coupling, mass, or field normalisation. An invertible local change of field variables that leaves the scattering matrix unchanged cannot remove this on-shell momentum dependence \cite{Chisholm1961,KamefuchiORaifeartaighSalam1961}. Thus different non-zero values of $s_0$ give different matched one-loop scattering amplitudes within this equivalence class.

Equation~\eqref{eq:FPTExactTower} is the low-momentum expansion of the full finite-$s_0$ one-loop amplitude, not a definition of the theory. Its coefficients are all fixed by the same endpoint. In the massless on-shell expansion, any two consecutive non-zero coefficients $g_n$ and $g_{n+1}$ determine $s_0$ through Eq.~\eqref{eq:FPTs0Identifiable}; all later ratios must then satisfy Eq.~\eqref{eq:FPTParameterFreeRatios}.

The perturbative distinction also cannot be erased by adding a finite higher-order correction. Writing
\begin{equation}
\Delta\mathcal M_{\rm on}
=
\lambda_R^2F_1(s,t;s_0)+\lambda_R^3F_2(s,t;s_0)+\order(\lambda_R^4),
\label{eq:FPTPerturbativePersistence}
\end{equation}
Eq.~\eqref{eq:FPTOnShellLeadingPhysical} gives a non-zero, non-constant $F_1$ for $s_0>0$. Equality of the two matched scattering amplitudes order by order would already require $F_1=0$, independently of the value of the finite function $F_2$. The explicit two-loop functional in Eq.~\eqref{eq:FPTTwoLoop} determines the next-order history domain and introduces no new endpoint parameter; a separate full two-loop $2\to2$ numerical evaluation is not used in establishing the one-loop distinction.}

\subsection{Differential renormalisation}

Differential renormalisation provides a useful comparison because it renders singular coordinate-space amplitudes finite without introducing a physical short-distance cutoff. Freedman, Johnson, and Latorre developed the method in massless field theory, and Haagensen and Latorre extended it to massive $\phi^4$ theory and QED \cite{FreedmanJohnsonLatorre1992,HaagensenLatorre1992}. Singular functions are replaced by derivatives of less singular distributions, introducing a renormalisation scale whose dependence is governed by ordinary RG equations. The existence of such schemes shows that ultraviolet finiteness, compact coordinate-space formulae, or preservation of Ward identities do not by themselves imply a minimum physical scale. In the present construction the distinctive claim must therefore be the residual, correlated $s_0$ dependence after local matching, not merely the fact that proper-time integrals are finite. The matched one-loop result sharpens this to a non-constant on-shell difference after the renormalisable matching freedom is exhausted.

\subsection{\texorpdfstring{\textcolor{paperred}{Finite spectral resolution and exact interacting completion}}{Finite spectral resolution and exact interacting completion}}

{\color{paperred}
The retained spectral boundary $s_0$ excludes the collapse of complete virtual circulations. It is not removed when the spectral or volume approximations are improved. This solves a different problem from summing the interaction expansion. The former is a statement about which virtual histories exist. The latter is a statement about whether the infinitely many interaction orders define one numerical function of $\lambda$.

The all-order causal construction in Eq.~\eqref{eq:causalSourceSeries} fixes every perturbative coefficient using the same $s_0$, the same local matching data, and the same sewing rule. It also keeps the local spacetime symmetry and the physical cut structure. A numerical all-coupling completion is obtained when the exact connected source kernel obeys the uniform remainder and physical-boundary bounds in Eqs.~\eqref{eq:scalarUniformFactorial} and \eqref{eq:scalarBoundaryBorelBound}. This distinction keeps the physical meaning of $s_0$ clear: increasing the number of spectral modes or enlarging the spatial volume improves the mathematical representation of the fixed-$s_0$ theory, while formally removing the spectral boundary through $s_0\to0^+$ lies outside the defined fixed-$s_0$ theory and returns a different spectral problem.
}

\section{\texorpdfstring{\textcolor{paperred}{Relation between the scalar analysis and finite-proper-time QED}}{Relation between the scalar analysis and finite-proper-time QED}}

{\color{paperred}
The scalar and spinor theories use the same complete-history principle but test different parts of it. The charged construction is developed independently in Ref.~\cite{BakrFPTQED2026}. The scalar theory isolates the spectral operator, graph topology, local sewing, and physical state reconstruction without a gauge field. The spinor theory adds spin, the photon sector, identities that relate charged vertices, and the long-range structure associated with massless photons.

The common statement is that the retained lower endpoint belongs to an independent complete virtual history before algebraic differentiation divides that history into pieces. In the scalar graph this statement becomes the circuit condition
\begin{equation}
\sum_{e\in c}s_e\ge s_0.
\end{equation}
In the charged theory the corresponding complete worldline carries one total proper time, and insertions divide it into non-negative daughter intervals without assigning a new endpoint to each interval. Separately complete internal photon histories carry their own proper-time variables.

The scalar analysis also clarifies what should be kept separate. The open Euclidean history kernel is useful virtual spectral data but is not the physical particle state kernel. Physical poles, cuts, and positive channel measures are imposed after complete amplitudes are formed. The same separation is useful in the charged theory. The scalar inequality $\Kop_{\phib}\ge\Kop_0$ makes the all-order Euclidean control particularly transparent here. The spinor case requires additional cancellations and identities because its fluctuation operator contains spin-dependent terms and its photon sector is massless. The scalar theory is therefore the clean control problem for the complete-history architecture developed in this paper.
}

\section{\texorpdfstring{\textcolor{paperred}{Discussion and conclusion}}{Discussion and conclusion}}

{\color{paperred}
The finite-proper-time scalar theory is most naturally understood as a restriction on complete virtual histories, not as a modification of every propagator. At one loop an open interval and a closed circle are two forms of the same proper-time history. Choosing an origin on the closed history and differentiating it only partitions its pre-existing total proper time. At higher loop order local momentum conservation identifies the primitive closed virtual circulations of the graph. They are the graph circuits. The retained domain is therefore
\begin{equation}
s_e\ge0,
\qquad
\sum_{e\in c}s_e\ge s_0
\quad(c\in\mathcal C(G)).
\end{equation}
An individual segment may approach zero. A bridge may also have arbitrarily small proper time because it carries no independent closed circulation. What the theory excludes is the collapse of a complete primitive virtual loop below the common spectral resolution $s_0$.

This distinction resolves several questions at once. The rule is independent of loop-momentum routing. It is not equivalent to assigning $\Rop_{s_0}$ to every internal line. It gives a positive quadratic form for every fixed loop graph and therefore finite massive amplitudes in simultaneous, nested, and overlapping ultraviolet regions. It is also compatible with connected composition because bridges remain ordinary propagators. Under a physical cut, every circuit intersected by a cut line is opened. The circuits that remain are exactly those of the deleted graph, so the cut factors are the same finite-$s_0$ subamplitudes that would be built independently. The pole residues and positive scalar phase-space measure are unchanged.

The same history rule can be used beyond isolated graphs. Fields are first smeared with smooth source functions. Local products are defined for separated insertion points and then extended to coincidence. The FPT circuit factor is bounded between zero and one and cannot make the local short-distance singularity worse than in the ordinary scalar theory. The only local freedom is therefore the finite set associated with the vacuum term, mass, field normalisation, and quartic coupling. Once these are fixed by physical conditions, the higher-derivative dependence on $s_0$ is correlated rather than freely adjustable. The history factors satisfy $\chi_{G_2}\chi_{G_1}=\chi_{G_2}$ whenever $G_1$ is contained in $G_2$. This makes interaction sewing associative and gives a consistent all-order source expansion. Because these factors depend only on scalar proper-time sums, they preserve the local spacetime transformation laws of the scalar field.

For $\lambda\ge0$ the real-background fluctuation operator obeys
\begin{equation}
\Kop_{\phib}\ge-\partial^2+m^2.
\end{equation}
The resulting heat-kernel bound is uniform in the background and gives explicit spatial suppression in every proper-time shell. The retained $s_0$ sets the lower spectral boundary of the short-history domain, while the mass controls the large-history end. A finite spectral-mode truncation is only a numerical approximation to this fixed-$s_0$ operator and converges at fixed admissible $s_0>0$.

The open Euclidean history kernel has a different role from the physical particle propagator. Although it is a positive Euclidean operator, it does not admit the positive energy-spectral representation required of the physical state kernel. We therefore keep the ordinary mass-shell particle sector and place the finite-history dependence in complete interaction amplitudes. Positivity is imposed only after the full Euclidean amplitudes have been continued to their physical boundary. The deletion identity then gives the ordinary positive on-shell phase space, and the complete boundary amplitudes define a positive inner product directly as a sum of absolute squares over physical channels. Smooth source derivatives of that positive kernel generate the interacting observable states.

Equation~\eqref{eq:causalSourceSeries} determines every interaction order at fixed $s_0$. If the exact connected source kernels also obey the uniform factorial remainder and physical-boundary bounds in Eqs.~\eqref{eq:scalarUniformFactorial} and \eqref{eq:scalarBoundaryBorelBound}, the expansion has one numerical sum at the measured coupling. Under those bounds the physical boundary can be taken after summation and the fixed-order cut identity becomes the corresponding summed unitarity relation.

The resulting physics is economical. The theory introduces one retained non-zero spectral boundary for complete virtual circulations. It leaves local spacetime symmetry, ordinary bridges, physical pole residues, and positive on-shell phase space intact. It predicts correlated momentum dependence after the usual local parameters have been matched. At one loop the first non-constant on-shell four-point difference is proportional to $s_0^2(s^2+t^2+u^2)$. A measurement of one process can therefore determine or constrain the non-zero spectral boundary $s_0$, while independent processes test the same spectral datum rather than introducing new coefficients.
}

\appendix

\section{Derivation of the six-field pairings}

For two cubic vertices, the six fields are grouped as $(a,b,c)$ and $(d,e,f)$. A Wick pairing can contain either three cross-contractions or one cross-contraction.

For three cross-contractions, choose the image of $a$ in three ways, the image of $b$ in two ways, and the last pairing is fixed:
\begin{equation}
3!=6.
\end{equation}
Every such pairing gives a term equivalent by symmetry to
\begin{equation}
G_{ad}G_{be}G_{cf}.
\end{equation}
For one cross-contraction, choose one field at the first vertex and one at the second:
\begin{equation}
3\times3=9.
\end{equation}
The two remaining fields at each vertex must contract locally, giving
\begin{equation}
G_{ab}G_{cf}G_{de}
\end{equation}
up to relabelling. Therefore
\begin{equation}
\vev{h_a h_b h_c h_d h_e h_f}
\longrightarrow
6G_{ad}G_{be}G_{cf}+9G_{ab}G_{cf}G_{de}
\end{equation}
after contraction with two fully symmetric cubic tensors.

\section{General connected-cumulant organisation}

The exact identity may be written formally as
\begin{align}
\Gamma[\phib]
={}&S[\phib]+\frac\hbar2\Tr\ln\Kop
-\hbar\ln\left\langle
\exp\left[-\sum_{n\ge3}\hbar^{n/2-1}\frac1{n!}S^{(n)}h^n
+\sum_{L\ge1}\hbar^{L-1/2}\Gamma_{L,i}h_i
\right]\right\rangle_0.
\end{align}
Expanding the logarithm in connected cumulants generates connected vacuum graphs, while the $\Gamma_{L,i}h_i$ terms recursively cancel one-particle-reducible contributions. This is the all-loop background-field organisation. In $\lambda\phi^4$ the bare derivatives $S^{(n)}$ vanish for $n>4$, but arbitrarily high loop orders arise from cumulants containing many cubic and quartic vertices.

\section{Ordinary Gaussian covariance model versus complete-history spectral model}

A mathematically conventional nonlocal scalar model can be defined by the Gaussian covariance
\begin{equation}
C_{s_0}=\Rop_{s_0}(\Kop_0)
\end{equation}
and local interaction $\lambda\phi^4/4!$. Its source functional is
\begin{equation}
Z^{\mathrm{cov}}_{s_0}[J]
=\exp\left[-\frac\lambda{4!}\hbar^3\int\dd^dx\frac{\delta^4}{\delta J(x)^4}\right]
\exp\left(\frac1{2\hbar}JC_{s_0}J\right).
\end{equation}
Every Wick line is independently modified. The corresponding free inverse covariance is
\begin{equation}
C_{s_0}^{-1}=\Kop_0\ee^{s_0\Kop_0},
\end{equation}
and its Gaussian determinant is
\begin{equation}
\frac12\Tr\ln C_{s_0}^{-1}
=\frac12\Tr\ln\Kop_0+\frac{s_0}{2}\Tr\Kop_0.
\end{equation}
This is not the complete-history closed functional $-\frac12\Tr\Eone(s_0\Kop_0)$. Their formal boundary-removal correspondence limits agree as $s_0\to0^+$ after the appropriate local subtractions; $s_0=0$ is not part of the defined fixed-$s_0$ FPT theory. This appendix makes explicit why ``modify every propagator'' and ``modify the complete heat-kernel history'' are distinct finite theories.

\section{Variational calculus as the continuous limit of linear algebra}

The variational calculus used above can be viewed as the continuum limit of ordinary multivariable linear algebra, with one field degree of freedom associated with each spacetime point. Coordinate arguments are treated as continuous labels rather than matrix indices
\begin{equation}
\mathcal{A}(x_i)=\mathcal{A}_i,
\qquad
\mathcal{B}(x_i,x_j)=\mathcal{B}_{ij}
\end{equation}
where $\mathcal{A}$ and $\mathcal{B}$ are two operators. In standard linear algebra, we take inner products by summing over a dummy index, for example
\begin{equation}
    (MN)_{ij}=\sum_k M_{ik}N_{kj}.
\end{equation}
The analogous operation in the continuous case would be
\begin{equation}
(\mathcal{M}\mathcal{N})(x_i,x_j)
=
\int \dd^4x_k\,
\mathcal{M}(x_i,x_k)\mathcal{N}(x_k,x_j)
\end{equation}
Partition four-dimensional Euclidean space into small hypercubes of volume $\delta V^{(4)}$ and label them by $i$. In the continuum limit,
\begin{equation}
    \int \dd^4x \rightarrow\sum \delta V^{(4)}
\end{equation}
Then
\begin{equation}
   (\mathcal{M}\mathcal{N})_{ik}
= \int \dd^4x_j\,
\mathcal{M}(x_i,x_j)\mathcal{N}(x_j,x_k)=\sum_j \delta V^{(4)}\mathcal{M}_{ij}\mathcal{N}_{jk}
\end{equation}
and
\begin{equation}
\delta(x_i-x_j) \rightarrow\frac{\delta_{ij}}{\delta V^{(4)}}
\end{equation}
\begin{equation}
    \frac{\delta}{\delta \phi(x_i)}\rightarrow\frac{1}{\delta V^{(4)}} \frac{\partial}{\partial \phi_i}
\end{equation}
The factors of $\delta V^{(4)}$ are bookkeeping factors associated with the chosen discretisation; they should not literally be set equal to one because $\delta V^{(4)}$ is dimensionful. Instead one may absorb appropriate powers of $\delta V^{(4)}$ into the definitions of the discrete vectors, matrices, and derivatives. The compact condensed-index rules are recovered when the continuum limit $\delta V^{(4)}\to0$ is taken with these scalings kept consistently.

As an example, in multivariable calculus we have
\begin{equation}
\sum_k\frac{\partial Q_i}{\partial P_k} \frac{\partial P_k}{\partial Q_j}=\delta_{ij}.
\end{equation}
The analogous object in the continuum limit, which can be read directly from the above prescription, would be
\begin{equation}
    \int \dd^4x_k\,\frac{\delta \mathcal{Q}(x_i)}{\delta \mathcal{P}(x_k)} \frac{\delta \mathcal{P}(x_k)}{\delta \mathcal{Q}(x_j)}=\delta(x_i-x_j).
\end{equation}
Likewise, in discrete theory we have
\begin{equation}
    \mathrm{Tr}(M)=\sum_jM_{jj}.
\end{equation}
In the continuum case we instead have
\begin{equation}
    \mathrm{Tr}(\mathcal{M})=\int \dd^4x\,\mathcal{M}(x,x).
\end{equation}

\section{\texorpdfstring{\textcolor{papergreen}{Momentum-space factors for one complete history}}{Momentum-space factors for one complete history}}
\label{app:oneHistoryMomentum}

{\color{papergreen}
For one complete history partitioned into $n$ Schwinger segments, define
\begin{equation}
\mathcal I_n(a_1,\ldots,a_n)
=
\int_{\substack{s_i\ge0\\s_1+\cdots+s_n\ge s_0}}
\left(\prod_{i=1}^n\dd s_i\right)
\exp\left(-\sum_{i=1}^n a_i s_i\right),
\qquad
\operatorname{Re}a_i>0.
\label{Eq.integrationIn}
\end{equation}
Insert a total proper time $t$,
\begin{equation}
f_n(t)
=
\int_{s_i\ge0}
\left(\prod_{i=1}^n\dd s_i\right)
\delta\left(t-\sum_{i=1}^ns_i\right)
\exp\left(-\sum_{i=1}^n a_i s_i\right),
\end{equation}
so that
\begin{equation}
\mathcal I_n=\int_{s_0}^\infty f_n(t)\,\dd t.
\end{equation}
Its Laplace transform is
\begin{equation}
\widetilde f_n(p)
=
\int_0^\infty\dd t\,\ee^{-pt}f_n(t)
=
\frac1{\prod_{i=1}^n(p+a_i)}.
\end{equation}
For pairwise distinct $a_i$, partial fractions give
\begin{equation}
f_n(t)
=
\sum_{i=1}^n
\frac{\ee^{-a_it}}
{\displaystyle\prod_{j\neq i}(a_j-a_i)},
\end{equation}
and therefore
\begin{equation}
\boxed{
\mathcal I_n(a_1,\ldots,a_n)
=
\sum_{i=1}^n
\frac{\ee^{-a_is_0}}
{a_i\displaystyle\prod_{j\neq i}(a_j-a_i)}.}
\label{eq:historyClosedForm}
\end{equation}

More useful for cutting is the exact decomposition obtained directly from the integration domain:
\begin{equation}
\boxed{
\mathcal I_n(a_1,\ldots,a_n)
=
\prod_{i=1}^n\frac1{a_i}
-
\mathcal Q_{n,s_0}(a_1,\ldots,a_n),}
\label{eq:appendixHistoryPoleEntire}
\end{equation}
where
\begin{equation}
\mathcal Q_{n,s_0}
=
\int_{\substack{s_i\ge0\\\sum_i s_i<s_0}}
\left(\prod_{i=1}^n\dd s_i\right)
\exp\left(-\sum_{i=1}^n a_i s_i\right).
\label{eq:appendixCompactRemainder}
\end{equation}
The excluded simplex in Eq.~\eqref{eq:appendixCompactRemainder} is compact. Hence $\mathcal Q_{n,s_0}$ is entire in all $a_i$. Equation~\eqref{eq:historyClosedForm}, initially written for distinct arguments, therefore has a unique meromorphic continuation to coincident arguments with only the ordinary poles $a_i=0$; the apparent factors $(a_j-a_i)^{-1}$ are removable in the full sum.

If $C\subseteq\{1,\ldots,n\}$ is non-empty, multiplication by the pole variables gives
\begin{equation}
\boxed{
\lim_{\{a_j\to0\}_{j\in C}}
\left(\prod_{j\in C}a_j\right)
\mathcal I_n(a_1,\ldots,a_n)
=
\prod_{j\notin C}\frac1{a_j}.}
\label{eq:oneHistoryMultipleResidue}
\end{equation}
Thus cutting any non-empty subset of segments of one complete history opens that history and removes its finite endpoint from the pole residue. This one-history identity is the local building block of the graph deletion identity in Appendix~\ref{app:circuitDerivations}.
}

\textcolor{red}{For a single complete circuit, the forbidden Schwinger-parameter region is always compact. Its contribution is therefore entire, so the finite-endpoint expression consists of the ordinary Feynman pole structure plus an entire correction and consequently has the same pole residues. For a graph containing several primitive-circuit constraints, however, the forbidden region need not be compact in every Schwinger direction. The excluded contribution may then itself contain propagator poles, so the residue of the full graph need not coincide with that of the corresponding unrestricted Feynman graph.}

\section{\texorpdfstring{\textcolor{papergreen}{Circuit topology, ultraviolet finiteness, and cutting}}{Circuit topology, ultraviolet finiteness, and cutting}}
\label{app:circuitDerivations}

{\color{papergreen}
This appendix derives the graph identities used in the main text. The graph is allowed to be a multigraph, so self-loops and parallel propagators are included.

\subsection{Loop space and the Schwinger quadratic form}

Let $G=(V,E)$ be connected and orient its internal edges. With $B$ the incidence matrix, local momentum conservation is
\begin{equation}
Bq+p=0.
\end{equation}
Choose one solution $q^{(0)}$. If the columns of the $|E|\times L$ matrix $Z$ form a basis of $\ker B$, every solution is
\begin{equation}
q=q^{(0)}+Z\ell,
\qquad
\ell\in(\mathbb R^d)^L,
\qquad
L=|E|-|V|+1.
\label{eq:generalMomentumRouting}
\end{equation}
Writing $S=\operatorname{diag}(s_e)$, the loop-dependent part of the Euclidean Schwinger exponent is
\begin{equation}
\ell^TM(s)\ell+\text{terms at most linear in }\ell,
\qquad
M(s)=Z^TSZ.
\label{eq:generalLoopMatrix}
\end{equation}
A change of loop basis $Z\mapsto ZA$, $A\in GL(L,\mathbb R)$, sends $M\mapsto A^TMA$ and does not change the graph domain.

{\color{paperred}\paragraph{Circuit inside every non-zero circulation.}
Every non-zero $z\in\ker B$ contains a graph circuit in its support. To see this, choose an edge with $z_e\neq0$. At each incident vertex, $Bz=0$ prevents the non-zero flow from terminating. Continue through non-zero edges. Finiteness of the graph forces a repeated vertex, which closes a circuit contained in the support.
\label{res:circulationContainsCircuit}}

{\color{paperred}\paragraph{Uniform loop-space positivity.}
Fix a graph $G$, a loop basis $Z$, and $s_0>0$. On the circuit domain $\mathcal D_G(s_0)$ there exists a constant $\kappa_G>0$ such that
\begin{equation}
\boxed{
M(s)\succeq \kappa_Gs_0\,I_L
\qquad
\text{for all }s\in\mathcal D_G(s_0).}
\label{eq:uniformLoopPositivity}
\end{equation}
Here $I_L$ is the $L\times L$ identity matrix.

\paragraph{Derivation.}
First prove strict positivity. If $x\neq0$ and $x^TM(s)x=0$, then
\begin{equation}
0
=
x^TZ^TSZx
=
\sum_{e\in E}s_e[(Zx)_e]^2.
\end{equation}
All terms are non-negative, so
\begin{equation}
(Zx)_e\neq0
\quad\Longrightarrow\quad
s_e=0.
\end{equation}
But $z=Zx$ is a non-zero circulation. By the circuit-in-support result above, its support contains a circuit $c$. Every edge of that circuit would then have $s_e=0$, giving
\[
\sum_{e\in c}s_e=0,
\]
contrary to the defining inequality $\sum_{e\in c}s_e\ge s_0$. Hence $M(s)\succ0$.

For uniformity, define the clipped parameters
\begin{equation}
\bar s_e=\min(s_e,s_0).
\end{equation}
If $s\in\mathcal D_G(s_0)$, then $\bar s\in\mathcal D_G(s_0)$. Indeed, on a given circuit either every $s_e<s_0$, so nothing changes, or at least one edge is clipped to exactly $s_0$, which by itself keeps the circuit sum at least $s_0$. Moreover
\[
M(s)\succeq M(\bar s).
\]
The set
\begin{equation}
K_G=
\left\{
\bar s\in\mathcal D_G(s_0):
0\le\bar s_e\le s_0
\right\}
\end{equation}
is compact. On $K_G\times S^{L-1}$, where $S^{L-1}$ is the unit sphere in $\mathbb R^L$, the continuous function
\[
(\bar s,x)\longmapsto x^TM(\bar s)x
\]
is strictly positive by the preceding argument. It therefore has a positive minimum, which has the form $\kappa_Gs_0$ after scaling all proper times by $s_0$. This gives Eq.~\eqref{eq:uniformLoopPositivity}.
}

For real Euclidean external momenta, completing the square in Eq.~\eqref{eq:generalLoopMatrix} leaves a non-negative external quadratic form. Therefore
\begin{equation}
\int_{\mathbb R^{dL}}\dd^{dL}\ell\,
\exp[-\ell^TM(s)\ell+\cdots]
\end{equation}
is bounded by a graph-dependent Gaussian proportional to
\begin{equation}
\exp[-\kappa_Gs_0|\ell|^2].
\end{equation}
For $m^2>0$, the factor $\exp[-m^2\sum_es_e]$ controls the large-proper-time region. Thus every fixed massive graph is ultraviolet finite. The same argument applies after a physical cut to every surviving connected component because the child graph obeys its own circuit inequalities.

\subsection{Bridge momenta are not loop variables}

{\color{paperred}\paragraph{Bridge criterion.}
For an internal edge $b$, the following statements are equivalent: $b$ is a bridge; $b$ belongs to no graph circuit; and every circulation $z\in\ker B$ has $z_b=0$. The first two statements are the standard graph definition of a bridge. If a circulation had $z_b\neq0$, following its non-zero support through the incident vertices would close a circuit containing $b$, which is impossible for a bridge. Conversely, an edge on a circuit supports the corresponding circuit circulation with non-zero coefficient on that edge.}
\label{res:bridgeCriterion}

The momentum on a bridge is fixed by local conservation. Removing $b$ separates the graph into vertex sets $V_L$ and $V_R$. Sum the vertex momentum-conservation equations over $V_L$. Every internal edge wholly inside $V_L$ occurs twice with opposite orientations and cancels. Only the bridge and the external or cut legs attached to $V_L$ remain:
\begin{equation}
\epsilon_b q_b
+
\sum_{a\in\partial V_L}\epsilon_a p_a
=0.
\label{eq:bridgeMomentumFixed}
\end{equation}
Hence $q_b$ is a function of the external and cut momenta. It may vary over physical cut phase space, but it is not an additional loop integration variable.

Because a bridge belongs to no circuit, its Schwinger parameter appears in no condition defining $\mathcal D_G(s_0)$ and
\begin{equation}
\int_0^\infty\dd s_b\,\ee^{-a_bs_b}
=
\frac1{a_b}.
\label{eq:appendixBridgeIntegral}
\end{equation}
Thus ordinary bridge propagation follows from the same circuit topology that sets the spectral boundary on closed loop directions.

\subsection{Deletion consistency}

Let $C\subseteq E$ be any set of edges. The circuit set satisfies
\begin{equation}
\boxed{
\mathcal C(G\setminus C)
=
\{c\in\mathcal C(G):c\cap C=\varnothing\}.}
\label{eq:appendixCircuitDeletion}
\end{equation}
Indeed, deleting edges cannot create a new closed path, and a parent circuit survives exactly when none of its edges is deleted.

For the graph Laplace kernel in Eq.~\eqref{eq:graphCircuitLaplace}, multiply by $\prod_{r\in C}a_r$ and set $t_r=a_rs_r$. The rescaled integral is Eq.~\eqref{eq:cutRescaledIntegral}. For almost every $t_r>0$, $s_r=t_r/a_r\to\infty$ as $a_r\to0^+$. Every circuit meeting $C$ then has total proper time tending to infinity and its step function tends to one. Every circuit disjoint from $C$ is unchanged. Equation~\eqref{eq:appendixCircuitDeletion} identifies the surviving factors with the standalone domain of $G\setminus C$. Since the step functions are bounded by one, dominated convergence gives the residue identity
\begin{equation}
\boxed{
\lim_{\{a_r\to0^+\}_{r\in C}}
\left(\prod_{r\in C}a_r\right)J_G(a)
=
J_{G\setminus C}(a_{E\setminus C}).}
\label{eq:appendixDeletionResidue}
\end{equation}
When $G\setminus C=G_L\sqcup G_R$, the surviving circuits are the disjoint union of $\mathcal C(G_L)$ and $\mathcal C(G_R)$, so
\begin{equation}
J_{G\setminus C}=J_{G_L}J_{G_R}.
\end{equation}
This establishes parent-independent composition at the proper-time level before the ordinary Lorentzian contour-pinch argument is applied.

\subsection{Worked examples}

\paragraph{One-loop three-segment history.}
A triangle or any three-segment descendant of one closed history has
\begin{equation}
s_1+s_2+s_3\ge s_0.
\end{equation}
Appendix~\ref{app:oneHistoryMomentum} gives
\begin{equation}
\mathcal I_3
=
\frac1{a_1a_2a_3}
-
\mathcal Q_{3,s_0}.
\end{equation}
Since $\mathcal Q_{3,s_0}$ is entire,
\begin{equation}
\lim_{a_1\to0}a_1\mathcal I_3
=
\frac1{a_2a_3},
\qquad
\lim_{a_1,a_2\to0}a_1a_2\mathcal I_3
=
\frac1{a_3}.
\label{eq:triangleCutExamples}
\end{equation}
After two cuts, the surviving edge is a tree bridge and is therefore ordinary.

\paragraph{Figure-eight.}
Each of the two internal edges is a self-loop circuit. Hence
\[
s_1\ge s_0,
\qquad
s_2\ge s_0.
\]
The loop matrix is diagonal,
\begin{equation}
M_{\mathrm{fig8}}
=
\begin{pmatrix}
s_1&0\\
0&s_2
\end{pmatrix}
\succeq s_0I_2.
\end{equation}
This reproduces the product of the two one-loop finite histories.

\paragraph{Sunset or theta graph.}
For three parallel edges between two vertices, choose
\begin{equation}
q_1=\ell_1,
\qquad
q_2=\ell_2,
\qquad
q_3=\ell_1+\ell_2.
\end{equation}
Then
\begin{equation}
M_{\theta}
=
\begin{pmatrix}
s_1+s_3&s_3\\
s_3&s_2+s_3
\end{pmatrix},
\qquad
\det M_{\theta}
=
s_1s_2+s_1s_3+s_2s_3.
\label{eq:thetaMatrixExample}
\end{equation}
The three primitive circuits give
\begin{equation}
s_1+s_2\ge s_0,
\qquad
s_1+s_3\ge s_0,
\qquad
s_2+s_3\ge s_0.
\end{equation}
Thus $s_1=0$, $s_2=s_3=s_0$ is allowed even though one edge has zero Schwinger length; the matrix remains positive:
\begin{equation}
M_{\theta}
=
s_0
\begin{pmatrix}
1&1\\
1&2
\end{pmatrix}.
\end{equation}

Cut edge $e_1$. The circuits $c_{12}$ and $c_{13}$ are opened and their endpoint factors disappear in the residue. The circuit $c_{23}$ survives, so the child domain is
\begin{equation}
s_2+s_3\ge s_0.
\end{equation}
This is exactly the standalone two-edge one-loop domain. If $e_1$ and $e_2$ are both cut, no circuit remains and Eq.~\eqref{eq:appendixDeletionResidue} gives
\begin{equation}
\lim_{a_1,a_2\to0}
a_1a_2J_{\theta}(a_1,a_2,a_3)
=
\frac1{a_3}.
\label{eq:thetaTwoCut}
\end{equation}

\paragraph{$K_4$ composition test.}
Let the parent graph have vertices $1,2,3,4$ and one edge between every pair. Cut the four edges crossing the partition
\begin{equation}
\{1,2\}\mid\{3,4\},
\qquad
C=\{13,14,23,24\}.
\end{equation}
The only uncut edges are $12$ and $34$. They are bridges in the deleted graph, and
\begin{equation}
\mathcal C(K_4\setminus C)=\varnothing.
\end{equation}
Every parent triangle or four-edge circuit intersects $C$, so every parent circuit endpoint becomes inactive in the pole residue. Therefore
\begin{equation}
\boxed{
\lim_{\{a_r\to0\}_{r\in C}}
\left(\prod_{r\in C}a_r\right)
J_{K_4}
=
\frac1{a_{12}a_{34}}.}
\label{eq:K4CutExample}
\end{equation}
The result is exactly the product of the two standalone tree bridge factors. No information about the parent graph is required once the cut is made.

\begin{figure}[H]
\centering
\begin{tikzpicture}[scale=1.25]
\node[circle,draw=papergreen,text=papergreen,minimum size=6mm] (v1) at (0,1.3) {$1$};
\node[circle,draw=papergreen,text=papergreen,minimum size=6mm] (v2) at (0,-1.3) {$2$};
\node[circle,draw=papergreen,text=papergreen,minimum size=6mm] (v3) at (3,1.3) {$3$};
\node[circle,draw=papergreen,text=papergreen,minimum size=6mm] (v4) at (3,-1.3) {$4$};
\draw[papergreen,thick] (v1)--(v2);
\draw[papergreen,thick] (v3)--(v4);
\draw[papergreen,dashed] (v1)--(v3);
\draw[papergreen,dashed] (v1)--(v4);
\draw[papergreen,dashed] (v2)--(v3);
\draw[papergreen,dashed] (v2)--(v4);
\node[text=papergreen] at (1.5,1.65) {cut};
\end{tikzpicture}
\caption{\textcolor{papergreen}{$K_4$ deletion test. The four dashed cross edges are cut. Every parent circuit uses at least one dashed edge. The two surviving solid edges are bridges, so the cut residue contains ordinary propagators $1/a_{12}$ and $1/a_{34}$.}}
\label{fig:K4Cut}
\end{figure}
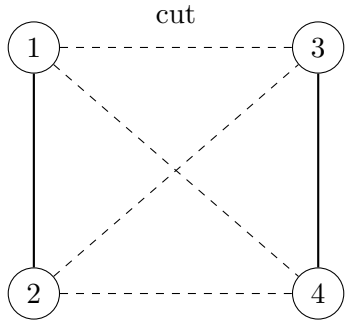

These examples illustrate the same general mechanism. The endpoint follows closed virtual circulation. Opening a circulation by a physical cut removes its endpoint from the residue, while every closed circulation that remains in a child graph retains precisely the same condition it has when that child is constructed independently.
}

\end{document}